%% file: MainTeX.tex
\documentclass[%
superscriptaddress,
nofootinbib,
 amsmath,amssymb,
 prd,
]{revtex4-2}
\usepackage{graphicx,array,makecell}
\usepackage[para,online,flushleft]{threeparttable}
\usepackage{fontawesome}
\usepackage{dcolumn}
\usepackage{bm}
\usepackage{hyperref}
\hypersetup{colorlinks=true
  ,urlcolor=blue
  ,anchorcolor=blue
  ,citecolor=blue
  ,filecolor=blue
  ,linkcolor=blue
  ,menucolor=blue
  ,linktocpage=true
}
\usepackage[mathlines]{lineno}
\usepackage{orcidlink}  

\graphicspath{{./}{Figs/}}

\begin{document}


\title{Cosmological Viability of Exponential Infrared $f(T)$ Gravity}

\author{Mahmoud Hashim\,\orcidlink{0000-0003-0691-9264}}\email{mahmoud.hashim@bue.edu.eg}
\affiliation{Centre for Theoretical Physics, The British University in Egypt, P.O. Box 43, El Sherouk City, Cairo 11837, Egypt}
\author{Eleonora Di Valentino\,\orcidlink{0000-0001-8408-6961}}\email{e.divalentino@sheffield.ac.uk}
\affiliation{School of Mathematical and Physical Sciences, University of Sheffield, Hounsfield Road, Sheffield S3 7RH, United Kingdom}
\author{Jackson Levi Said\,\orcidlink{0000-0002-7835-4365}}\email{jsaid01@um.edu.mt}
\affiliation{Institute of Space Sciences and Astronomy, University of Malta, Malta, MSD 2080 Department of Physics, University of Malta, Malta}
\author{Waleed El Hanafy\,\orcidlink{0000-0002-0097-6412}}\email{waleed.elhanafy@bue.edu.eg}
\affiliation{Centre for Theoretical Physics, The British University in Egypt, P.O. Box 43, El Sherouk City, Cairo 11837, Egypt}


\begin{abstract}
We investigate the cosmological viability of exponential infrared $f(T)$ teleparallel gravity using current cosmological observations. This framework realizes late-time cosmic acceleration through torsional modifications of gravity without enlarging the six-parameter cosmological parameter space of spatially flat $\Lambda$CDM, and admits two distinct solution branches: a phantom-like model (Model~I) and a model featuring a negative-to-positive transition in the effective torsional dark-energy density (Model~II). We constrain both branches using CMB observations from Planck, ACT, and SPT together with DESI BAO and Pantheon+ Type~Ia supernovae. We find that the principal branch (Model~I) alleviates the Hubble tension relative to $\Lambda$CDM, but remains statistically disfavoured by the combined dataset. The secondary branch (Model~II) is decisively ruled out. We show that the failure of Model~II originates from the interplay between background and perturbation constraints: once late-time distance measurements constrain the expansion history, the model becomes overconstrained, forcing correlated shifts in $\Omega_{\rm m}h^2$, $A_s$, $n_s$, and $\tau_{\rm reio}$, degrading the fit to the CMB damping tail and driving the optical depth to unphysical values. Our results demonstrate that perturbation observables provide stringent and complementary tests of teleparallel gravity beyond the background expansion history.
\end{abstract}

\maketitle

\section{Introduction}

The General Relativity (GR) theory of gravitation describes gravity as the manifestation of spacetime curvature generated by the distribution of matter and energy. Since its formulation, GR has successfully passed a wide range of experimental and observational tests, including Solar System experiments, the orbital decay of binary pulsars~\cite{Hulse:1974eb}, and, more recently, the direct detection of gravitational waves by the LIGO/Virgo collaboration~\cite{LIGOScientific:2016aoc}. Among its most remarkable predictions is the existence of black holes, whose astrophysical nature has been spectacularly confirmed by the first images of black-hole shadows obtained by the Event Horizon Telescope~\cite{EventHorizonTelescope:2019dse,EventHorizonTelescope:2019pgp}.

On cosmological scales, where the Friedmann--Lema\^{\i}tre--Robertson--Walker (FLRW) metric provides an accurate description of the large-scale Universe, GR supplemented by a cosmological constant ($\Lambda$) and cold dark matter (CDM) constitutes the standard $\Lambda$CDM cosmological model. This framework successfully reproduces the observed abundance of light elements through Big Bang Nucleosynthesis (BBN), while the discovery of the accelerated expansion of the Universe from observations of Type Ia supernovae by the High-$z$ Supernova Search Team~\cite{SupernovaSearchTeam:1998fmf} and the Supernova Cosmology Project~\cite{SupernovaCosmologyProject:1998vns} established the need for a dark-energy component, naturally identified with the cosmological constant.

Current observations of the cosmic microwave background (CMB), together with large-scale structure and Type Ia supernova data~\cite{Scolnic:2021amr,DES:2025sig,Hoyt:2026fve}, indicate that only about $\sim5\%$ of the total energy density of the Universe is in the form of baryonic matter, while the remaining $\sim95\%$ is composed of two poorly understood components: approximately $\sim25\%$ cold dark matter and $\sim70\%$ dark energy, the latter being consistent with an equation-of-state parameter $w_{\rm DE}=-1$ within the minimal $\Lambda$CDM model~\cite{Planck:2018vyg, Planck:2018nkj,ACT:2025fju,SPT-3G:2025bzu}. Despite its remarkable success across an impressive range of observations, the physical nature of both dark matter and dark energy remains unknown.

Moreover, the standard cosmological model faces several theoretical and observational challenges:

\begin{enumerate}

\item The nature of the dark sector remains unknown. While the existence of cold dark matter is strongly supported by cosmological and astrophysical observations, its particle nature has not yet been established. Weakly interacting massive particles (WIMPs), with masses in the GeV--TeV range, have long been regarded as among the leading dark-matter candidates~\cite{Jungman:1995df}. However, despite decades of experimental effort, no evidence for WIMPs has been found, and their allowed parameter space is becoming increasingly constrained by searches at the Large Hadron Collider (LHC)~\cite{Craig:2013cxa} and by direct-detection experiments~\cite{COSINE-100:2021xqn}. At the same time, the physical origin of the dark-energy component responsible for the late-time acceleration of the Universe remains one of the outstanding open questions in cosmology.

\item {\bf The $H_0$ tension:} CMB observations interpreted within the $\Lambda$CDM model predict $H_0 = 67.19 \pm 0.38$ km\,s$^{-1}$\,Mpc$^{-1}$~\cite{SPT-3G:2025bzu}, in significant ($\gtrsim5\sigma$) tension with local distance-ladder measurements~\cite{Freedman:2020dne,Birrer:2020tax,Riess:2021jrx,Anderson:2023aga,Scolnic:2023mrv,Jones:2022mvo,Anand:2021sum,Freedman:2021ahq,Uddin:2023iob,Huang:2023frr,Li:2024yoe,Pesce:2020xfe,Kourkchi:2020iyz,Schombert:2020pxm,Blakeslee:2021rqi,deJaeger:2022lit,Murakami:2023xuy,Breuval:2024lsv,Freedman:2024eph,Riess:2024vfa,Vogl:2024bum,Scolnic:2024hbh,Said:2024pwm,Boubel:2024cqw,Scolnic:2024oth,Li:2025ife,Jensen:2025aai,Riess:2025chq,Benisty:2025tct,Newman:2025gwg,Stiskalek:2025ktq,H0DN:2025lyy,Agrawal:2025tuv,Bhardwaj:2025kbw}. More recently, the Hubble Distance Network has reported $H_0 = 73.50 \pm 0.81$ km\,s$^{-1}$\,Mpc$^{-1}$, increasing the discrepancy to approximately $7\sigma$~\cite{H0DN:2025lyy} (see also Refs.~\cite{Verde:2019ivm,DiValentino:2020zio,DiValentino:2021izs,Perivolaropoulos:2021jda,Abdalla:2022yfr,DiValentino:2022fjm,Kamionkowski:2022pkx,Hu:2023jqc,Verde:2023lmm,DiValentino:2024yew,CosmoVerseNetwork:2025alb}).

\item {\bf Observational indications for evolving dark energy:} Recent cosmological observations provide increasing evidence that the late-time expansion history may depart from that expected in a Universe dominated by a cosmological constant. In particular, the latest DESI DR2 BAO measurements, when interpreted within dynamical dark-energy models and combined with complementary datasets such as CMB observations and/or Type Ia supernovae, consistently show a preference for an evolving dark-energy component at the $\sim3\sigma$ level~\cite{DES:2025sig,Hoyt:2026fve,DESI:2025zgx}. Although the exact significance depends on the assumed cosmological model and dataset combination, these results challenge the cosmological-constant interpretation of dark energy~\cite{DESI:2025fii,DESI:2024mwx,DESI:2025zgx,DESI:2024kob,Cortes:2024lgw,Shlivko:2024llw,Luongo:2024fww,Gialamas:2024lyw,Wang:2024dka,Ye:2024ywg,Tada:2024znt,Carloni:2024zpl,Chan-GyungPark:2024mlx,Bhattacharya:2024hep,Reboucas:2024smm,Najafi:2024qzm,Giare:2024gpk,Giare:2024ocw,Jiang:2024xnu,RoyChoudhury:2024wri,Giare:2024oil,Giare:2025pzu,Kessler:2025kju,RoyChoudhury:2025dhe,Scherer:2025esj,Wolf:2025jlc,Santos:2025wiv,Specogna:2025guo,Cheng:2025lod,Cheng:2025hug,Ozulker:2025ehg,Li:2025vuh,Lee:2025pzo,Fazzari:2025lzd,Smith:2025icl,Herold:2025hkb,Cheng:2025yue,Gokcen:2026pkq,Ishak:2025cay,Najafi:2026kxs,Yang:2026yaq,Kessler:2026dbi,Lee:2026yzs}.

\item {\bf The $A_{\rm lens}$ anomaly:} CMB observations exhibit a preference for a lensing amplitude approximately $3\sigma$ larger than expected within the $\Lambda$CDM model~\cite{Planck:2018vyg,SPT-3G:2025bzu}. In extended cosmological models, this anomaly can also manifest itself as an apparent preference for unphysical negative values of the total neutrino mass, $\sum m_\nu<0$~\cite{Craig:2024tky,Green:2024xbb,Elbers:2024sha,Elbers:2025vlz,Graham:2025dqn,Pulido-Hernandez:2026hcs,Yang:2026yaq,Loverde:2024nfi,Kibris:2026cqq}, highlighting possible internal inconsistencies within the standard framework.

\item {\bf Early structure formation:} Recent observations with the James Webb Space Telescope (JWST)~\cite{Adams_2022}, supported by spectroscopic observations from ALMA~\cite{2023MNRAS.519.5076B}, have revealed an unexpectedly abundant population of massive galaxies at high redshift. Although the interpretation of these observations is still under active debate, they may point towards more efficient early structure formation than predicted by the standard $\Lambda$CDM scenario.

\end{enumerate}

These theoretical and observational challenges motivate the exploration of alternatives beyond the standard $\Lambda$CDM model. One possibility is to extend the dark sector by introducing new dynamical components. Another, pursued in this work, is to modify gravity itself, so that the observed cosmic acceleration arises as an effective geometric phenomenon rather than from a fundamental dark-energy fluid. This approach is particularly appealing because both dark matter and dark energy are inferred primarily through their gravitational effects, suggesting that modifications of gravity may provide a more fundamental description of the dark sector.

Different cosmological tensions may point towards modifications of the expansion history at different cosmic epochs. For example, early-time extensions, such as Early Dark Energy~\cite{Poulin:2018cxd, Smith:2019ihp, Niedermann:2019olb, Krishnan:2020obg, Schoneberg:2021qvd, Ye:2021iwa,Poulin:2021bjr, Niedermann:2021vgd, Poulin:2023lkg, Vagnozzi:2023nrq, Efstathiou:2023fbn, Simon:2024jmu, Giare:2024akf, Giare:2024syw, Poulin:2024ken, Pedrotti:2024kpn, Kochappan:2024jyf,Poulin:2025nfb,Smith:2025zsg,Bella:2026zuk,Carloni:2026yut,Jhaveri:2026bla} or modifications at recombination~\cite{Hart:2017ndk, Hart:2019dxi, Sekiguchi:2020teg, Hart:2021kad, Lee:2022gzh, Chluba:2023xqj, Greene:2023cro, Greene:2024qis, Seto:2024cgo, Lynch:2024hzh, Toda:2024ncp, Schoneberg:2024ynd,Smith:2025uaq,GarciaEscudero:2025lef,Toda:2025kcq,Wang:2025dzn,Smith:2025icl,Garramone:2026evc}, have been extensively explored as possible solutions to the $H_0$ tension, whereas the recent DESI measurements have renewed interest in models that modify the late-time evolution of the Universe~\cite{DiValentino:2017iww,Krolewski:2024jwj, Yang:2018euj,Bousis:2024rnb, Tang:2024gtq, Jiang:2024xnu, Manoharan:2024thb, DiValentino:2019ffd, Specogna:2025guo,Ozulker:2025ehg,Lee:2025pzo,Benisty:2024lmj,Wang:2024vmw,Giare:2024ytc,Silva:2024ift,Li:2024qso,Pooya:2024wsq,Halder:2024uao,Silva:2025hxw,Yang:2025uyv,DiValentino:2020naf,vanderWesthuizen:2025rip,Zhang:2025dwu,Li:2025muv,Li:2025owk,Li:2026xaz,DiValentino:2020vnx,Zhai:2026uwr,Ruchika:2024ymt,Akarsu:2022typ,Akarsu:2021fol,Akarsu:2024qsi, Halder:2024uao, Pan:2019hac, Anchordoqui:2024gfa, Akarsu:2024eoo, Yadav:2024duq,Paraskevas:2024ytz,Gomez-Valent:2024tdb,Toda:2024ncp,Gomez-Valent:2024ejh, Akarsu:2025gwi, Souza:2024qwd,Soriano:2025gxd,Akarsu:2025ijk, Escamilla:2025imi,Bouhmadi-Lopez:2025ggl,Ghafari:2025eql}. The latter include dark-energy scenarios with interacting, evolving, emerging, phantom, phantom-crossing or sign-changing dark-energy behaviour. However, such phenomenology can be difficult to interpret if dark energy is regarded as a fundamental physical fluid. In contrast, modified gravity provides a natural framework in which these behaviours arise as effective geometric phenomena, without introducing exotic dark-energy components.

A particularly interesting example is provided by exponential infrared (IR) $f(T)$ teleparallel gravity, where $T$ denotes the torsion scalar. General aspects of $f(T)$ gravity are reviewed in Ref.~\cite{Bahamonde:2021gfp}, while the exponential infrared model was originally introduced in Ref.~\cite{Awad:2017yod}. The model reproduces General Relativity at early times while generating an effective dynamical dark-energy sector at late times through geometric torsion. Depending on the branch of the solution, it can produce either a phantom-like effective equation of state or a transition from negative to positive effective dark-energy density, without introducing additional cosmological parameters beyond those of flat $\Lambda$CDM. Previous studies investigated the background evolution of the model and confronted the principal branch with Planck CMB observations~\cite{Hashim:2020sez,Hashim:2021pkq}. In this work, we revisit the model using the latest cosmological observations from Planck, ACT, SPT, DESI DR2 and Type Ia supernovae, and perform the first observational study of the secondary Lambert-$W$ branch. A comprehensive review of $f(T)$ gravity and its cosmological applications, together with a broader discussion of current challenges in modern cosmology, can be found in Ref.~\cite{CosmoVerseNetwork:2025alb}.

The paper is organized as follows. In Sec.~\ref{Sec:fT_gravity}, we review the cosmological equations of $f(T)$ teleparallel gravity at both the background and linear perturbation levels. We then introduce the exponential infrared $f(T)$ model and discuss its phase-space structure, highlighting the two physically viable branches of the solution. In Sec.~\ref{Sec:Data}, we describe the observational datasets, the parameter space, and the statistical methodology adopted in our analysis. The observational constraints and parameter degeneracies are presented and discussed in Sec.~\ref{Sec:Results}. Finally, we summarize our main results and conclusions in Sec.~\ref{Sec:Conclusion}.

\section{Exponential Infrared $f(T)$ Teleparallel Gravity}\label{Sec:fT_gravity}

In a modified-gravity (MG) framework, the cosmological field equations can be recast into a form formally analogous to the Friedmann equations of General Relativity,
\begin{eqnarray}
H^2 &=& \frac{\kappa^2}{2}\left(\rho+\rho_{\rm MG}\right),\\
2\dot H+3H^2 &=& -\kappa^2\left(p+p_{\rm MG}\right),
\end{eqnarray}
where $\rho$ and $p$ denote the total energy density and pressure of the standard matter sector, including radiation, baryons, and cold dark matter, while $\rho_{\rm MG}$ and $p_{\rm MG}$ represent the effective contribution from the modified gravitational sector. The corresponding effective equation-of-state parameter is
\begin{equation}
w_{\rm MG}\equiv\frac{p_{\rm MG}}{\rho_{\rm MG}}.
\end{equation}

We first review the general cosmological equations of $f(T)$ teleparallel gravity and subsequently specialize to the exponential infrared model considered in this work.

\subsection{$f(T)$ cosmological equations}

The action of $f(T)$ teleparallel gravity is
\begin{equation}
\mathcal{S}=\frac{1}{2\kappa^2}\int d^{4}x\,|e|f(T)+\mathcal{S}_{M},
\label{action}
\end{equation}
where $T$ is the torsion scalar, $|e|=\sqrt{-g}=\det(e_\mu{}^a)$, $\kappa=\sqrt{8\pi G_N}$ with $G_N$ the Newtonian gravitational constant, and $\mathcal{S}_{M}$ is the action of the matter sector. Throughout this work we adopt natural units with $c=1$.

We consider a homogeneous, isotropic, and spatially flat cosmological background described by the FLRW geometry. We adopt the diagonal tetrad compatible with this symmetry, written in Cartesian coordinates $(t;x,y,z)$ as
\begin{equation}
   e_\mu{}^{a}=\mathrm{diag}\left(1,a(t),a(t),a(t)\right),
\label{eq:tetrad}
\end{equation}
where $a(t)$ is the cosmic scale factor. For the $f(T)$ equations of motion, this choice of tetrad has been shown to be consistent with the so-called Weitzenb\"{o}ck gauge, meaning that the teleparallel spin connection vanishes in this case \cite{Bahamonde:2021gfp}. The spacetime metric is obtained from the tetrad through
$g_{\mu\nu}=\eta_{ab}e^a{}_\mu e^b{}_\nu$, giving the spatially flat FLRW line element
\begin{equation}
   ds^2=-dt^{2}+a(t)^{2}\delta_{ij} dx^{i} dx^{j},
\label{FRW-metric}
\end{equation}
with $\eta_{ab}=(-,+,+,+)$. With the conventions adopted in this work, the corresponding torsion scalar is
\begin{equation}
T=6H^2 .
\label{eq:Torsion_sc}
\end{equation}

Varying the action~\eqref{action} with respect to the tetrad gives the field equations, which can be written in an effective Einstein-like form,
\begin{equation}
    \frac{1}{\kappa^2_{\rm eff}} \mathfrak{G}_{\mu\nu}
    =
    \mathfrak{T}^{(M)}_{\mu\nu}
    +
    \mathfrak{T}^{(DE)}_{\mu\nu},
\label{eq:field_eqns}
\end{equation}
where $\kappa^2_{\rm eff}=\kappa^2/f_T$, $f_T\equiv df/dT$, and $\mathfrak{T}^{(M)}_{\mu\nu}$ is the matter energy-momentum tensor. The contribution from the modified torsional sector is
\begin{equation}
\label{eq:torsion-Tmn}
\mathfrak{T}^{(DE)}_{\mu\nu}
=
\frac{1}{\kappa^2}
\left[
\frac{1}{2}g_{\mu\nu}\left(Tf_T-f\right)
-
f_{TT}S_{\nu\mu\rho}\nabla^{\rho}T
\right],
\end{equation}
where $f_{TT}\equiv d^2f/dT^2$ and $S_{\nu\mu\rho}$ is the teleparallel superpotential. This term encodes the deviation from TEGR and acts as an effective geometric dark-energy component.

For the tetrad choice in Eq.~\eqref{eq:tetrad}, the field equations reduce to the modified Friedmann and Raychaudhuri equations,
\begin{align}
\frac{3}{\kappa^2}H^2
&=
\rho_m+\rho_r+\rho_T
\equiv \rho_{\rm eff},
\label{eq:FR1T}\\
-\frac{1}{\kappa^2}\left(3H^2+2\dot H\right)
&=
p_r+p_T
\equiv p_{\rm eff}.
\label{eq:FR2T}
\end{align}
Here $\rho_T$ and $p_T$ denote the effective torsional energy density and pressure, while the matter component is assumed to be pressureless. Using $T=6H^2$, the torsional contribution can be written as
\begin{align}
\rho_T
&=
\frac{1}{2\kappa^2}
\left(6H^2-f+Hf_H\right),
\label{eq:Tor-density}\\
p_T
&=
-\frac{1}{6\kappa^2}\dot H\left(12+f_{HH}\right)-\rho_T,
\label{eq:Tor-press}
\end{align}
where $f=f(H)$, $f_H=df/dH$, and $f_{HH}=d^2f/dH^2$.

A useful property of $f(T)$ gravity is that the cosmological field equations remain second order. For a universe dominated by a fluid with equation-of-state parameter $w$, the background evolution can be written as a one-dimensional autonomous system~\cite{Awad:2013tha,Awad:2017yod},
\begin{equation}
\dot H
=
3(1+w)\frac{f-Hf_H}{f_{HH}}
\equiv \mathcal{F}(H).
\label{eq:phase_portrait_fT}
\end{equation}
This form is particularly useful for studying the phase-space structure of the theory.

The conservation equations for pressureless matter (including baryons and cold dark matter), radiation, and the effective torsional sector are
\begin{align}
\dot{\rho}_m+3H\rho_m &=0,
\label{eq:continuity_matter}\\
\dot{\rho}_r+4H\rho_r &=0,
\label{eq:continuity_radiation}\\
\dot{\rho}_T+3(1+w_T)H\rho_T &=0,
\label{eq:continuity_torsion}
\end{align}
where the effective torsion equation-of-state parameter is
\begin{equation}
w_T
=
\frac{p_T}{\rho_T}
=
-1+
\frac{\left(f_{HH}+12\right)\left(f-Hf_H\right)}
{f_{HH}\left(f-6H^2-Hf_H\right)}.
\label{eq:torsion_EoS}
\end{equation}
In deriving the above expression, we have used Eqs.~\eqref{eq:Tor-density}, \eqref{eq:Tor-press}, and \eqref{eq:phase_portrait_fT}.

It is also convenient to define the effective equation-of-state parameter of the Universe,
\begin{equation}
w_{\rm eff}
\equiv
\frac{p_{\rm eff}}{\rho_{\rm eff}}
=
-1-\frac{2}{3}\frac{\dot H}{H^2},
\label{eq:eff_EoS0}
\end{equation}
which is related to the deceleration parameter through
\begin{equation}
q
\equiv
-1-\frac{\dot H}{H^2}
=
\frac{1}{2}\left(1+3w_{\rm eff}\right).
\label{eq:deceleration}
\end{equation}

The above equations completely determine the background cosmological evolution in $f(T)$ gravity. To confront the theory with observations, however, one must also consider the evolution of linear cosmological perturbations, which govern the growth of large-scale structure and the anisotropies of the cosmic microwave background. We review the perturbation equations in the following subsection.

\subsection{$f(T)$ cosmological linear perturbation}

We formulate perturbation theory in conformal time, defined through $a(\tau)\,d\tau=dt$, such that the background vierbein is given by
${e_{\mu}}^{a}=a(\tau)\,\mathrm{diag}(1,1,1,1)$.

When analyzing perturbations, it is important to note that, in addition to the metric degrees of freedom, the vierbein possesses six extra degrees of freedom associated with local Lorentz transformations. These correspond to three boosts and three rotations, which can be decomposed into scalar, vector, and pseudoscalar modes. Nevertheless, the diffeomorphism invariance remains identical to that in GR, allowing one to fix two scalar modes and one vector mode, as in the standard case (see Ref.~\cite{Golovnev:2018wbh} for details).

For scalar perturbations we adopt the Newtonian gauge, while tensor modes are gauge invariant. As in GR, vector perturbations are typically negligible~\cite{Golovnev:2018wbh} and will not be considered here.

Within the Newtonian gauge, the scalar perturbations of the metric take the form
\begin{equation}
dS^2=a^2(\tau)\left[-(1+2\psi)d\tau^2+(1-2\phi)\delta_{ij}dx^idx^j\right],
\end{equation}
where $\psi$ and $\phi$ denote the scalar gravitational potentials and $x^i$ are comoving spatial coordinates.

The corresponding linear scalar perturbations of the vierbein can be written as
\begin{eqnarray*}
e^0_0 &=& a(\tau)(1+\psi),\\
e^0_i &=& a(\tau)\,\partial_i\zeta,\\
e^a_0 &=& a(\tau)\,\partial_a\zeta,\\
e^a_j &=& a(\tau)\left[(1-\phi)\delta^a_j+\epsilon_{ajk}\partial_ks\right],
\end{eqnarray*}
where $\zeta$ corresponds to the scalar part of a Lorentz boost, and $s$ represents the scalar component of spatial rotations. Compared to Ref.~\cite{Golovnev:2018wbh}, we directly impose the Newtonian gauge, neglect vector and tensor perturbations, and relabel the gravitational potentials as $(\phi,\psi)$ to match the convention adopted in Ref.~\cite{Ma:1995ey}.

It can be shown that $s$ does not contribute to the field equations at linear order. Therefore, in addition to the two standard scalar gravitational potentials, the only extra scalar degree of freedom is $\zeta$.

The evolution of $\zeta$ is determined by the antisymmetric part of the field equations. The purely spatial components are identically satisfied, leaving $s$ unconstrained, while the mixed components give
\begin{equation}
\label{asymm}
k^2 \zeta = 3\left(\phi' + {\cal H}\psi - \frac{{\cal H}' - {\cal H}^2}{{\cal H}}\phi\right),
\end{equation}
where ${\cal H}=aH$, the prime denotes differentiation with respect to conformal time $\tau$, and $k$ is the comoving wavenumber.

Deviations from GR are conveniently parameterized by introducing
$Q=1/f_T$ and $\Xi=12({\cal H}'-{\cal H}^2)Qf_{TT}$, with the GR limit recovered for $Q\rightarrow1$ and $\Xi\rightarrow0$.

The symmetric part of the field equations retains a structure similar to that of GR. The temporal component reads
\begin{equation}
\label{eq:pert1}
k^2\phi + 3{\cal H}(\phi' + {\cal H}\psi) + \frac{3{\cal H}^2 \Xi}{a^2}\phi = -4\pi Q G_N a^2 \delta\rho,
\end{equation}
where $\delta\rho$ denotes the total density perturbation, and Eq.~\eqref{asymm} has been used to eliminate $\zeta$.

Similarly, the mixed components become
\begin{equation}
\label{smix}
\phi' = 4\pi Q G_N \left(\frac{a}{k}\right)^2 (\rho + p)\theta - {\cal H}\psi - \frac{{\cal H}\Xi}{a^2}\phi,
\end{equation}
where $\theta$ is the divergence of the fluid velocity. Ref.~\cite{Golovnev:2018wbh} instead adopts the velocity potential $u$, related to our convention through $\theta=-k^2u$.

The off-diagonal spatial components yield
\begin{equation}
\label{eq:pert2}
\psi = \phi - 12\pi Q G_N \left(\frac{a}{k}\right)^2 (\rho + p)\sigma - {\cal H}\Xi\,\zeta,
\end{equation}
showing that $f(T)$ gravity can generate a gravitational slip even in the absence of anisotropic stress, $\sigma$. Combining Eqs.~\eqref{asymm} and \eqref{smix}, the additional scalar mode can be written as
\begin{equation}
\label{eq:grav_slip}
\zeta = \frac{3}{k^2}\left[\left(\frac{{\cal H}}{a^2}(a^2-\Xi)-\frac{{\cal H}'}{{\cal H}}\right)\phi
+4\pi Q G_N\left(\frac{a}{k}\right)^2(\rho+p)\theta\right].
\end{equation}

The gravitational slip is sourced by both the gravitational potential $\phi$ and the velocity divergence $\theta$. In the de Sitter limit, the contribution from the modified gravity sector becomes suppressed, recovering the standard GR behaviour.

For practical applications, it is convenient to rewrite the perturbation equations in terms of the effective functions $Q$, $\mu$, $\Sigma$, and $R$, which parameterize deviations from GR:
\begin{align}
-k^2 \phi &= 4 \pi G_N a^2  \sum_i \rho_i \Delta_i \, Q,\label{eq:Poisson-like}\\
-k^2 \psi &= 4 \pi G_N a^2  \sum_i \rho_i \Delta_i \, \mu +12 \pi G_N a^2 \sum_i \rho_i (1+w_i) \sigma_i \, Q,\label{eq:Poisson2-like}\\
-k^2\left(\psi-R \phi\right) &= 12\pi G_N a^2  \sum_i \rho_i (1+w_i) \left(\sigma_i+\frac{\mathcal{H}\Xi\theta_i}{k^2}\right) \, Q,\label{eq:gslip}\\
-k^2 (\phi+\psi)&= 8 \pi G_N a^2 \sum_i \rho_i \Delta_i \Sigma + 12 \pi G_N a^2 \sum_i \rho_i (1+w_i) \sigma_i Q,\label{eq:ISW}
\end{align}
where
\begin{align}
\Delta_i &= \delta_i + \frac{3\mathcal{H}(1+w_i)}{k^2}\theta_i,\label{eq:fTDelta}\\
\mu &= Q\Gamma,\qquad
\Gamma=\frac{\mu}{Q}=R+\frac{3\mathcal{H}\Xi\sum_i \rho_i (1+\omega_i)\theta_i}{k^2\sum_i \rho_i \Delta_i},\label{eq:fTmu}\\
\Sigma &= \frac{1}{2}(Q+\mu),\label{eq:fTSigma}\\
R &= 1+\frac{3\Xi}{k^2a^2}\left[\mathcal{H}^2\Xi+(\mathcal{H}'-\mathcal{H}^2)a^2\right].\label{eq:fTR}
\end{align}
We note that Eqs.~\eqref{eq:Poisson-like} and \eqref{eq:gslip} reduce to the generic parameterized framework introduced in Refs.~\cite{Bertschinger:2008zb,Bean:2010zq}, and subsequently adopted in Refs.~\cite{Dossett:2011tn,Dossett:2014oia,Dossett:2015nda}, apart from an additional relativistic correction proportional to the velocity divergence $\theta$. Strictly speaking, Eq.~\eqref{eq:Poisson-like} is not a Poisson equation, since it relates the spatial metric potential $\phi$ to the comoving overdensity $\Delta$, as already discussed in Ref.~\cite{Dossett:2014oia}. Instead, the potential $\psi$, which governs the motion of non-relativistic matter, satisfies Eq.~\eqref{eq:Poisson2-like}. We therefore refer to Eq.~\eqref{eq:Poisson2-like} as the modified Poisson equation. Consequently, the parameter $\mu$ quantifies deviations from GR in observables probing structure growth, such as galaxy clustering and redshift-space distortions.

Relativistic observables depend on both metric potentials. In particular, weak gravitational lensing is sensitive to the lensing potential $\phi+\psi$, while the Integrated Sachs--Wolfe (ISW) effect probes its time derivative. Accordingly, the parameter $\Sigma$ characterizes deviations from GR in these observables.

In the absence of anisotropic stress at late times, the perturbation equations reduce to
\begin{align}
-k^2 \phi &= 4 \pi G_N a^2  \rho_m \Delta_m \, Q,\label{eq:fs_Poisson-like}\\
-k^2 \psi &= 4 \pi G_N a^2  \rho_m \Delta_m \, \mu,\label{eq:fs_Poisson2-like}\\
-k^2\left(\psi-R \phi\right) &= 12\pi G_N a^2 \rho_m \frac{\mathcal{H}\Xi\theta_m}{k^2} \, Q,\label{eq:fs_gslip}\\
-k^2 (\phi+\psi)&= 8 \pi G_N a^2 \rho_m \Delta_m \Sigma.\label{eq:fs_ISW}
\end{align}

These equations show that $f(T)$ gravity predicts a non-vanishing gravitational slip even in the absence of anisotropic stress. On sub-horizon scales, where terms proportional to $k^{-2}$ become negligible, one finds
\begin{equation}
\Gamma=\frac{\mu}{Q}\simeq\frac{\psi}{\phi}\simeq R,
\qquad
\Sigma\simeq\frac{Q}{2}(1+R).
\end{equation}

In the GR limit, one has $Q\to1$, $\mu\to1$, $\Sigma\to1$, and $\Xi\to0$, which consequently implies $R\to1$ and $\Gamma\to1$. The perturbed Einstein equations are then recovered,
\begin{eqnarray}
-k^2 \phi &=& 4 \pi G_N a^2  \sum_i \rho_i \Delta_i ,\label{eq:GRPoisson-like}\\
-k^2 \psi &=& 4 \pi G_N a^2  \sum_i \rho_i \Delta_i +12 \pi G_N a^2 \sum_i \rho_i (1+w_i) \sigma_i,\label{eq:GRPoisson2-like}\\
-k^2\left(\psi-\phi\right) &=& 12\pi G_N a^2  \sum_i \rho_i (1+w_i) \sigma_i,\label{eq:GRgslip}\\
-k^2 (\phi+\psi)&=& 8 \pi G_N a^2 \sum_i \rho_i \Delta_i + 12 \pi G_N a^2 \sum_i \rho_i (1+w_i) \sigma_i.\label{eq:GRISW}
\end{eqnarray}

In the absence of anisotropic stress, these equations reduce to
\begin{eqnarray}
-k^2 \psi &=& 4 \pi G_N a^2  \sum_i \rho_i \Delta_i,\label{eq:fs_GRPoisson2-like}\\
\psi &=& \phi.\label{eq:fs_GRgslip}
\end{eqnarray}

For the exponential infrared $f(T)$ model considered here, deviations from GR are strongly suppressed at high Hubble rates, corresponding to the early Universe, where the model closely reproduces the standard cosmological evolution. Consequently, departures from GR become more relevant only at late times, affecting predominantly large-scale modes entering the horizon at relatively low redshift. Although these effects can, in principle, be sizeable, their observational detectability is expected to be limited by cosmic variance.

Finally, the diagonal spatial component of the perturbed field equations yields an equation involving the pressure perturbation $\delta p$. However, this equation is not independent, as it follows from the remaining perturbation equations through the Bianchi identities~\cite{Golovnev:2018wbh}.

Tensor perturbations are straightforward to derive and, for a perfect-fluid source, satisfy
\begin{equation}
h_{ij}^{\prime\prime}
+2{\cal H}\left(1+\frac{\Xi}{2a^2}\right)h_{ij}^{\prime}
+k^2h_{ij}=0.
\end{equation}
Compared to GR, the propagation speed of gravitational waves remains unchanged, while the friction term is modified through the background-dependent quantity $\Xi$.

To complete our review of linear perturbation theory, we adopt the standard perfect-fluid description for the matter sector,
\begin{equation}
\mathfrak{T}^{(M)\mu}{}_{\nu}
=
p\,g^\mu{}_\nu
+
(\rho+p)U^\mu U_\nu,
\end{equation}
where $\rho$, $p$, and
$U^\mu=dx^\mu/\sqrt{-ds^2}$ denote the energy density, pressure, and four-velocity of the fluid, respectively. Conservation of the matter energy-momentum tensor in Fourier space then yields the standard evolution equations~\cite{Ma:1995ey}
\begin{eqnarray}
\delta' &=& -(1+w)\left(\theta-3\phi'\right)
+3{\cal H}\left(w-\frac{\delta p}{\delta\rho}\right)\delta,\\
\theta' &=& -{\cal H}(1-3w)\theta
-\frac{w'}{1+w}\theta
+\frac{\delta p/\delta\rho}{1+w}k^2\delta
-k^2\sigma
+k^2\psi.
\end{eqnarray}
These equations apply to a single, uncoupled fluid.

\subsection{Phase portrait of the exponential infrared $f(T)$ gravity}

Motivated by phase-portrait analyses of viable cosmological models in which late-time acceleration arises from geometric contributions, the exponential infrared (IR) $f(T)$ model was proposed in Ref.~\cite{Awad:2017yod},
\begin{equation}
\label{eq:ExpIRfT}
f(T)=Te^{\beta(T_0/T)},
\end{equation}
where $T_0=6H_0^2$ and $\beta$ is a dimensionless parameter. At early times ($H\gg H_0$), the exponential factor approaches unity and the theory naturally reduces to GR, while deviations become relevant only at late times. The GR limit is recovered for $\beta=0$. Although no value of $\beta$ reproduces the $\Lambda$CDM model exactly, the latter emerges as the leading-order approximation of the theory in the infrared expansion.

Indeed, for $|\beta T_0/T|<1$, the action can be expanded as
\begin{equation}
\label{eq:fT_expansion}
f(T)\approx
T
+6\beta H_0^2
+18\beta^2H_0^4\left(\frac{1}{T}\right)
+36\beta^3H_0^6\left(\frac{1}{T^2}\right)
+O\!\left(\frac{1}{T^3}\right),
\end{equation}
where the inverse powers of $T$ represent genuine infrared corrections that become relevant only at late times, while they are strongly suppressed in the early Universe. By comparison, the $\Lambda$CDM model corresponds to
\[
f(T)=T+2\Lambda=T+6\Omega_\Lambda H_0^2,
\]
with $\Lambda=3\Omega_\Lambda H_0^2$. To first order in the infrared expansion, this corresponds to the identification
$\Lambda\simeq3\beta H_0^2$, implying an effective positive (negative) cosmological constant for $\beta>0$ ($\beta<0$). Higher-order corrections naturally generate an evolving effective dark-energy component arising from the gravitational sector, rather than from a fundamental physical fluid.

In this model, the modified Friedmann equation reads~\cite{Hashim:2020sez}
\begin{equation}
\left(E^2-2\beta\right)e^{\beta/E^2}
=
\Omega_{\rm m}(1+z)^3
+
\Omega_{\rm r}(1+z)^4,
\label{FR-E-exp-IR}
\end{equation}
where $E\equiv H/H_0$, while $\Omega_{\rm m}$ and $\Omega_{\rm r}$ denote the present-day matter and radiation density parameters, respectively.

We remark that the model parameter $\beta$ is completely determined by the present-day density parameters. This follows by evaluating Eq.~\eqref{FR-E-exp-IR} at the present epoch, $z=0$ (or equivalently $E=1$), yielding
\begin{equation}
\label{eq:ExpIRbeta}
\beta
=
\frac{1}{2}
+
W_k\!\left(
\frac{\Omega_{\rm m}+\Omega_{\rm r}}
{-2e^{1/2}}
\right),
\end{equation}
where $W_k(x)$ denotes the Lambert-$W$ function. Consequently, the exponential infrared $f(T)$ model does not introduce any additional free parameters with respect to the flat $\Lambda$CDM model.

\begin{figure}
    \centering
    \includegraphics[width=0.3\linewidth]{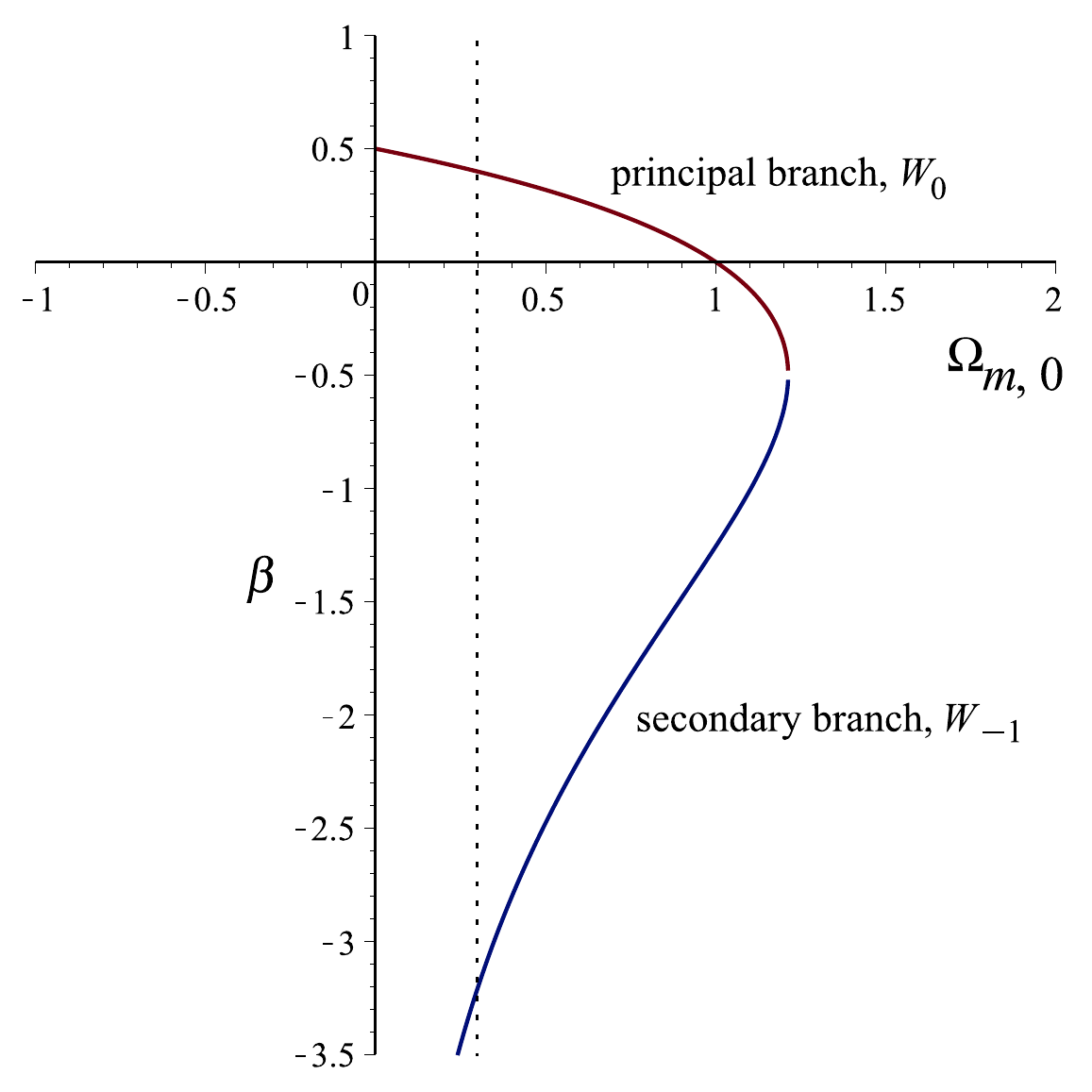}\hspace{1cm}
    \includegraphics[width=0.5\linewidth]{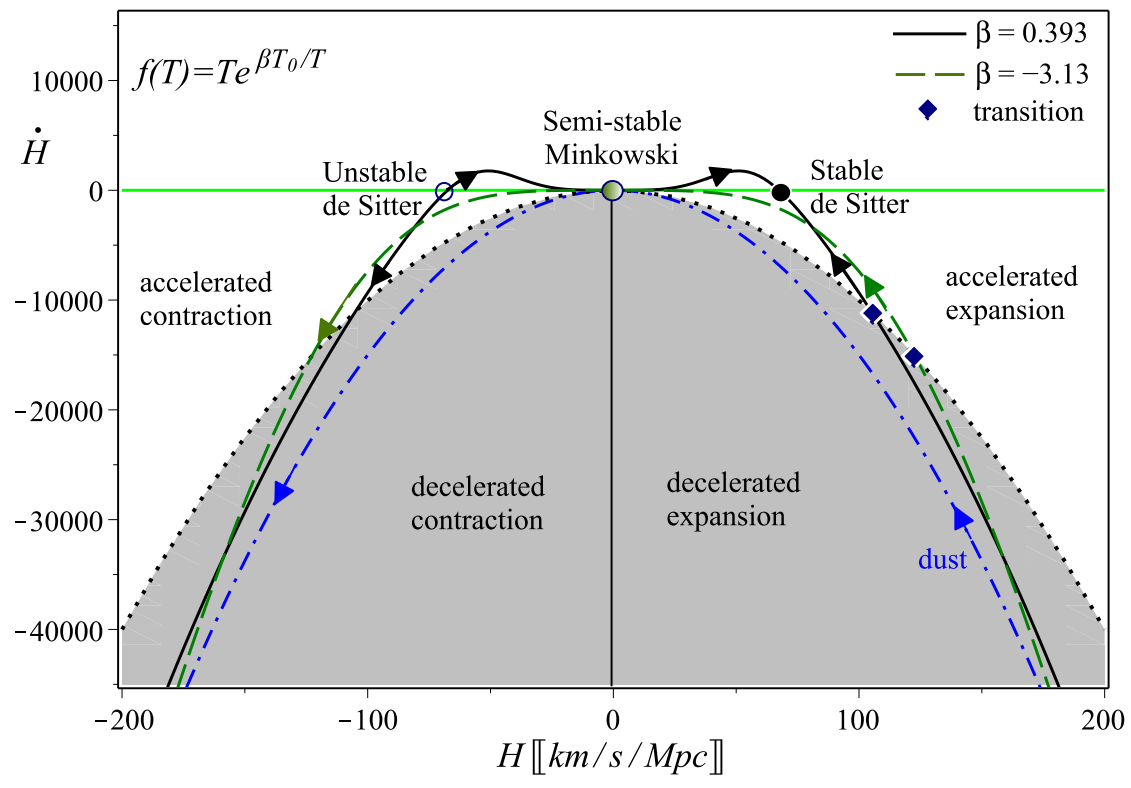}
    \caption{
\textit{Left panel:} The two branches of the Lambert-$W$ function in Eq.~\eqref{eq:ExpIRbeta} give rise to a double-valued relation in the $(\Omega_{\rm m},\beta)$ plane. For $\Omega_{\rm m}\simeq0.3$, the principal branch ($W_0$) gives $\beta\simeq0.4$, whereas the secondary branch ($W_{-1}$) gives $\beta\simeq-3.2$.
\textit{Right panel:} Phase portrait of the exponential infrared $f(T)$ model, Eq.~\eqref{eq:exp_fT_phase_portrait}, for the two Lambert-$W$ branches: $W_0$ ($\beta>0$) and $W_{-1}$ ($\beta<0$). The GR limit is recovered for $\beta=0$, corresponding to $\Omega_{\rm m}=1$, which coincides with the Einstein--de Sitter (sCDM) cosmology, i.e., a matter-dominated Universe with no dark-energy component.
}
    \label{fig:LambertW_phase_port1}
\end{figure}

The Lambert-$W$ function has two real branches relevant to the present analysis: the principal branch ($k=0$), denoted by $W_0$, and the secondary branch ($k=-1$), denoted by $W_{-1}$. Therefore, for a given value of $\Omega_{\rm m}$, the theory admits two distinct values of the parameter $\beta$. For example, adopting $\Omega_{\rm m}=0.318$ gives $\beta=0.393$ and $\beta=-3.127$ for the principal and secondary branches, respectively~\cite{Awad:2017yod}; see the left panel of Fig.~\ref{fig:LambertW_phase_port1}.

Substituting the exponential infrared $f(T)$ model, Eq.~\eqref{eq:ExpIRfT}, into the autonomous equation~\eqref{eq:phase_portrait_fT}, we obtain
\begin{equation}
\label{eq:exp_fT_phase_portrait}
\dot{H}
=
-\frac{3}{2}(1+w)
\frac{(H^2-2\beta H_0^2)H^4}
{H^4-\beta H_0^2H^2+2\beta H_0^4}.
\end{equation}
The corresponding phase portraits for the two Lambert-$W$ branches are shown in Fig.~\ref{fig:LambertW_phase_port1}. Both branches lead to viable late-time cosmological evolutions, exhibiting the transition from a decelerated to an accelerated expansion phase~\cite{Awad:2017yod}. Previous observational studies have focused exclusively on the principal branch ($W_0$), corresponding to $\beta>0$~\cite{Hashim:2020sez,Hashim:2021pkq}. However, the secondary branch ($W_{-1}$), corresponding to $\beta<0$, is also cosmologically viable according to the phase-space analysis, although it has not yet been confronted with cosmological observations. A theoretical investigation of this branch has recently been presented in Ref.~\cite{Akarsu:2024nas}. We emphasize that both branches are described by the same modified Friedmann equation, differing only in the value of the parameter $\beta$, which is determined by the chosen Lambert-$W$ branch through Eq.~\eqref{eq:ExpIRbeta}.

To complete the cosmological description of the exponential infrared $f(T)$ model, Eq.~\eqref{eq:ExpIRfT}, the modified Friedmann equations, Eqs.~\eqref{eq:FR1T} and \eqref{eq:FR2T}, can be written as~\cite{Awad:2017yod}
\begin{eqnarray}
\rho_m(H)&=&\frac{3}{\kappa^2}\left(H^2-2\beta H_0^2\right)e^{\beta \frac{H_0^2}{H^2}},\label{density}\\
p_m(H)&=&-\frac{2\dot{H}}{\kappa^2}\left[1-\beta\left(\frac{H_0}{H}\right)^2+2\beta^2\left(\frac{H_0}{H}\right)^4\right]e^{\beta \frac{H_0^2}{H^2}}-\rho_m(H).\label{pressure}
\end{eqnarray}

For the effective torsional dark-energy component, and restricting to the late-time matter-dominated Universe ($w=0$), substituting Eq.~\eqref{eq:exp_fT_phase_portrait} into Eqs.~\eqref{eq:Tor-density} and \eqref{eq:Tor-press} gives
\begin{eqnarray}
\rho_{T}(H)&=&\frac{3}{\kappa^2}\left[H^2-\left(H^2-2\beta H_0^2\right)e^{\beta \frac{H_0^2}{H^2}}\right],\label{rhoT}\\
p_{T}(H)&=&-\frac{3\beta H_0^2H^2}{\kappa^2}
\left[
\frac{H^2+2\beta H_0^2}
{H^4-\beta H_0^2H^2+2\beta^2H_0^4}
\right],\label{pTourmodel}
\end{eqnarray}
and consequently the effective torsional equation-of-state parameter, Eq.~\eqref{eq:torsion_EoS}, becomes
\begin{equation}
\label{Tor_EoS}
w_{T}(H)=
\frac{-\beta H_0^2H^2\left(H^2+2\beta H_0^2\right)}
{\left(H^4-\beta H_0^2H^2+2\beta^2H_0^4\right)
\left[H^2-\left(H^2-2\beta H_0^2\right)e^{\beta \frac{H_0^2}{H^2}}\right]}.
\end{equation}

Throughout the remainder of this work, we refer to the principal Lambert-$W$ branch as \textbf{Model I} and to the secondary branch as \textbf{Model II}. The observational properties of these two cosmological solutions are investigated separately in the following sections.

\subsubsection{Model I: Phantom dark energy model}

\begin{figure}
    \centering
    \includegraphics[width=0.3\linewidth]{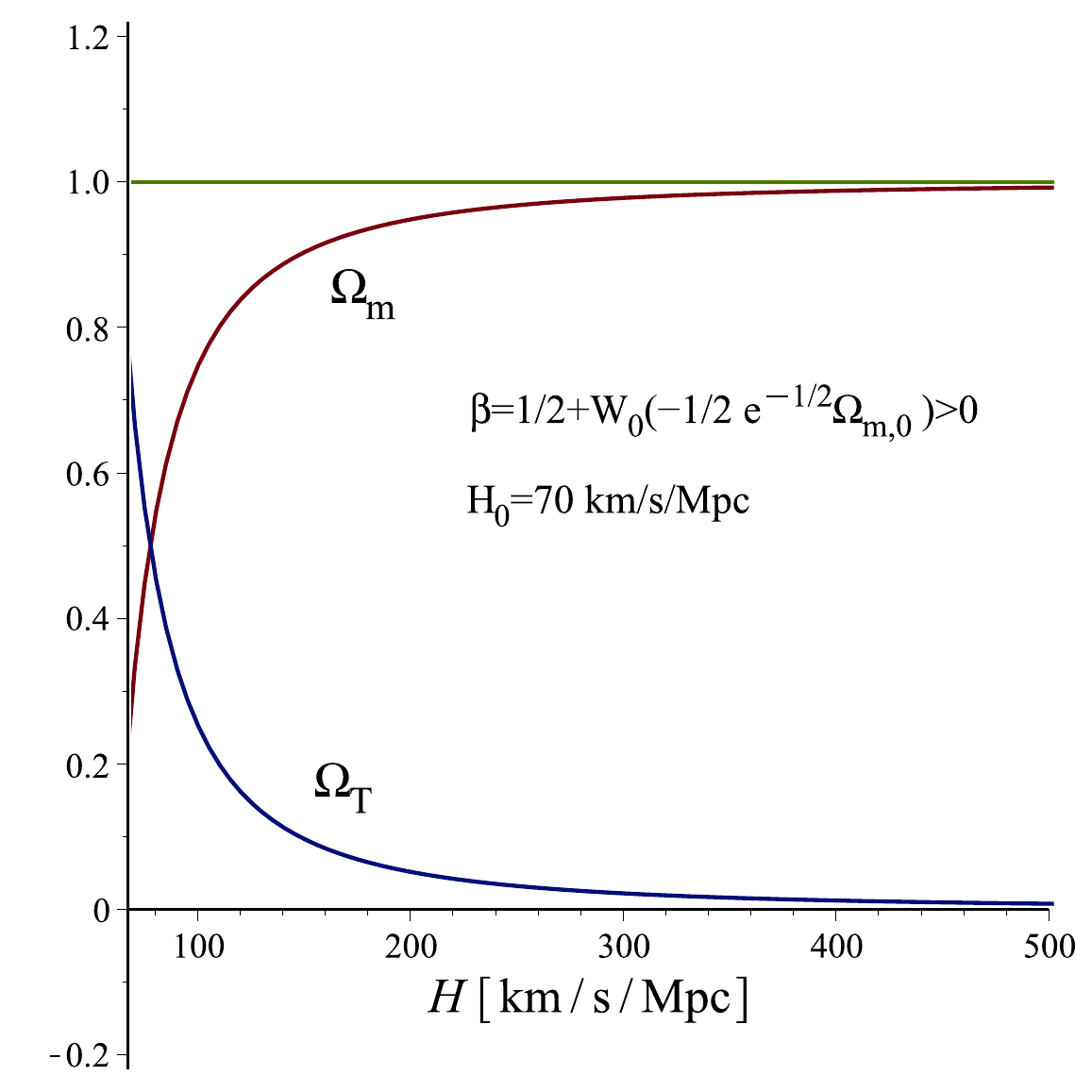}\hspace{1cm}
    \includegraphics[width=0.32\linewidth]{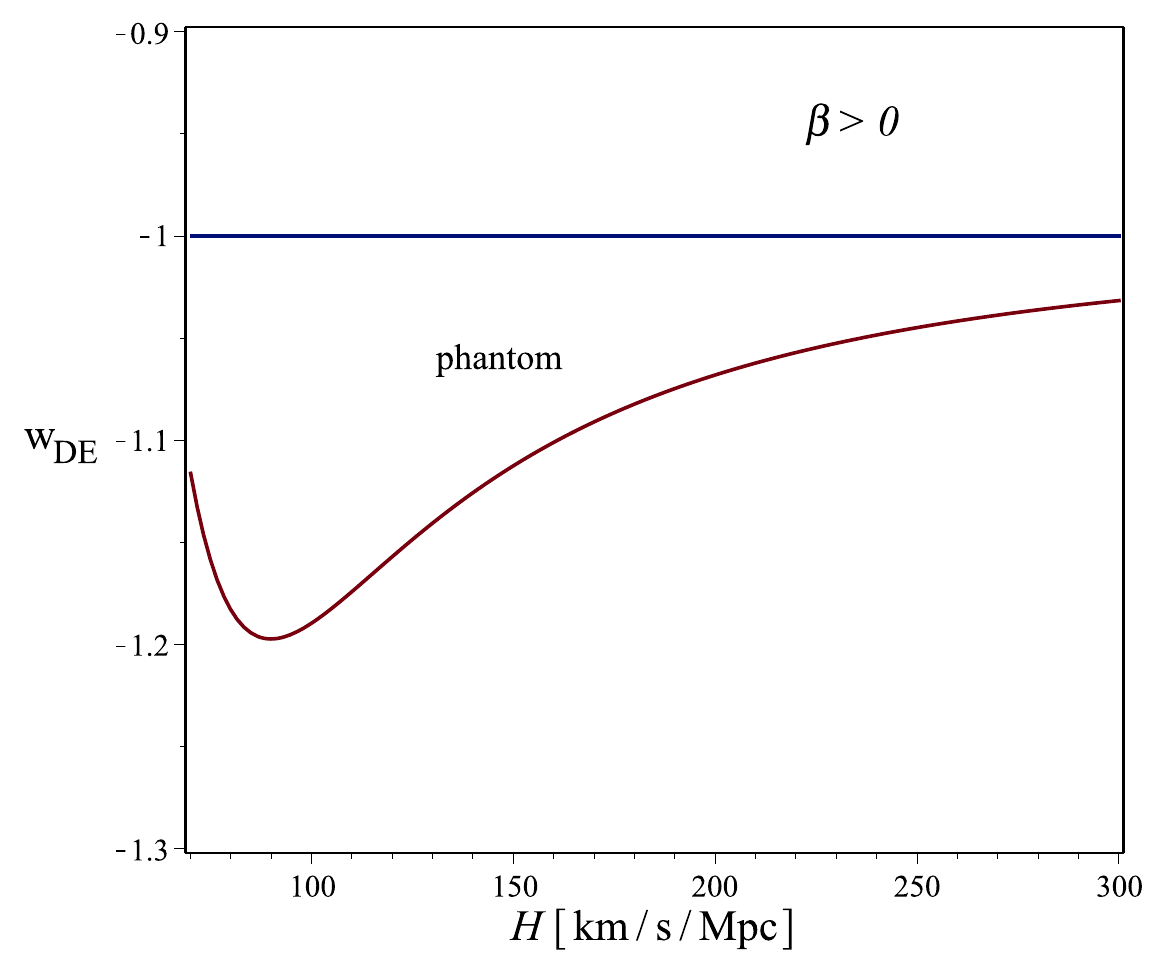}
    \caption{
\textbf{Model I.}
The left panel shows the evolution of the density parameters, while the right panel shows the evolution of the effective torsional equation-of-state parameter. The model exhibits an effective phantom behaviour at late times (small $H$), while asymptotically approaching a cosmological-constant behaviour ($w_T\rightarrow-1$) at early times (large $H$), where the effective torsional energy density vanishes and GR is recovered.
}
    \label{fig:modelI}
\end{figure}

It can be shown that, for the principal Lambert-$W$ branch, the exponential infrared $f(T)$ model behaves as an effective phantom dark-energy component at late times, while recovering the GR limit in the early Universe (large $H$); see Fig.~\ref{fig:modelI}. We emphasize that $w_T\rightarrow-1$ as $H\rightarrow\infty$, while both the effective torsional energy density and pressure satisfy $\rho_T\rightarrow0$ and $p_T\rightarrow0$. Thus, the recovery of GR at early times is associated with the vanishing of the effective torsional contribution, rather than with the value of its equation-of-state parameter.
Effective phantom behaviour has been shown to alleviate the $H_0$ tension by increasing the CMB-inferred value of the Hubble constant within extensions of the $\Lambda$CDM model~\cite{DiValentino:2016hlg,El-Zant:2018bsc}. Moreover, phantom-like evolution can increase the age of the Universe, thereby reducing potential tensions with the ages of the oldest Galactic globular clusters even for relatively large values of $H_0$~\cite{Cepa:2004bc}. In the present model, however, the effective phantom behaviour originates from the modified gravitational sector rather than from a physical dark-energy fluid, thereby avoiding the pathologies commonly associated with phantom-fluid models. Remarkably, as discussed above, the exponential infrared $f(T)$ theory introduces no additional free cosmological parameters beyond the six parameters of the spatially flat $\Lambda$CDM model.

\begin{figure}
    \centering
    \includegraphics[width=0.3\linewidth]{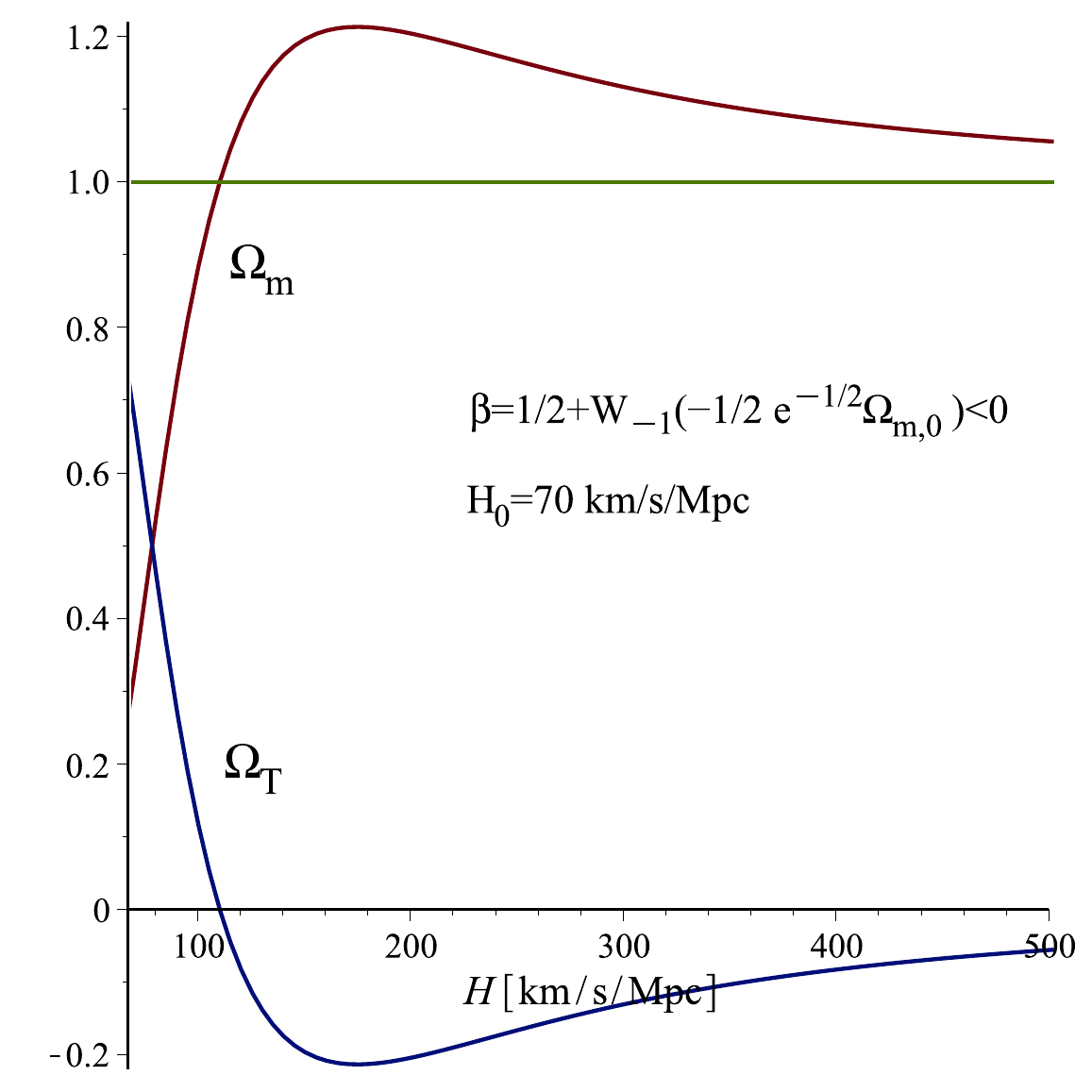}\hspace{1cm}
    \includegraphics[width=0.32\linewidth]{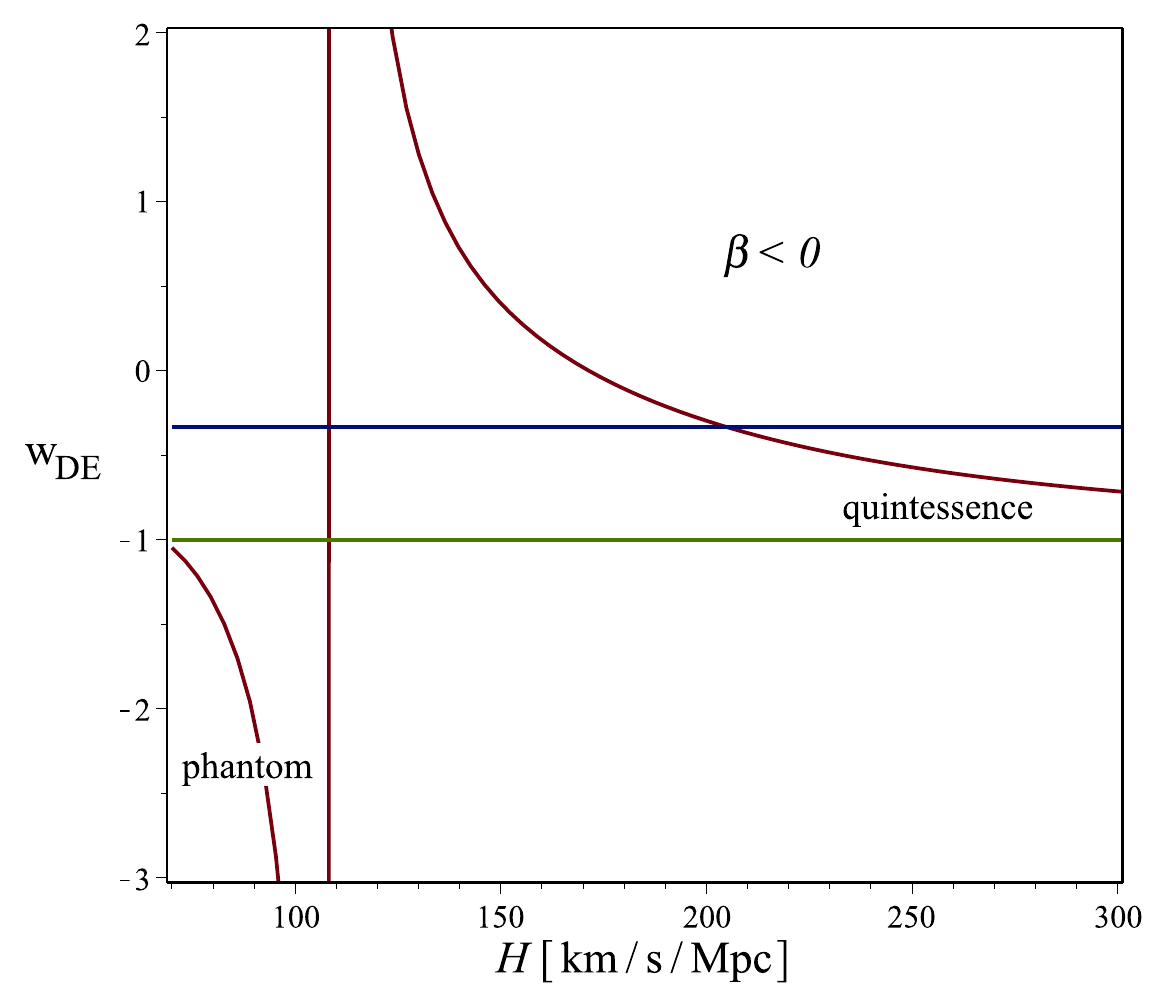}
    \caption{
\textbf{Model II.}
The left panel shows the evolution of the density parameters, while the right panel shows the evolution of the effective torsional equation-of-state parameter. The secondary branch exhibits a transition of the effective torsional energy density from negative to positive values at late times, accompanied by a pole in the effective equation of state. At early times, $w_T\rightarrow-1$, while the effective torsional energy density and pressure both vanish, recovering the GR limit.
}
    \label{fig:modelII}
\end{figure}

\subsubsection{Model II: Negative-to-positive dark energy density transition}

We note that this branch retains the same minimal six-parameter space as the spatially flat $\Lambda$CDM model. However, the effective torsional (dark-energy) density undergoes a transition from negative to positive values during the cosmological evolution. Similar to Model I, the effective torsional equation-of-state parameter satisfies $w_T\rightarrow-1$ at large $H$, while both $\rho_T$ and $p_T$ vanish. Therefore, the recovery of GR at early times is associated with the disappearance of the effective torsional contribution rather than with the value of its equation-of-state parameter.
At the transition, the effective torsional equation of state exhibits a pole, as shown in Fig.~\ref{fig:modelII}. This behaviour is not pathological, since $w_T$ is an effective equation-of-state parameter describing the geometrically induced dark-energy sector rather than the equation of state of a physical fluid. A detailed theoretical analysis of Model II can be found in Ref.~\cite{Akarsu:2024nas}, while in the present work we confront the secondary branch with cosmological observations for the first time.
Interestingly, a negative effective dark-energy density has recently attracted renewed attention in the context of model-independent dark-energy reconstructions and modified-gravity models. Such a behaviour has been suggested as a possible mechanism to enhance the growth of structure at high redshift, potentially alleviating the apparent overabundance of massive galaxies reported by JWST~\cite{Adams_2022,Menci:2022wia}. The secondary branch of the exponential infrared $f(T)$ model naturally realizes this phenomenology through a purely geometric mechanism.

In this work, we confront both branches of the exponential infrared $f(T)$ model---the principal branch $W_0$ ($\beta>0$) and the secondary branch $W_{-1}$ ($\beta<0$)---with the latest cosmological observations, as described in the next section.

\section{Observational Data and Methodology}\label{Sec:Data}

We perform a Bayesian parameter inference using the \texttt{Cobaya} package.\footnote{\href{https://cobaya.readthedocs.io/en/latest/}{https://cobaya.readthedocs.io/en/latest/}.} The theoretical predictions, including the CMB temperature and polarization power spectra, the matter power spectrum, cosmological distances, and the background expansion history, are computed with a modified version of the Boltzmann solver \texttt{CLASS}.\footnote{The modified version of \texttt{CLASS} used in this work is publicly available at \href{https://github.com/mwhashim/class_tmg}{\faGithub} \url{https://github.com/mwhashim/class_tmg}.}

We assume a spatially flat Universe containing baryons, cold dark matter, radiation, and an effective dark-energy component arising from the teleparallel $f(T)$ modification of gravity. Since the exponential infrared $f(T)$ model does not introduce any additional free cosmological parameters with respect to the spatially flat $\Lambda$CDM model, the parameter space is described by the six standard cosmological parameters,
\begin{equation}
\left\{
\Omega_{\rm b}h^2,\,
\Omega_{\rm c}h^2,\,
H_0,\,
\tau_{\rm reio},\,
\ln(10^{10}A_s),\,
n_s
\right\},
\end{equation}
where $\Omega_{\rm b}h^2$ and $\Omega_{\rm c}h^2$ denote the physical baryon and cold-dark-matter densities, respectively, $H_0$ is the Hubble constant today, $\tau_{\rm reio}$ is the optical depth to reionization, $A_s$ is the amplitude of the primordial scalar perturbations at the pivot scale $k=0.05~{\rm Mpc}^{-1}$, and $n_s$ is the corresponding scalar spectral index. Flat priors are adopted for all cosmological parameters, with the prior ranges summarized in Table~\ref{tab:pprior}.

\begin{table}[htp]
    \centering
    \caption{Prior ranges adopted in the Bayesian analysis.}
    \begin{tabular}{|c|c|}
    \Xhline{1.1pt}
    Parameter & Prior distribution \\
    \hline
         $\Omega_{\rm b}h^2$ & ${\cal U}(0.005,\,0.1)$ \\
         $\Omega_{\rm c}h^2$ & ${\cal U}(0.001,\,0.99)$ \\
         $H_0$ & ${\cal U}(40,\,100)$ \\
         $\log(10^{10}A_s)$ & ${\cal U}(1.61,\,3.91)$ \\
         $n_s$ & ${\cal U}(0.8,\,1.2)$ \\
         $\tau_{\rm reio}$ &
         \makecell[l]{Fixed to $0.06$ for late-time-only analyses;\\
         otherwise ${\cal N}(\mu=0.051,\sigma=0.006)$} \\
         \Xhline{1.1pt}
    \end{tabular}
    \label{tab:pprior}
\end{table}

For parameter sampling, we employ a Markov-Chain Metropolis-Hastings (MCMC) sampler with a convergence stopping criterion based on the Gelman--Rubin statistic ($R-1$), requiring $R-1\le0.02$ for four parallel chains. We consider the following data sets:

\begin{itemize}

\item \textbf{Cosmic Microwave Background (CMB):} This data set combines early-Universe observations from \textit{Planck}, the Atacama Cosmology Telescope (ACT), and the South Pole Telescope (SPT), comprising:
\begin{enumerate}
    \item \textbf{Planck low-T}: Planck-2018 low-$\ell$ temperature measurements based on the \texttt{Commander} likelihood, accounting for the TT power spectrum over the multipole range $2\le\ell<30$~\cite{Planck:2019nip}.

    \item \textbf{Planck (TT-TE-EE)}: Planck-2018 high-$\ell$ temperature and polarization measurements based on the \texttt{Plik} likelihood, using the TT, TE, and EE spectra with cuts at $\ell_{\max}=(1000,600,600)$, respectively~\cite{Planck:2019nip}.

    \item \textbf{ACT (TT-TE-EE)}: ACT DR6 high-multipole temperature and polarization measurements of the TT, TE, and EE spectra over the multipole range $\ell\ge600$, using the baseline foreground-marginalized likelihood~\cite{ACT:2025fju}.

    \item \textbf{SPT (TT-TE-EE)}: SPT-3G D1 high-multipole temperature and polarization measurements of the TT, TE, and EE spectra over the multipole range $\ell\ge400$, with TT measurements extending up to $\ell_{\max}=3000$ and TE/EE measurements up to $\ell_{\max}=4000$~\cite{SPT-3G:2025bzu}.

    \item \textbf{Planck-ACT-SPT Lensing}: Combined CMB lensing information obtained from the ACT DR6 lensing likelihood~\cite{ACT:2023dou,ACT:2023kun}, together with the Planck lensing likelihood~\cite{Carron:2022eyg} and the SPT-3G Year-2 \texttt{MUSE} lensing likelihood~\cite{SPT-3G:2024atg}.

    \item \textbf{Gaussian $\tau$ prior}: A Gaussian prior on the optical depth to reionization,
    $\tau_{\rm reio}=0.051\pm0.006$.
\end{enumerate}
We refer to this combined data set as \textbf{CMB-SPA}~\cite{SPT-3G:2025bzu}.

    \item \textbf{Baryon Acoustic Oscillations (BAO):} We use the Dark Energy Spectroscopic Instrument Data Release 2 (DESI DR2) BAO measurements~\cite{DESI:2025zgx}, hereafter referred to as \textbf{DESI}. This dataset provides sub-percent precision measurements of the BAO distance scale over the redshift range $0.1<z<2.1$, using more than 14 million galaxies, quasars, and Lyman-$\alpha$ forest tracers. These measurements provide powerful constraints on the late-time expansion history of the Universe and on the properties of dark energy.

    \item \textbf{Type Ia Supernovae (SNIa):} We use the Pantheon+ compilation~\cite{Brout:2022vxf}, hereafter referred to as \textbf{PP}, consisting of 1701 light curves from 1550 spectroscopically confirmed Type Ia supernovae collected from 18 different surveys. Pantheon+ spans the redshift range $0<z\lesssim2.3$, providing precise measurements of the luminosity-distance relation from the nearby Universe to cosmological distances. We also consider the Cepheid-calibrated Pantheon+ sample based on the SH0ES distance ladder~\cite{Riess:2021jrx}, hereafter referred to as \textbf{PPS}.

\end{itemize}

We consider five different combinations of these datasets. Two combinations probe the late Universe only, namely PP+DESI and PPS+DESI. The early-Universe dataset is represented by CMB-SPA. Finally, we consider two combinations including both early- and late-Universe observations, namely CMB-SPA+PP+DESI and CMB-SPA+PPS+DESI.

\section{Results and Discussion}\label{Sec:Results}

\begin{figure}
    \centering
    \includegraphics[width=0.8\linewidth]{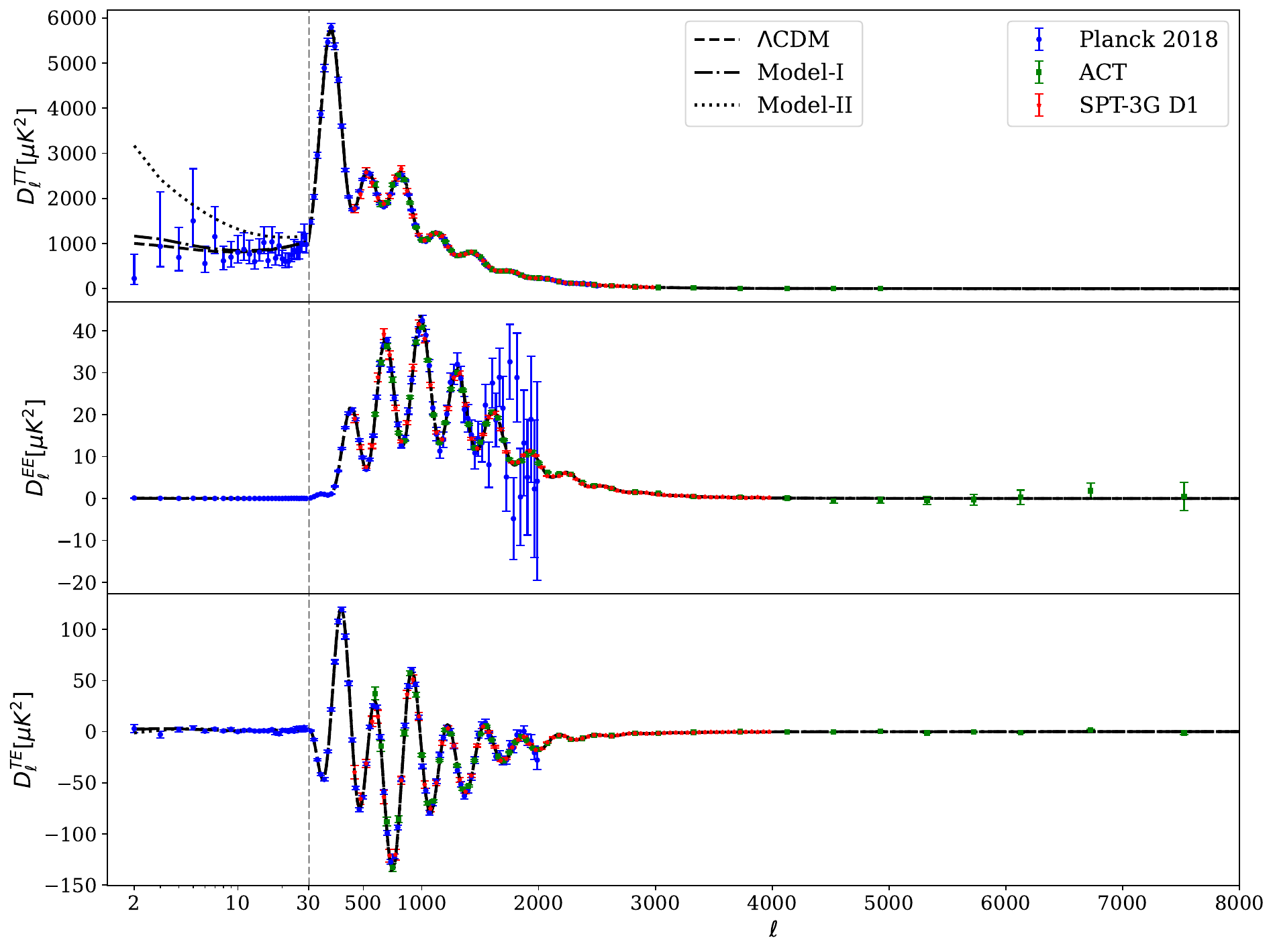}
    \caption{The TT, TE, and EE angular power spectra of the three models compared with the CMB measurements from \textit{Planck} 2018, ACT DR6, and SPT-3G. The dashed, dash-dotted, and dotted curves correspond to $\Lambda$CDM, Model I, and Model II, respectively. Model I remains nearly indistinguishable from $\Lambda$CDM, showing only a mild enhancement of the low-$\ell$ TT power spectrum, consistent with its quasi-de Sitter evolution at late times. In contrast, Model II predicts a significantly larger excess at low multipoles, driven by an enhanced late Integrated Sachs--Wolfe (ISW) contribution.}
    \label{fig:Plk18Cll}
\end{figure}

We show in Fig.~\ref{fig:Plk18Cll} the TT, TE, and EE angular power spectra predicted by the best-fit parameters obtained from the CMB-SPA dataset for $\Lambda$CDM, Model I, and Model II. As expected, the $\Lambda$CDM prediction is in excellent agreement with the \textit{Planck} measurements~\cite{Planck:2019nip}. The power spectra predicted by Model I remain very close to those of $\Lambda$CDM, with the only noticeable difference being a slight enhancement of the TT power spectrum at the largest angular scales ($\ell\lesssim30$). Since the exponential infrared $f(T)$ model reproduces GR at early times, this deviation is entirely a late-time effect. It originates from the modified perturbation equations, where the gravitational slip induced by the $f(T)$ modification enhances the late Integrated Sachs--Wolfe (ISW) contribution. However, this enhancement remains observationally consistent because of the large cosmic variance affecting these multipoles.

In contrast, Model II predicts a much larger enhancement of the low-$\ell$ TT power spectrum, leading to a clear excess with respect to the observations. As in Model I, this behaviour is entirely associated with the late-time ISW effect rather than modifications of the primordial CMB anisotropies. In this case, however, the gravitational slip becomes sufficiently large on super-horizon scales, as indicated by Eq.~\eqref{eq:grav_slip}, preventing the evolution from approaching the quasi-de Sitter behaviour required to suppress the ISW contribution. Consequently, Model II provides a significantly poorer fit to the CMB temperature anisotropies, which is reflected in the goodness-of-fit statistics discussed below.

We obtain the observational constraints on the cosmological parameters at 68\% CL for the three models, as reported in Tables~\ref{tab:LCDMTab}--\ref{tab:model2Tab}. Since all three models are described by the same cosmological parameter space, we separate the parameters into two categories: the six independent cosmological parameters sampled in the MCMC analysis and the corresponding derived parameters. For each dataset combination, we also compute the minimum $\chi^2$, and define the relative goodness-of-fit with respect to $\Lambda$CDM as
\begin{equation}
\Delta\chi^2=\chi^2_{\rm model}-\chi^2_{\Lambda{\rm CDM}}.
\end{equation}
In Figs.~\ref{fig:lcdm}--\ref{fig:model2}, we present the marginalized 68\% and 95\% confidence regions for the cosmological parameters obtained from the five dataset combinations considered in this work. These comparisons illustrate the robustness of the parameter constraints and highlight the different degeneracy directions exhibited by the three models. In the following, we discuss the results for each model separately.

\subsection{$\Lambda$CDM}\label{Sec:Results_LCDM}

\begin{figure}
    \includegraphics[width=0.4\linewidth]{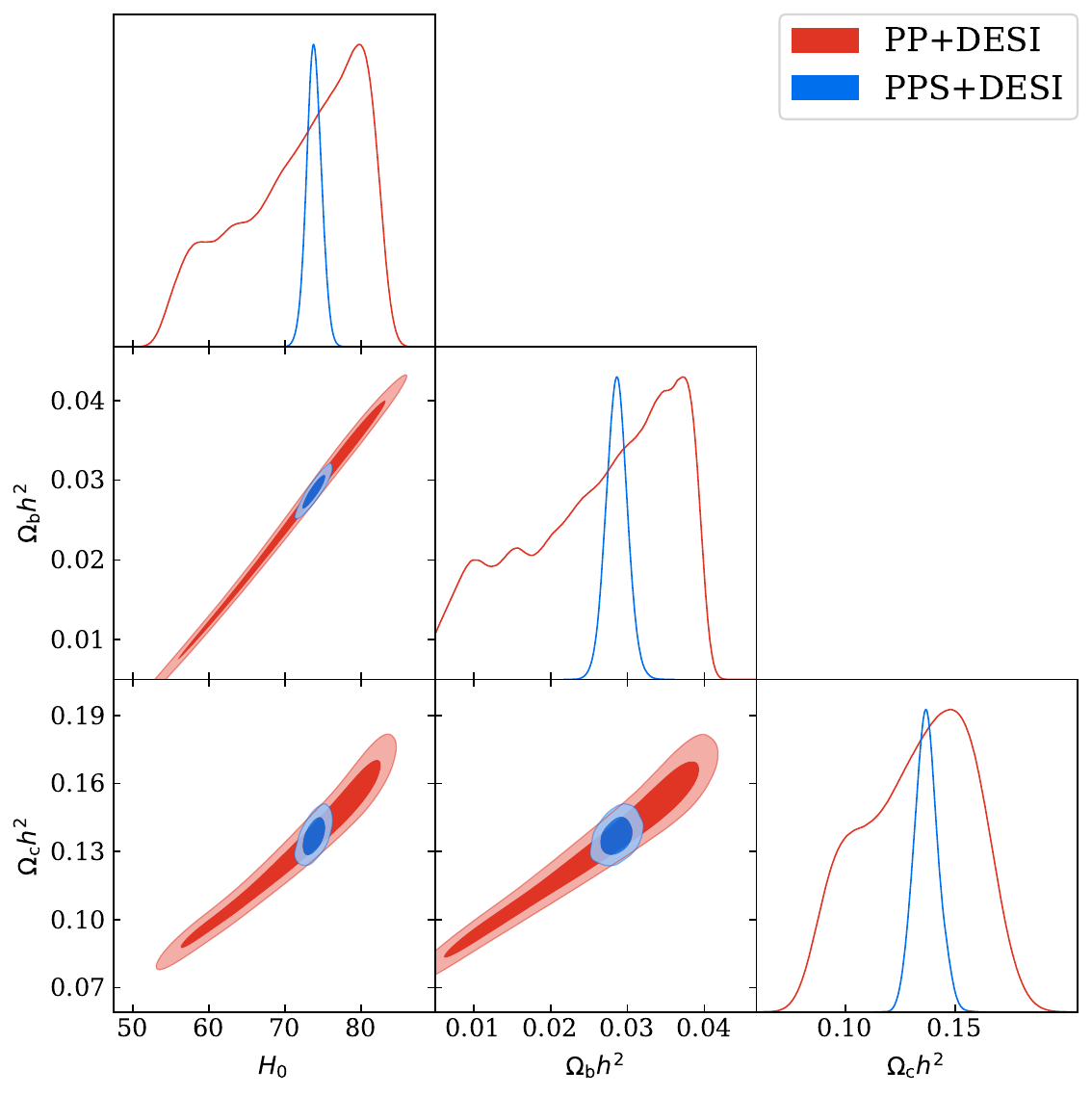}
    \includegraphics[width=0.8\linewidth]{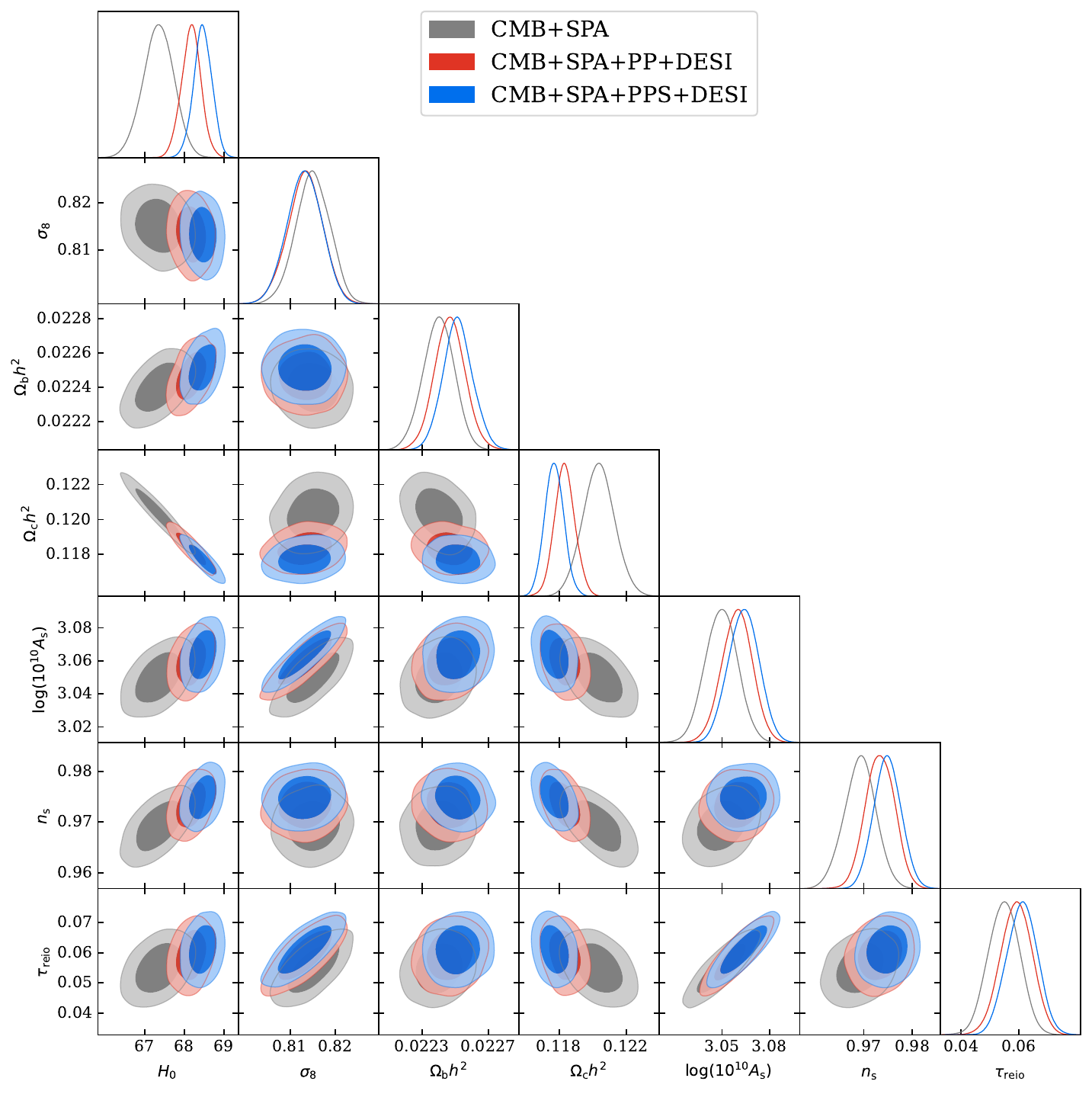}
    \caption{Observational constraints on the $\Lambda$CDM model. Marginalized 68\% and 95\% confidence regions for the cosmological parameters obtained from the different dataset combinations. The upper panels show the constraints from the purely late-Universe combinations PP+DESI and PPS+DESI. The lower panels show the constraints from the CMB-SPA dataset together with the combined analyses CMB-SPA+PP+DESI and CMB-SPA+PPS+DESI.}
    \label{fig:lcdm}
\end{figure}

We first discuss the $\Lambda$CDM model, which serves as the reference scenario for comparison with the two $f(T)$ models. For the purely late-time dataset combinations, PP+DESI and PPS+DESI, $\Lambda$CDM favours a smaller sound horizon at the baryon drag epoch than that inferred from CMB observations (see Table~\ref{tab:LCDMTab}). This is accompanied by a preference for larger values of $H_0$, reflecting the well-known degeneracy between the sound horizon and the Hubble constant. Since both Pantheon+ and DESI are uncalibrated distance probes, the PP+DESI combination cannot determine the absolute distance scale, leaving $H_0$ largely unconstrained, as expected. This degeneracy is generic to late-time probes and therefore extends to any dark-energy or modified-gravity model that modifies only the late-time expansion history. The same degeneracy also propagates to the physical baryon density, leading to $\Omega_{\rm b}h^2=0.0286\pm0.0014$, significantly higher than the value inferred from Big Bang Nucleosynthesis, $\Omega_{\rm b}h^2\simeq0.022-0.023$~\cite{Cooke:2017cwo,Schoneberg:2024ifp}. Once the SH0ES calibration is included, the absolute distance scale is fixed, breaking the degeneracy and yielding the expected constraint $H_0=73.83\pm0.99~{\rm km\,s^{-1}\,Mpc^{-1}}$. The corresponding late-time constraints are shown in the upper panels of Fig.~\ref{fig:lcdm}.

\begin{table}
    \centering
    \caption{Observational constraints on the $\Lambda$CDM cosmological parameters at 68\% CL. The upper panel reports the six independent cosmological parameters, while the lower panel lists the derived parameters. The minimum $\chi^2$ values for the different dataset combinations are reported in the last row. The subscript $\star$ denotes quantities evaluated at the photon decoupling epoch.}
    \label{tab:LCDMTab}
    \renewcommand{\arraystretch}{1.5}
    \resizebox{\textwidth}{!}{%
    \begin{tabular}{|l|c|c|c|c|c|}
    \Xhline{1.1pt} 
    \input{Tables_CCG/summarytable_LCDM.txt}\\
    \Xhline{1.1pt} 
    \end{tabular}}
\end{table}

For the purely early-Universe dataset, CMB-SPA, we recover cosmological constraints fully consistent with the baseline CMB-SPA~\cite{SPT-3G:2025bzu} analysis within $1\sigma$. In contrast to the purely late-time analysis, the inclusion of CMB observations calibrates the sound horizon, leading to the lower values of $H_0$ characteristic of the standard cosmological model. The best-fit CMB-SPA value corresponds to an $H_0$ tension of approximately $6.8\sigma$ with the recent Hubble Distance Network determination, $H_0=73.50\pm0.81~{\rm km\,s^{-1}\,Mpc^{-1}}$~\cite{H0DN:2025lyy}. After including the DESI and Pantheon+ datasets (CMB-SPA+DESI+PP), the preferred value of $H_0$ increases slightly, reducing the discrepancy to approximately $6.1\sigma$. We do not quote the corresponding tension for the PPS combination since the SH0ES calibration is included directly in the likelihood.
The addition of late-time observations also shifts several cosmological parameters relative to the CMB-only constraints. In particular, $\Omega_{\rm b}h^2$, $A_s$, $n_s$, and $\tau_{\rm reio}$ increase by approximately $2$--$3\sigma$, while the matter density parameter $\Omega_{\rm m}$ shifts to lower values. The latter is a direct consequence of the slight increase in $H_0$, since the CMB primarily constrains the physical matter density $\Omega_{\rm m}h^2$. These parameter shifts reflect the mild preference of current late-time observations for a cosmology different from that preferred by the CMB alone, as discussed in recent studies~\cite{Jhaveri:2025neg,Ferreira:2025lrd,Sailer:2025lxj,Huang:2025xyf}. The corresponding constraints are shown in Table~\ref{tab:LCDMTab} and Fig.~\ref{fig:lcdm}.

Within the $\Lambda$CDM framework, the mild mismatch between the CMB and recent DESI measurements can be understood in terms of well-defined degeneracy directions in parameter space. CMB anisotropies tightly constrain combinations such as $A_s e^{-2\tau_{\rm reio}}$ and $\Omega_{\rm m}h^2$, leaving residual freedom along directions where $\tau_{\rm reio}$, $A_s$, and the late-time expansion history vary coherently. When DESI data are included, the preferred distance measurements shift the fit along these degeneracies, leading to correlated changes in the cosmological parameters. Within the $\Lambda$CDM framework, this results in lower values of $\Omega_{\rm m}$, together with shifts in $A_s$, $n_s$, and $\tau_{\rm reio}$. In simple extensions of $\Lambda$CDM, the same mismatch may instead manifest itself as an apparent preference for dynamical dark energy or for values of the total neutrino mass below the minimal value allowed by oscillation experiments, even extending to formally unphysical negative values~\cite{Craig:2024tky,Green:2024xbb,Elbers:2024sha,Elbers:2025vlz,Graham:2025dqn,Pulido-Hernandez:2026hcs,Yang:2026yaq,Loverde:2024nfi,Kibris:2026cqq}. These features illustrate the limitations of the standard cosmological model in simultaneously accommodating current early- and late-time observations and motivate the exploration of alternative cosmological scenarios.

This behaviour is closely related to the well-known CMB lensing $A_{\rm lens}$ anomaly. An enhanced lensing amplitude smooths the acoustic peaks and modifies the damping tail of the CMB power spectra, thereby affecting the determination of the combination $A_s e^{-2\tau_{\rm reio}}$. Consequently, the calibration of structure-growth observables becomes highly sensitive to the inferred value of the optical depth. It has been shown that, for BAO combined with high-$\ell$ \textit{Planck} CMB data and CMB lensing, allowing a larger optical depth, $\tau_{\rm reio}\approx0.09$, restores the standard neutrino mass, $\sum m_\nu=0.06$ eV, and largely removes the apparent preference for dynamical dark energy. However, this solution introduces a $3$--$5\sigma$ disagreement with the low-$\ell$ polarization measurements from \textit{Planck}, as well as tension with independent probes of the reionization history~\cite{Giare:2023ejv,Jhaveri:2025neg,Sailer:2025lxj}.

Recent studies have further suggested that the low-$\ell$ polarization signal itself may be biased by a suppression of the primordial power spectrum on large scales, which can partially compensate the effect of $\tau_{\rm reio}$ and therefore bias its inferred value~\cite{Huang:2025xyf}. Nevertheless, explicit realizations of this mechanism, such as punctuated inflation, produce only marginal changes in the recovered optical depth and fail to resolve the CMB--DESI discrepancy, indicating that the degeneracy structure within $\Lambda$CDM remains remarkably robust~\cite{Huang:2025xyf}. As a consequence, the combined likelihood continues to explore correlated variations of $\tau_{\rm reio}$, $A_s$, $n_s$, and $\Omega_{\rm m}$, illustrating how uncertainties in the optical-depth calibration propagate to several cosmological parameters when combining CMB and DESI observations~\cite{Jhaveri:2025neg,Sailer:2025lxj,Ferreira:2025lrd}.

\subsection{Model I}\label{Sec:Results_ModelI}

\begin{figure}
    \centering
    \includegraphics[width=0.4\linewidth]{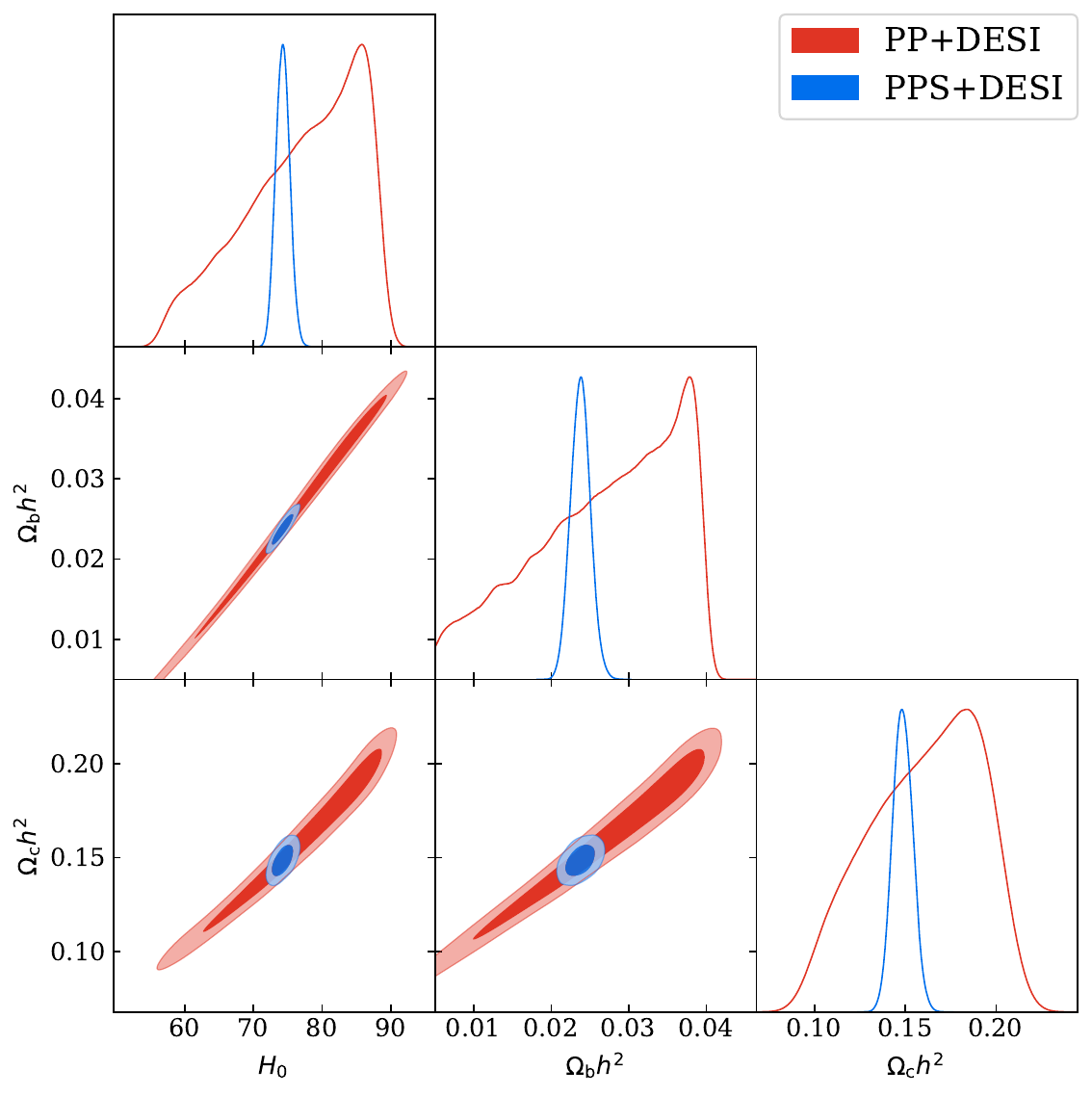}
    \includegraphics[width=0.8\linewidth]{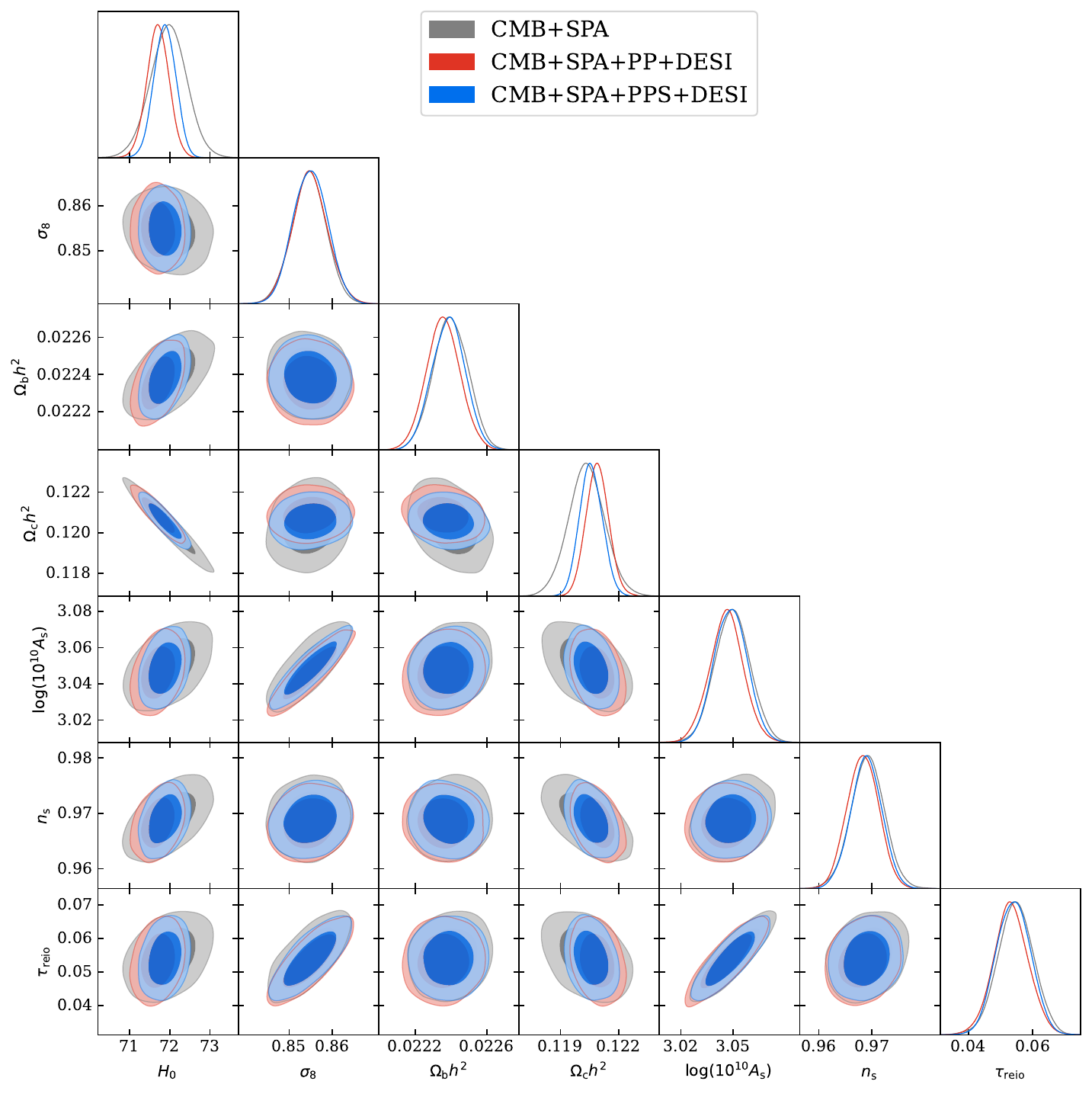}
    \caption{Observational constraints on Model~I. Marginalized 68\% and 95\% confidence regions for the cosmological parameters obtained from the different dataset combinations. The upper panels show the constraints from the purely late-Universe combinations PP+DESI and PPS+DESI. The lower panels show the constraints from the CMB-SPA dataset together with the combined analyses CMB-SPA+PP+DESI and CMB-SPA+PPS+DESI.}
    \label{fig:model1}
\end{figure}

For the purely late-Universe dataset combinations, PP+DESI and PPS+DESI, Model~I exhibits the same qualitative behaviour as $\Lambda$CDM. Since the late-time probes remain uncalibrated, the sound horizon is left largely unconstrained, leading to a broad posterior for $H_0$ in the PP+DESI case. After including the SH0ES calibration, the model yields $H_0=73.83\pm0.99~{\rm km\,s^{-1}\,Mpc^{-1}}$, fully consistent with the calibrated late-time determination. A notable difference with respect to $\Lambda$CDM is found for the physical baryon density: Model~I prefers $\Omega_{\rm b}h^2=0.0238\pm0.0013$, bringing it into much better agreement with the value inferred from Big Bang Nucleosynthesis. The corresponding constraints from late-time data are shown in the upper panels of Fig.~\ref{fig:model1}.

\begin{table}
    \centering
     \caption{Observational constraints on the cosmological parameters of Model~I at 68\% CL. The upper panel reports the six independent cosmological parameters, while the lower panel lists the derived parameters. The minimum $\chi^2$ values for the different dataset combinations are reported in the last row, together with $\Delta\chi^2=\chi^2_{f(T)}-\chi^2_{\Lambda{\rm CDM}}$.}
    \label{tab:model1Tab}
    \renewcommand{\arraystretch}{1.5}
    \resizebox{\textwidth}{!}{%
    \begin{tabular}{|l|c|c|c|c|c|}
    \Xhline{1.1pt} 
    \input{Tables_CCG/summarytable_Model-I.txt}\\
    \Xhline{1.1pt} 
    \end{tabular}}
\end{table}

We note that the GR limit (without a cosmological constant) is recovered by setting the model parameter $\beta=0$ in Eq.~\eqref{eq:ExpIRfT}, corresponding to $f(T)=T$. Therefore, no choice of $\beta$ reproduces $\Lambda$CDM exactly. Instead, as discussed in Sec.~II, $\Lambda$CDM emerges as the first-order approximation of the infrared expansion given in Eq.~\eqref{eq:fT_expansion}, while the higher-order terms introduce genuine infrared modifications to gravity.

The preference of Model~I for a larger value of $H_0$ originates from the effective phantom behaviour generated by these infrared corrections~\cite{Ludwick:2017tox}, as illustrated in Fig.~\ref{fig:modelI}. At late times, the expansion history can be written as
\begin{equation}
E^2(z)=\Omega_{\rm m}(1+z)^3+\Omega_{\rm de}\,y(z),
\label{eq:E_evolution}
\end{equation}
where
\begin{equation}
\Omega_{\rm de}=1-\Omega_{\rm m},\qquad
y(z)=\exp\left[3\int_0^z\frac{1+w_{\rm de}(z')}{1+z'}\,dz'\right].
\end{equation}
For a cosmological constant, $w_{\rm de}=-1$, yielding $y(z)=1$ at all redshifts. In contrast, an effective phantom regime with $w_{\rm de}<-1$ gives $y(0)=1$ and $y(z)<1$ for $z>0$, resulting in a lower expansion rate than in $\Lambda$CDM over the redshift range where dark energy is dynamically relevant.

Since the exponential infrared $f(T)$ model reproduces GR at early times, the sound horizon at photon decoupling, $r_s(z_\star)$, remains essentially unchanged with respect to $\Lambda$CDM. Consequently, the reduced late-time expansion lowers the angular diameter distance to the last-scattering surface, $D_A(z_\star)$. To preserve the precisely measured angular acoustic scale,
\begin{equation}
\theta_\star=\frac{r_s(z_\star)}{D_A(z_\star)},
\end{equation}
the fit naturally shifts toward a larger present-day Hubble constant $H_0$, as discussed in Refs.~\cite{El-Zant:2018bsc,Hashim:2020sez}.

Using the CMB-SPA dataset alone, Model~I yields $H_0=71.97\pm0.45~{\rm km\,s^{-1}\,Mpc^{-1}}$, reducing the discrepancy with the Hubble Distance Network measurement, $H_0=73.50\pm0.81~{\rm km\,s^{-1}\,Mpc^{-1}}$~\cite{H0DN:2025lyy}, from approximately $6.8\sigma$ in $\Lambda$CDM to only $1.65\sigma$. In this sense, replacing $\Lambda$CDM with Model~I largely alleviates the $H_0$ tension while preserving an excellent fit to the CMB data.
Remarkably, unlike $\Lambda$CDM, Model~I does not exhibit the $2$--$3\sigma$ shifts in the cosmological parameters $\Omega_{\rm b}h^2$, $\Omega_{\rm c}h^2$, $A_s$, $n_s$, and $\tau_{\rm reio}$ when DESI and Pantheon+ data are combined with CMB-SPA, as shown in the lower panels of Fig.~\ref{fig:model1}. This demonstrates that the model provides a much more stable cosmological solution across different dataset combinations.
However, this success comes at the price of making the model highly constrained. Since the exponential infrared $f(T)$ theory introduces no additional cosmological parameters beyond those of flat $\Lambda$CDM, increasing $H_0$ simultaneously fixes the entire late-time expansion history. Consequently, once the model is adjusted to reproduce both the CMB acoustic scale and the larger local value of $H_0$, there is no remaining freedom to modify the shape of $H(z)$ preferred by DESI and Type Ia supernova observations. The model therefore becomes overconstrained when early- and late-Universe datasets are analysed jointly, leading to the significantly larger $\chi^2$ values reported in Table~\ref{tab:model1Tab}.

This behaviour is consistent with the general argument that late-time modifications of the expansion history alone cannot fully reconcile the CMB with local measurements of $H_0$ while simultaneously fitting BAO observations~\cite{Lemos:2018smw}. Indeed, although DESI data favour dynamical dark-energy models such as the CPL parametrization over $\Lambda$CDM at approximately $3.5\sigma$, these models still predict $H_0\simeq68~{\rm km\,s^{-1}\,Mpc^{-1}}$ when combined with CMB observations. Conversely, models such as Model~I, which successfully raise $H_0$ through late-time modifications, inevitably predict an expansion history that becomes incompatible with the precise DESI and supernova measurements once all datasets are combined.

\subsection{Model II}\label{Sec:Results_ModelII}

As discussed in Sec.~II, the secondary branch of the Lambert-$W$ function in Eq.~\eqref{eq:ExpIRbeta} provides an alternative cosmological solution capable of producing late-time accelerated expansion~\cite{Awad:2017yod}. More recently, this branch has been investigated in detail~\cite{Akarsu:2024nas}, where the effective $f(T)$ component behaves as a negative dark-energy density over part of the cosmic history, as illustrated in the left panel of Fig.~\ref{fig:modelII}. In the present work, we perform the first comprehensive observational analysis of this branch using current cosmological datasets. Although the model provides a viable background evolution from a theoretical perspective, it is strongly disfavoured by all dataset combinations, as reflected by the large $\chi^2$ values reported in Table~\ref{tab:model2Tab}. In the following, we first discuss its general phenomenology and then investigate the physical origin of its poor agreement with observations.

As in Model~I, the exponential infrared $f(T)$ theory modifies gravity only at late times, while recovering GR at early times, as follows directly from the infrared expansion in Eq.~\eqref{eq:fT_expansion}. Consequently, the background evolution remains virtually indistinguishable from $\Lambda$CDM until dark energy becomes dynamically relevant. The main deviation appears at late times, where the effective torsional dark-energy component departs from a cosmological constant, as illustrated in the right panel of Fig.~\ref{fig:modelII}.
Similarly, at the perturbation level, the model approaches GR on small scales (large $T$), so that the CMB temperature and polarization power spectra at high multipoles are nearly identical to those predicted by $\Lambda$CDM, as shown in Fig.~\ref{fig:Plk18Cll}. Unlike Model~I, however, the effective torsional dark-energy density undergoes a smooth transition from negative to positive values at a transition redshift $z_t$. During this transition, the effective equation-of-state parameter evolves continuously from a quintessence-like regime to a phantom-like regime and necessarily develops a pole when the effective dark-energy density crosses zero, as shown in Fig.~\ref{fig:modelII}. This divergence is purely a consequence of the vanishing denominator in $w_T=p_T/\rho_T$ and therefore does not signal any physical pathology.

In principle, a negative effective dark-energy density naturally increases the inferred value of $H_0$. This can be understood from Eq.~\eqref{eq:E_evolution}, where the effective dark-energy contribution $y(z)\propto\rho_T$ becomes negative over part of the cosmic history. As in Model~I, the exponential infrared $f(T)$ theory leaves the early Universe essentially unchanged, preserving both the sound horizon at photon decoupling, $r_s(z_\star)$, and the measured angular acoustic scale, $\theta_\star$. The temporary suppression of the late-time expansion produced by $\rho_T<0$ reduces the angular diameter distance to the last-scattering surface. Consequently, the CMB fit naturally shifts toward a larger value of $H_0$ in order to preserve the observed acoustic scale.
Following Ref.~\cite{Awad:2017yod}, the transition redshift can be obtained by solving $\rho_T=0$ using Eq.~\eqref{eq:Tor-density}, yielding
\begin{equation}
    z_t=-1+\left[\left(\frac{1}{2}+W_{-1}\left(-\frac{1}{2} e^{-1/2}\right)\right)\frac{1-\Omega_\textrm{T}}{\beta}\right]^{-1/3},
\end{equation}
where $\beta$ is determined by the secondary Lambert-$W$ branch through Eq.~\eqref{eq:ExpIRbeta}. Therefore, the transition redshift is not an independent parameter but is completely determined by the cosmological density parameters. Using the best-fit cosmological parameters reported in Table~\ref{tab:model2Tab}, we find $0.81\lesssim z_t\lesssim1.41$.
Despite this appealing background behaviour, Model~II provides a very poor description of the data. As discussed above, the transition from negative to positive effective dark-energy density is accompanied by a large gravitational slip on super-horizon scales, which significantly enhances the late Integrated Sachs--Wolfe contribution to the low-multipole CMB temperature spectrum. This excess large-scale power is incompatible with current CMB observations and results in the substantially larger $\chi^2$ values reported in Table~\ref{tab:model2Tab}. The failure of Model~II therefore originates primarily from its perturbation behaviour rather than from its background expansion history.

\begin{table}
    \centering
    \caption{Observational constraints on the cosmological parameters of Model~II at 68\% CL. The upper panel reports the six independent cosmological parameters, while the lower panel lists the derived parameters. The minimum $\chi^2$ values for the different dataset combinations are reported in the last row, together with $\Delta\chi^2=\chi^2_{f(T)}-\chi^2_{\Lambda{\rm CDM}}$.}
    \label{tab:model2Tab}
    \renewcommand{\arraystretch}{1.5}
    \resizebox{\textwidth}{!}{%
    \begin{tabular}{|l|c|c|c|c|c|}
    \Xhline{1.1pt} 
    \input{Tables_CCG/summarytable_Model-II.txt}\\
    \Xhline{1.1pt} 
    \end{tabular}}
\end{table}

For the purely late-Universe datasets, Model~II exhibits the same qualitative behaviour as $\Lambda$CDM and Model~I. As expected, PP+DESI alone does not provide a tight constraint on $H_0$, while the inclusion of the SH0ES calibration yields a much narrower posterior centred around $H_0=73.8\pm1.0~{\rm km\,s^{-1}\,Mpc^{-1}}$, corresponding to a smaller effective sound horizon than the value calibrated by early-Universe observations. A striking difference, however, is the preferred baryon density, with $\Omega_{\rm b}h^2=0.01346\pm0.00098$, which is significantly below the value inferred from Big Bang Nucleosynthesis.

\begin{figure}
    \centering
    \includegraphics[width=0.35\linewidth]{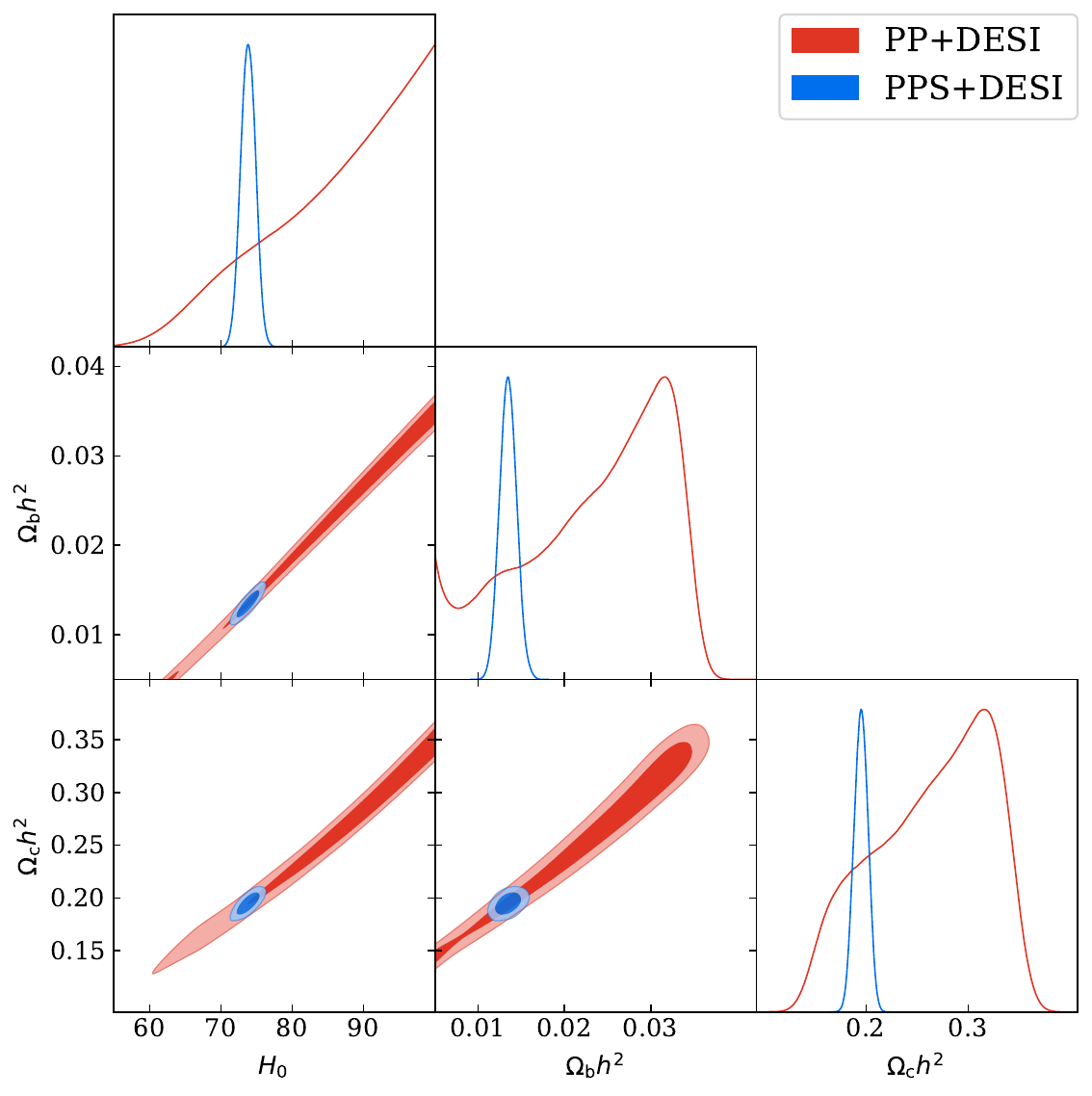}
    \includegraphics[width=0.9\linewidth]{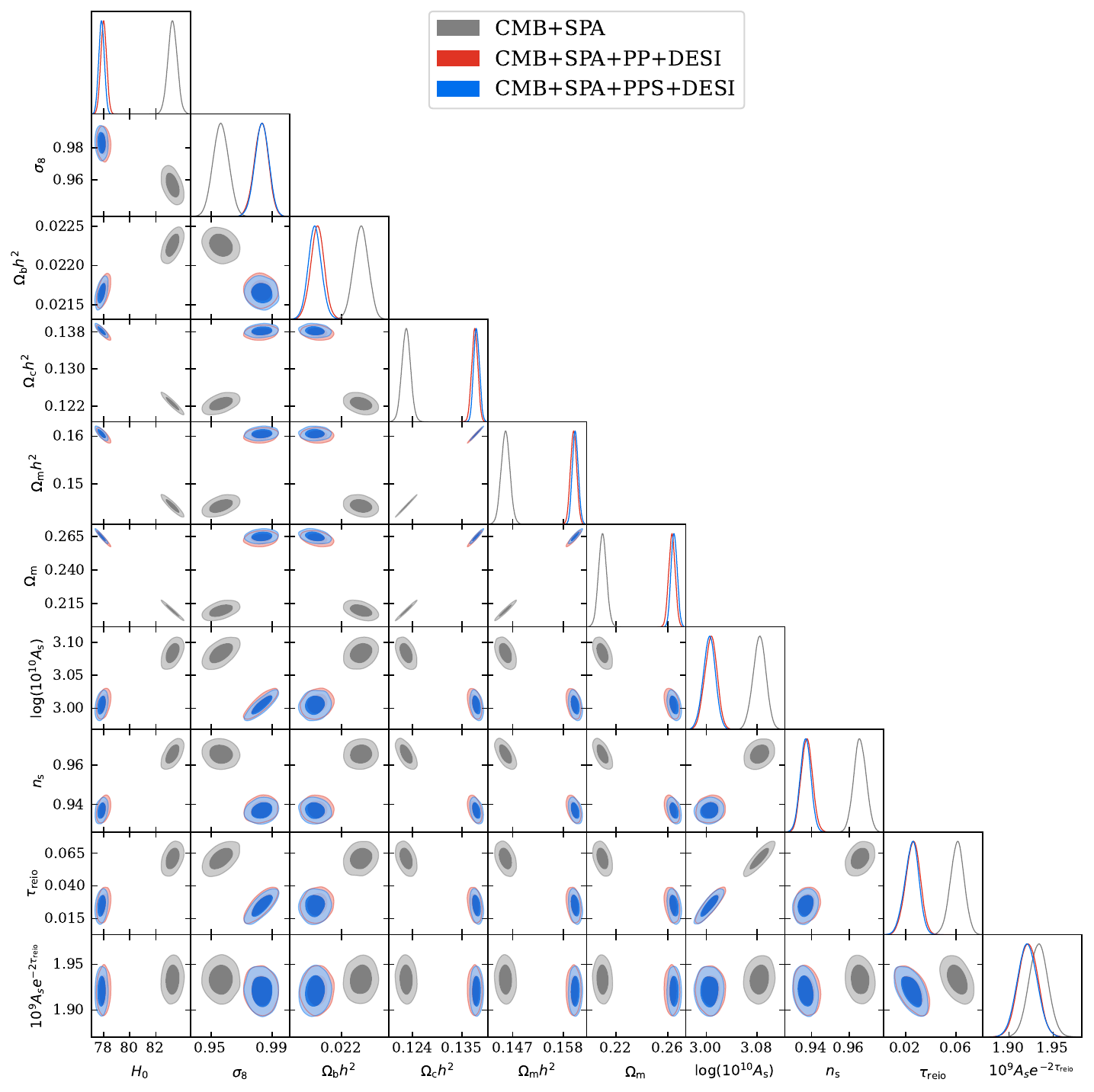}
    \caption{Observational constraints on Model~II: 2D joint marginalized constraints at 68\% and 95\% CL for the cosmological parameters. The upper panel shows the constraints obtained from the purely late-Universe combinations PP+DESI and PPS+DESI. The lower panel shows the constraints from the early-Universe dataset CMB-SPA together with the combined analyses CMB-SPA+PP+DESI and CMB-SPA+PPS+DESI.}
    \label{fig:model2}
\end{figure}

We note that Model~II leaves the CMB power spectra essentially unchanged on small angular scales, while significantly enhancing the TT power spectrum on large angular scales, particularly at low multipoles ($\ell\lesssim30$), as shown in Fig.~\ref{fig:Plk18Cll}. This enhancement explains the substantially larger best-fit $\chi^2$ obtained with the CMB-SPA dataset compared to both $\Lambda$CDM and Model~I (Table~\ref{tab:model2Tab}). Furthermore, when DESI or Type~Ia supernova data are combined with CMB-SPA, the cosmological parameters shift in directions opposite to those found in $\Lambda$CDM (Fig.~\ref{fig:model2}), with displacements reaching approximately $6$--$13\sigma$. This indicates that Model~II follows degeneracy directions markedly different from those of the standard cosmological model, driving some parameters towards extreme values. In particular, the optical depth shifts to the remarkably low value $\tau_{\rm reio}\approx0.026$ when DESI data are included. A coherent explanation of the parameter shifts observed in Model~II emerges from the interplay between the altered expansion history, the CMB constraints, and the late-Universe measurements.

In this model, the MG sector behaves as an effective dark-energy component whose density changes sign at $z=z_t$, transitioning from a negative contribution at $z>z_t$ to a phantom-like behaviour at late times. As a consequence, the expansion rate $H(z)$ is suppressed relative to $\Lambda$CDM at intermediate redshifts before becoming enhanced at late times, naturally favouring larger values of $H_0$, as follows from the modified Friedmann equation~\eqref{eq:FR1T}. Compared to Model~I, however, the negative effective torsional dark-energy density is significantly larger (see Fig.~\ref{fig:modelII}), leading to a much stronger preference for high values of $H_0$.

In the CMB-SPA analysis (Table~\ref{tab:model2Tab}), the modified expansion history allows a very large present-day Hubble constant, $H_0\simeq83.3~{\rm km\,s^{-1}\,Mpc^{-1}}$. This is compensated by a substantially lower matter density, $\Omega_{\rm m}\simeq0.21$, reflecting the well-known geometrical degeneracy of CMB observations, whereby $H_0$ and $\Omega_{\rm m}$ adjust coherently to reproduce the observed angular acoustic scale.

At the perturbation level, Model~II closely reproduces the $\Lambda$CDM CMB power spectra on small angular scales, while significantly enhancing the TT spectrum at large scales ($\ell\lesssim30$). This enhancement is driven by a strong late-time ISW effect, which becomes prominent in $f(T)$ gravity when the background evolution departs from a quasi-de Sitter regime owing to the large gravitational slip on super-horizon scales (see Eqs.~\eqref{eq:pert2} and \eqref{eq:grav_slip}). Consequently, the CMB-only analysis favours a slightly larger amplitude of primordial scalar perturbations, with $\log(10^{10}A_s)\approx3.084$, compared to the $\Lambda$CDM value of $\log(10^{10}A_s)\approx3.050$. At this stage, however, both the physical matter density $\Omega_{\rm m}h^2$ and the optical depth $\tau_{\rm reio}$ remain close to their standard values, indicating that the CMB damping tail is still well reproduced. Therefore, despite the excess TT power at low multipoles, the CMB-SPA dataset alone provides an acceptable fit.

The situation changes once the low-redshift data (DESI BAO together with Type~Ia supernovae) are included. These datasets tightly constrain the late-time expansion history through measurements of the comoving angular diameter distance $D_M(z)$ and the Hubble expansion rate $H(z)$, driving the matter density towards $\Omega_{\rm m}\simeq0.26$. Consequently, the preferred Hubble constant decreases from $H_0\simeq83.3$ to $H_0\simeq78~{\rm km\,s^{-1}\,Mpc^{-1}}$. However, the CMB acoustic peaks still require the angular acoustic scale to remain fixed, preventing $H_0$ from decreasing sufficiently to preserve the CMB-preferred value of the physical matter density. As a result, the combination $\Omega_{\rm m}h^2$ increases (Fig.~\ref{fig:model2}), worsening the agreement with the observed CMB damping tail. To recover the observed amplitude of the damping tail, the likelihood compensates by shifting towards smaller values of the primordial amplitude and scalar spectral index, with $\log(10^{10}A_s)$ decreasing from $3.084$ to $3.0072$. However, the damping tail tightly constrains the combination $A_s e^{-2\tau_{\rm reio}}$. Consequently, the reduction in $A_s$ is accompanied by a corresponding decrease in the preferred optical depth, driving $\tau_{\rm reio}$ from its standard value of $\sim0.06$ to $\sim0.02$, in strong disagreement with the adopted Gaussian prior. Throughout this process, the enhancement of the TT spectrum at low multipoles remains, further degrading the overall CMB fit.

This behaviour reflects the fact that Model~II introduces no additional cosmological parameters beyond the six of $\Lambda$CDM. While the CMB-only analysis can exploit the geometrical degeneracy to accommodate a very large value of $H_0$, the inclusion of BAO and Type~Ia supernovae effectively removes this freedom by fixing the late-time expansion history. The model is therefore unable to simultaneously preserve the CMB-preferred value of $\Omega_{\rm m}h^2$, reproduce the observed damping tail, satisfy the Gaussian prior on $\tau_{\rm reio}$, and fit the low-redshift distance measurements. In this sense, the model becomes overconstrained by the combined datasets. The resulting shifts in $A_s$, $n_s$, and $\tau_{\rm reio}$ are not preferred by the data themselves, but are forced by the model in an attempt to reconcile incompatible constraints, leading to a substantial increase in the total $\chi^2$. This ultimately rules out Model~II with current cosmological observations.

While in $\Lambda$CDM an increase in $\tau_{\rm reio}$ (and consequently $A_s$) can partially improve the consistency between CMB and BAO by enhancing the CMB lensing signal~\cite{Jhaveri:2025neg,Sailer:2025lxj}, this compensation mechanism is no longer available in Model~II. As discussed above, the inclusion of BAO and SNIa data drives the fit towards larger values of $\Omega_{\rm m}h^2$, worsening the agreement with the CMB damping tail. The likelihood therefore compensates by reducing the primordial amplitude, with $\Delta\log(10^{10}A_s)\simeq-0.08$. Since the damping tail tightly constrains the combination $A_se^{-2\tau_{\rm reio}}$, this shift naturally translates into $\Delta\tau_{\rm reio}\simeq-0.04$, consistent with the approximate Fisher-matrix relation $\delta\ln A_s\simeq2\,\delta\tau_{\rm reio}$. As a result, the preferred optical depth is driven from its standard value of $\tau_{\rm reio}\simeq0.06$ to the unphysically small value of $\tau_{\rm reio}\simeq0.02$. Therefore, in addition to the degradation of the damping-tail fit, the model incurs a significant penalty from the Gaussian prior on $\tau_{\rm reio}$, further increasing the total $\chi^2$.

This extremely low optical depth should not be interpreted as a physical prediction for the reionization history. Rather, it reflects the inability of the model to simultaneously reproduce the CMB damping tail and the late-time expansion history once the low-redshift data are included. We therefore interpret the low value of $\tau_{\rm reio}$ as a consequence of the parameter degeneracies induced by the modified expansion history, rather than as evidence for a non-standard reionization scenario.

For completeness of the discussion, Appendix~\ref{App:Comparison} presents complementary figures comparing the behaviour of the three cosmological models for each dataset considered in this work.

\section{Summary and Conclusion}\label{Sec:Conclusion}

In this work, we have presented a comprehensive observational analysis of the exponential infrared $f(T)$ teleparallel gravity model, investigating both branches of its cosmological solutions. This framework provides a purely geometrical explanation for late-time cosmic acceleration without introducing any additional cosmological parameters beyond those of spatially flat $\Lambda$CDM. The principal branch (Model~I) gives rise to an effective phantom-like dark energy component, whereas the secondary branch (Model~II) predicts a transition from negative to positive effective dark-energy density.

Both branches naturally predict larger values of the Hubble constant when constrained by CMB observations alone. Since the modifications affect only the late Universe, the CMB geometrical degeneracy shifts towards higher $H_0$ and lower $\Omega_{\rm m}$, while preserving the excellent description of the acoustic peak structure. However, the inclusion of DESI BAO and Type~Ia supernovae measurements substantially reduces this freedom by tightly constraining the late-time expansion history.

Among the two branches, only Model~I remains compatible with the full cosmological dataset. The effective phantom behaviour increases the CMB-inferred Hubble constant to $H_0=71.97\pm0.45~{\rm km\,s^{-1}\,Mpc^{-1}}$, reducing the tension with the Hubble Distance Network measurement from approximately $6.8\sigma$ in $\Lambda$CDM to about $1.6\sigma$. Moreover, unlike $\Lambda$CDM, the inclusion of DESI BAO and Type~Ia supernovae data does not induce significant shifts in the remaining cosmological parameters, demonstrating that the model provides a consistent description of both early- and late-Universe observations. Nevertheless, because the model contains no additional free parameters, it cannot fully reproduce the expansion history preferred by DESI while maintaining such a high value of $H_0$, resulting in a degradation of the global fit.

Model~II, on the other hand, is decisively ruled out by current observations. Although the model also predicts very large values of $H_0$ in CMB-only analyses, the inclusion of BAO and Type~Ia supernovae data fixes the late-time matter density, forcing the physical matter density $\Omega_{\rm m}h^2$ away from its CMB-preferred value. This degrades the fit to the CMB damping tail and triggers a chain of correlated parameter shifts: the primordial amplitude $A_s$ and scalar spectral index $n_s$ decrease, while the tight CMB constraint on the combination $A_se^{-2\tau_{\rm reio}}$ drives the optical depth towards the unphysical value $\tau_{\rm reio}\simeq0.02$. We interpret this behaviour not as evidence for an exotic reionization history, but as the consequence of attempting to reconcile incompatible early- and late-Universe constraints within a model that possesses no additional cosmological degrees of freedom. The model therefore becomes overconstrained by the combined datasets, leading to a dramatic increase in the total $\chi^2$.

Finally, our analysis highlights the importance of jointly analysing cosmological background and perturbation observables when testing modified gravity. Although both branches successfully modify the late-time expansion history and naturally accommodate larger values of $H_0$, only Model~I remains compatible with current cosmological observations once CMB anisotropies, BAO, and Type~Ia supernovae are simultaneously considered. This demonstrates that perturbation observables, and in particular the CMB damping tail, provide stringent and complementary tests of modified gravity models beyond their background evolution.

The present exponential infrared $f(T)$ theory contains no additional cosmological degrees of freedom beyond the six parameters of spatially flat $\Lambda$CDM. While this high level of predictivity allows a stringent test of the theory, our analysis of Model~II shows that it also leaves insufficient freedom to control the evolution of the effective torsional dark-energy density once early- and late-Universe observations are combined. This suggests that the failure of Model~II should not be interpreted as evidence against a negative-to-positive transition of the effective torsional density itself, but rather as an indication that a more general realization of this scenario is required.

A natural extension is the generalized exponential model proposed in Ref.~\cite{Santos:2022atq},
\begin{equation}
f(T)=T\,e^{\beta(T/T_0)^b},
\end{equation}
which introduces one additional free parameter with respect to the present analysis. So far, this model has only been confronted with background observations, while a full analysis including CMB anisotropies and cosmological perturbations is still lacking. Moreover, although the theory also admits multiple Lambert-$W$ branches, only the principal branch, corresponding to the phantom-crossing solution, has been investigated. The observational viability of the remaining branches therefore remains an open question.

Another interesting possibility is the generalized exponential infrared model
\begin{equation}
f(T)=\left(T+\alpha\sqrt{TT_0}\right)e^{\beta T/T_0},
\end{equation}
which likewise introduces one additional degree of freedom while preserving the infrared exponential structure of the theory. In particular, the critical case $\alpha=-1$ appears especially interesting and deserves a dedicated investigation.

Future high-precision observations will provide stringent tests of these extended scenarios. Improved measurements of large-scale CMB polarization will further constrain the optical depth and the damping-tail degeneracies identified in this work, while forthcoming galaxy clustering and weak-lensing surveys will test the modified growth history predicted by teleparallel gravity. Together, these observations will provide powerful and complementary tests of geometrical alternatives to dark energy.



\begin{acknowledgments}
W.E. and M.H. would like to thank Amr El-Zant for many discussions about the role of negative dark energy density to address $H_0$ tension and early galaxy formation. EDV is supported by a Royal Society Dorothy Hodgkin Research Fellowship. JLS would also like to acknowledge funding from ``Xjenza Malta'' as part of the ``Technology Development Programme'' DTP-2024-014 (CosmicLearning) Project. This article is based upon work from COST Action CA21136 \emph{Addressing observational tensions in cosmology with systematics and fundamental physics} (CosmoVerse), supported by COST (European Cooperation in Science and Technology).
\end{acknowledgments}

\appendix

\section{Comparison between models}\label{App:Comparison}

In this appendix, we compare the observational behaviour of the three cosmological models, namely $\Lambda$CDM, Model~I, and Model~II. We present the two-dimensional marginalized posterior distributions at 68\% and 95\% confidence levels for the cosmological parameters using the five dataset combinations considered in this work. These include purely late-Universe constraints (PP+DESI and PPS+DESI), purely early-Universe constraints (CMB-SPA), and combined early- and late-Universe constraints (CMB-SPA+PP+DESI and CMB-SPA+PPS+DESI). The purpose of these figures is to facilitate a direct comparison of the parameter degeneracies and the impact of each dataset on the three models.

\subsection{Constraints from purely late universe data}

\begin{figure}[h!]
    \centering
    \includegraphics[width=0.45\linewidth]{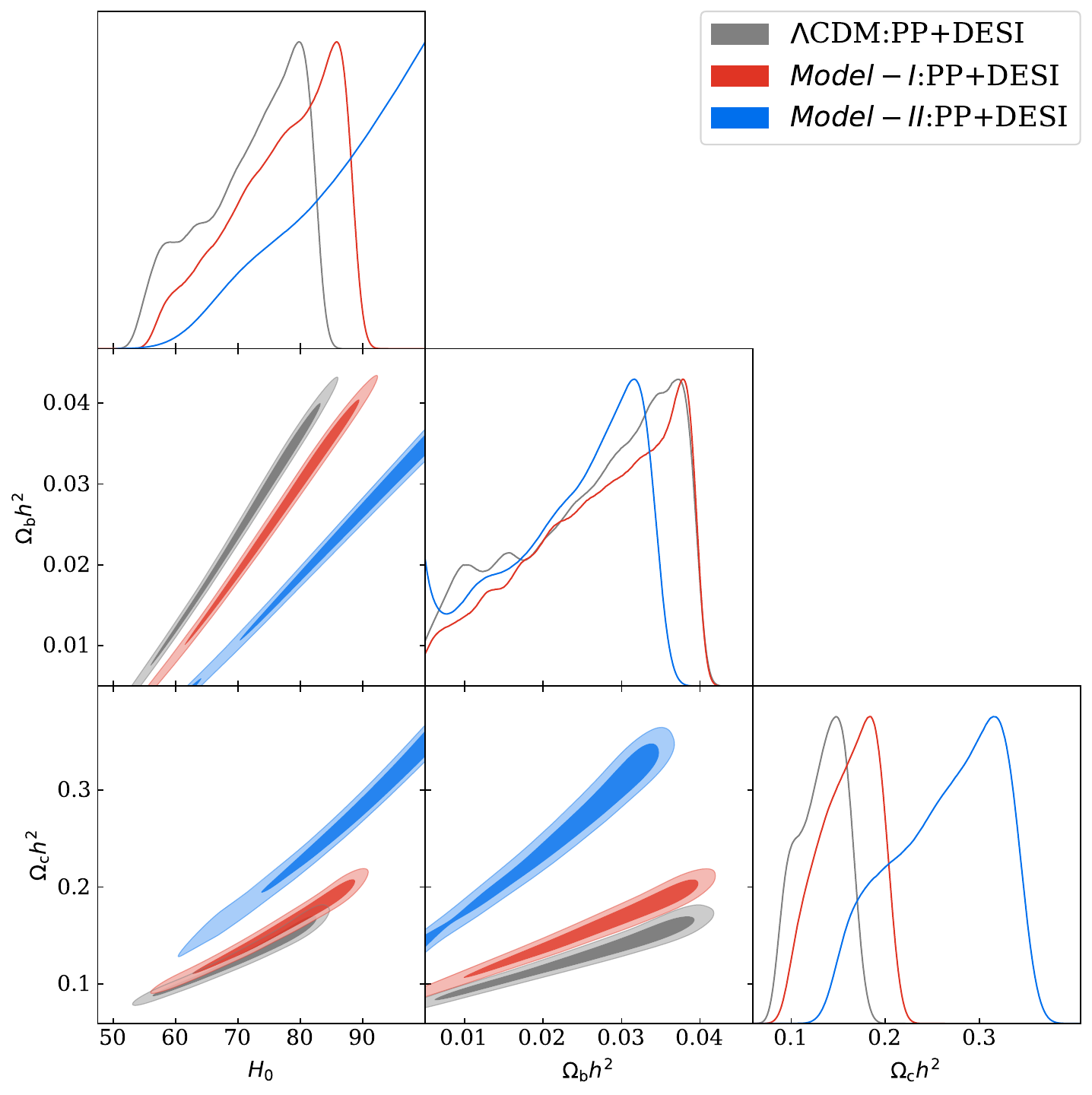}
    \includegraphics[width=0.45\linewidth]{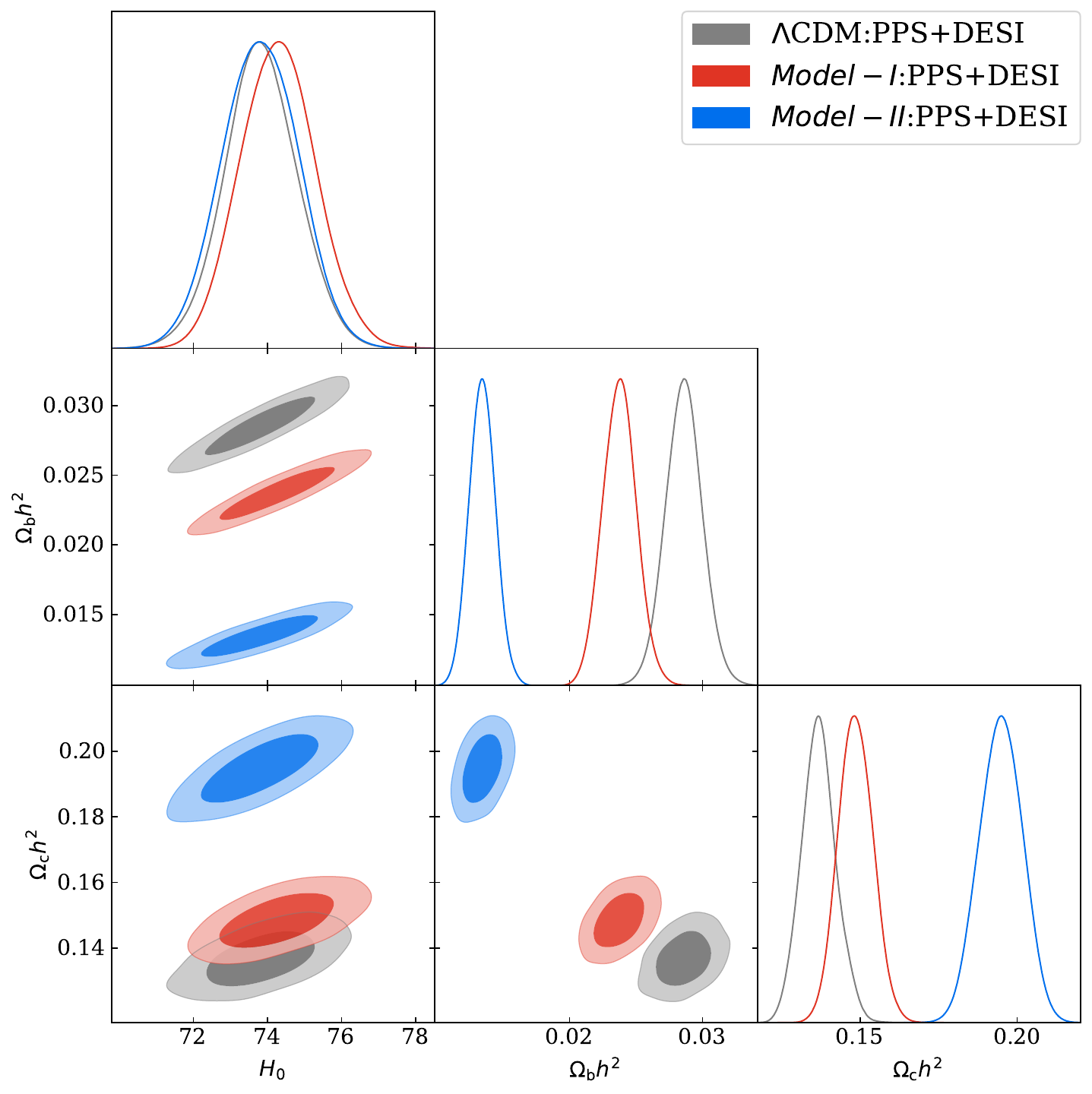}
    \caption{Purely late-Universe constraints (PP/PPS+DESI) for $\Lambda$CDM, $f(T)$-CDM Model~I, and $f(T)$-CDM Model~II. The figure shows the one-dimensional marginalized posterior distributions and the two-dimensional joint constraints at 68\% and 95\% confidence levels for the Hubble constant $H_0$, the baryon density $\Omega_{\rm b}h^2$, and the cold dark matter density $\Omega_{\rm c}h^2$.}
    \label{fig:PP+DESI}
\end{figure}

In the left panel of Fig.~\ref{fig:PP+DESI}, we show the observational constraints obtained from the purely late-Universe dataset PP+DESI. Since both SNIa and BAO measurements are uncalibrated, they do not determine the absolute distance scale and therefore cannot tightly constrain the Hubble constant. This freedom propagates through the correlations among the cosmological parameters, resulting in broad constraints on $H_0$, $r_d$, $\Omega_{\rm b}h^2$, and $\Omega_{\rm m}h^2$. In all three models, the preference for larger values of $H_0$ is accompanied by a correspondingly smaller sound horizon at the drag epoch than that inferred from the CMB, as reported in Tables~\ref{tab:LCDMTab}--\ref{tab:model2Tab}.
The inclusion of the SH0ES Cepheid calibration (PPS+DESI; right panel of Fig.~\ref{fig:PP+DESI}) fixes the absolute distance scale, leading to a much tighter determination of $H_0$, while maintaining the preference for a smaller sound horizon. Interestingly, even without early-Universe information, Model~I predicts a baryon density of $\Omega_{\rm b}h^2\simeq0.0238$, in excellent agreement with the BBN determination, $\Omega_{\rm b}h^2\simeq0.022{-}0.023$~\cite{Cooke:2017cwo,Driver:2021cde,Schoneberg:2024ifp}. By contrast, $\Lambda$CDM prefers a significantly larger value, $\Omega_{\rm b}h^2\simeq0.0286$, whereas Model~II predicts a substantially lower value, $\Omega_{\rm b}h^2\simeq0.0135$.

\subsection{Constraints from purely early universe data}

\begin{figure}
    \centering
    \includegraphics[width=0.9\linewidth]{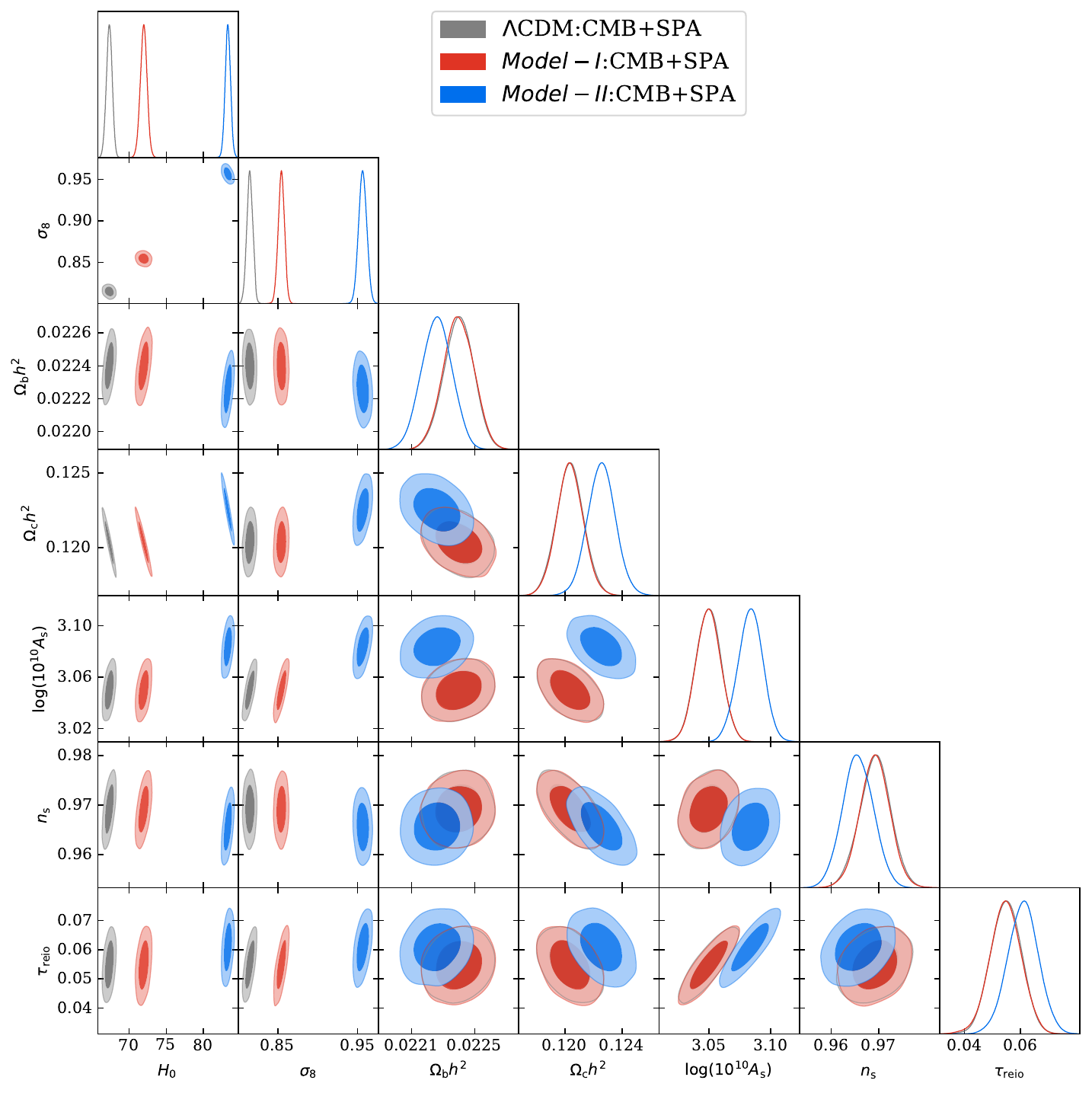}
    \caption{Purely early-Universe constraints (CMB-SPA) for $\Lambda$CDM, $f(T)$-CDM Model~I, and $f(T)$-CDM Model~II. The figure shows the one-dimensional marginalized posterior distributions and the two-dimensional joint constraints at 68\% and 95\% confidence levels for the six independent cosmological parameters, together with the derived parameters $H_0$ and $\sigma_8$.}
    \label{fig:CMB-SPA}
\end{figure}

In Fig.~\ref{fig:CMB-SPA}, we compare the constraints obtained using the purely early-Universe dataset CMB-SPA. For the primary cosmological parameters, $\Omega_{\rm b}h^2$, $\Omega_{\rm c}h^2$, $A_s$, $n_s$, and $\tau_{\rm reio}$, Model~I remains in excellent agreement with $\Lambda$CDM, reflecting the fact that the modifications are confined to the late Universe. The main difference is the shift towards larger values of $H_0$, accompanied by a corresponding increase in $\sigma_8$, driven by the effective phantom-like torsional dark-energy component at low redshift.
Model~II exhibits a markedly different behaviour. Although the physical matter density remains close to its standard value, the geometrical degeneracy allows a substantially larger Hubble constant, $H_0$, together with a much smaller matter density parameter, $\Omega_{\rm m}$. In addition, the enhanced late-time ISW effect produces excess TT power at low multipoles owing to the large gravitational slip that develops when the background evolution departs from a quasi-de Sitter regime. Consequently, the CMB-only fit favours a larger primordial amplitude, with $A_s$ shifted by approximately $2.4\sigma$ relative to $\Lambda$CDM, while the remaining primary parameters remain broadly consistent with their standard values.

\subsection{Constraints from both early and late universe data}

\begin{figure}[h!]
    \centering
    \includegraphics[width=\linewidth]{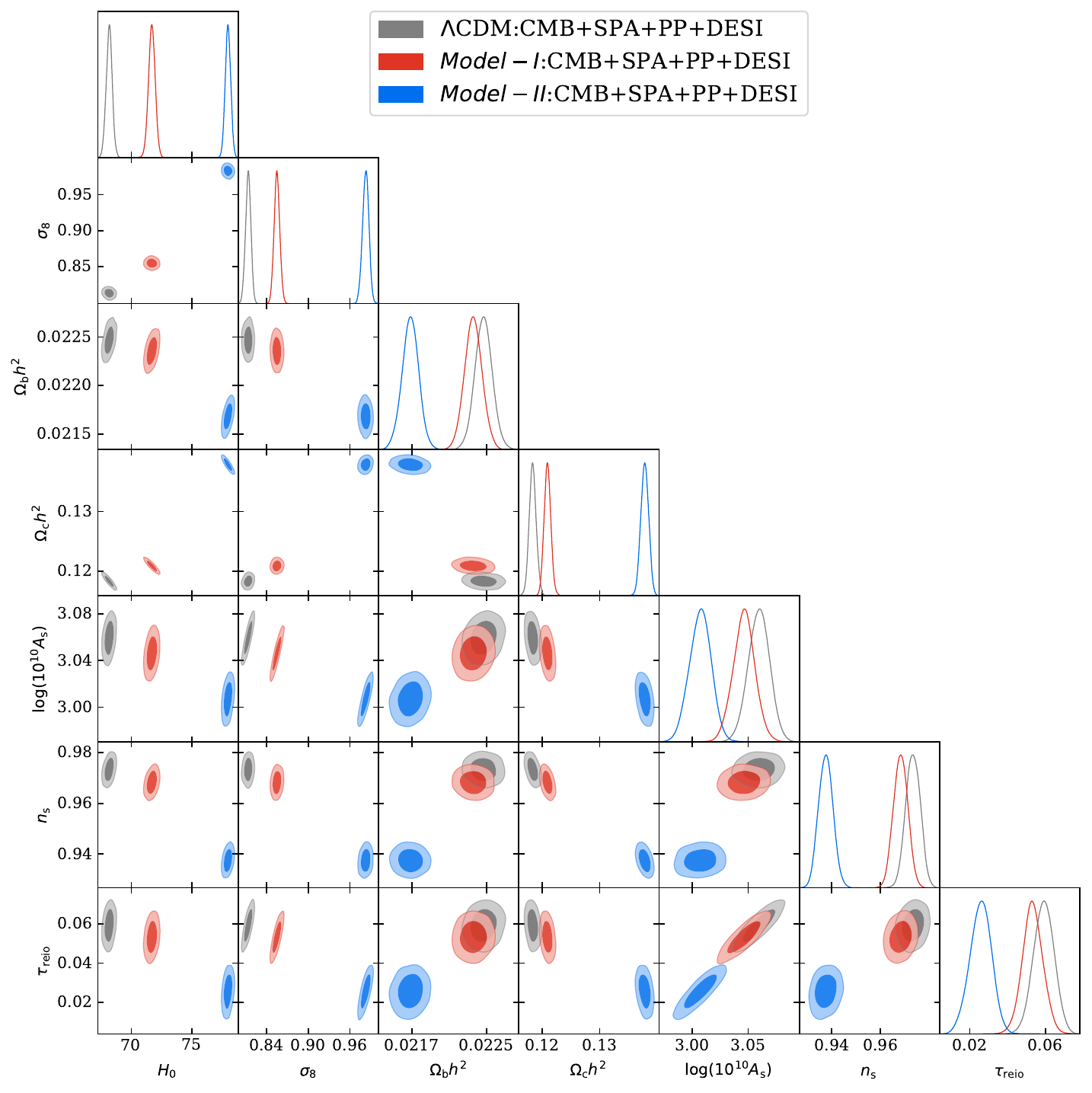}
    \caption{Combined early- and late-Universe constraints (CMB-SPA+PP+DESI) for $\Lambda$CDM, $f(T)$-CDM Model~I, and $f(T)$-CDM Model~II. The figure shows the one-dimensional marginalized posterior distributions and the two-dimensional joint constraints at 68\% and 95\% confidence levels for the six independent cosmological parameters, together with the derived parameters $H_0$ and $\sigma_8$.}
    \label{fig:CMB+PP+DESI}
\end{figure}

\begin{figure}[h!]
    \centering
    \includegraphics[width=\linewidth]{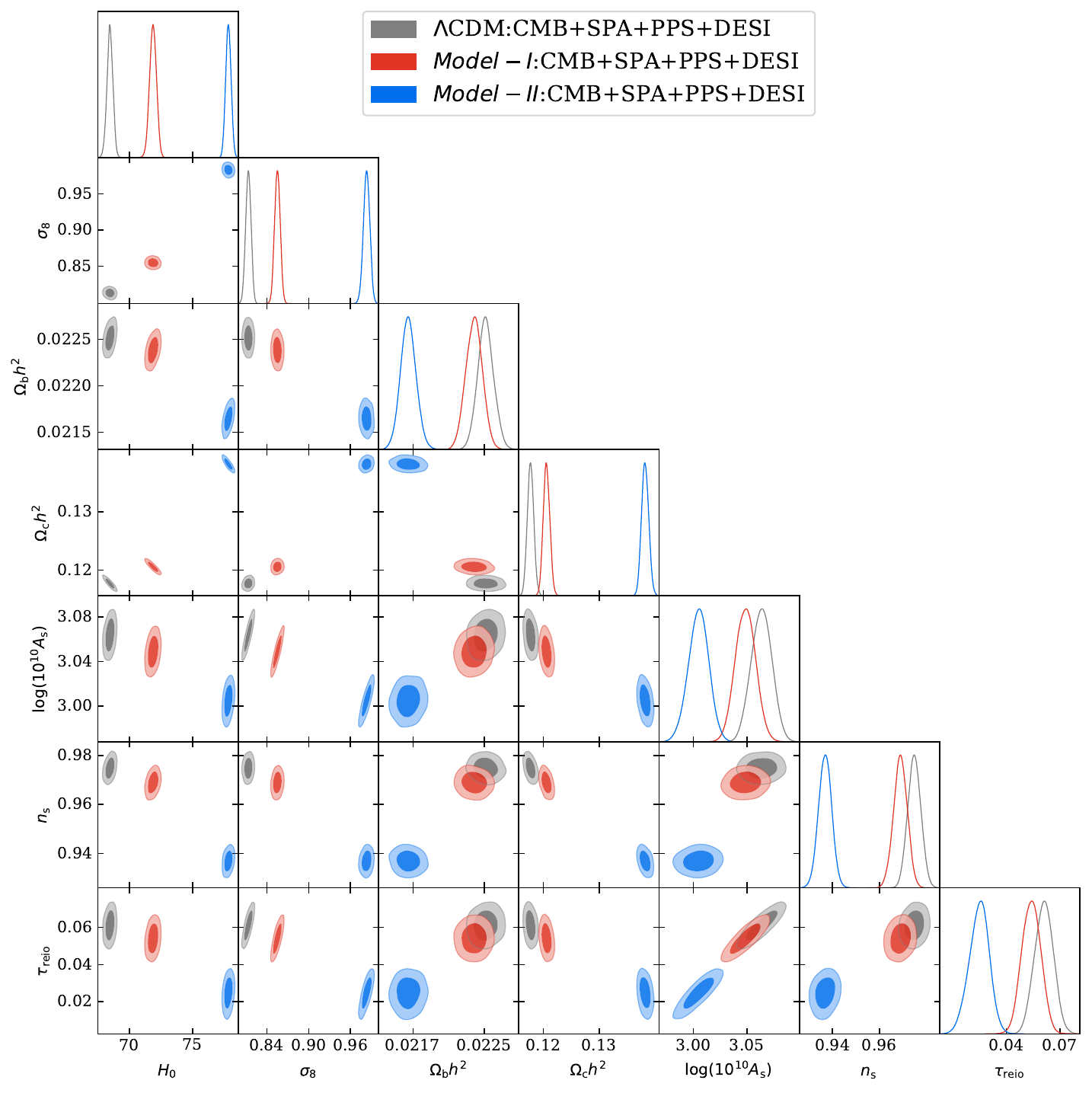}
    \caption{Combined early- and late-Universe constraints (CMB-SPA+PPS+DESI) for $\Lambda$CDM, $f(T)$-CDM Model~I, and $f(T)$-CDM Model~II. The figure shows the one-dimensional marginalized posterior distributions and the two-dimensional joint constraints at 68\% and 95\% confidence levels for the six independent cosmological parameters, together with the derived parameters $H_0$ and $\sigma_8$.}
    \label{fig:CMB+PPS+DESI}
\end{figure}

In Figs.~\ref{fig:CMB+PP+DESI} and \ref{fig:CMB+PPS+DESI}, we compare the three models using the combined early- and late-Universe datasets. Within the $\Lambda$CDM framework, the inclusion of DESI BAO and Type~Ia supernovae data shifts several independent cosmological parameters towards larger values, reflecting the well-known tension between CMB and BAO observations. By contrast, Model~II follows a markedly different degeneracy direction. The combined datasets drive the fit towards nonphysical regions of parameter space, most notably yielding an optical depth at reionization of $\tau_{\rm reio}\approx0.025$, together with significant shifts in $A_s$ and $n_s$, as discussed in Sec.~\ref{Sec:Results_ModelII}. These behaviours reflect the inability of Model~II to simultaneously reproduce the CMB damping tail and the late-time expansion history, resulting in a substantially poorer fit to the combined observations. In contrast, the inferred values of the independent cosmological parameters in Model~I remain remarkably stable across all dataset combinations, demonstrating the robustness of the model.

\clearpage

\bibliographystyle{apsrev4-1}
\bibliography{biblio}

\end{document}

%% file: Tables_CCG/summarytable_LCDM.txt
$\Lambda$CDM parameters  & PP+DESI & CMB-SPA & CMB-SPA+PP+DESI & PPS+DESI & CMB-SPA+PPS+DESI \\
\hline
$\Omega_\mathrm{b} h^2 $ & $ 0.0265^{+0.013}_{-0.0051}$ & $ 0.022399\pm 0.000094$ & $ 0.022466\pm 0.000093$ & $ 0.0286\pm 0.0014$ & $ 0.022516\pm 0.000090$ \\
$\Omega_\mathrm{c} h^2 $ & $ 0.133^{+0.032}_{-0.020}$ & $ 0.12035\pm 0.00095$ & $ 0.11832\pm 0.00061$ & $ 0.1368^{+0.0051}_{-0.0057}$ & $ 0.11770\pm 0.00057$ \\
$H_0 $ & $ 72^{+10}_{-4}$ & $ 67.34\pm 0.39$ & $ 68.17\pm 0.25$ & $ 73.83\pm 0.99$ & $ 68.47\pm 0.23$ \\
$\tau_\mathrm{reio} $ & $--$ & $ 0.0549\pm 0.0054$ & $ 0.0591\pm 0.0054$ & $--$ & $ 0.0611\pm 0.0052$ \\
$\log(10^{10} A_\mathrm{s}) $ & $ ---$ & $ 3.050\pm 0.010$ & $ 3.0596\pm 0.0095$ & $ ---$ & $ 3.0637\pm 0.0095$ \\
$n_\mathrm{s} $ & $ ---$ & $ 0.9693\pm 0.0032$ & $ 0.9734\pm 0.0030$ & $ ---$ & $ 0.9749\pm 0.0028$ \\
\hline
$\Omega_\mathrm{m} $ & $ 0.3046\pm 0.0083$ & $ 0.3162\pm 0.0056$ & $ 0.3043\pm 0.0034$ & $ 0.3045\pm 0.0080$ & $ 0.3005\pm 0.0031$ \\
$\sigma_8 $ & $ 0.78^{+0.16}_{-0.36}$ & $ 0.8150\pm 0.0037$ & $ 0.8133\pm 0.0038$ & $ 0.79^{+0.21}_{-0.36}$ & $ 0.8132\pm 0.0038$ \\
$z_\star $ & $ 1085.371^{+0.044}_{-12}$ & $ 1085.10\pm 0.56$ & $ 1084.45\pm 0.53$ & $ 1079.5^{+1.0}_{-1.2}$ & $ 1084.15\pm 0.51$ \\
$r_\star $ & $ 139.8^{+6.5}_{-17}$ & $ 144.68\pm 0.25$ & $ 145.18\pm 0.18$ & $ 136.3\pm 1.8$ & $ 145.32\pm 0.17$ \\
$100\theta_\star $ & $ 1.052^{+0.029}_{-0.0069}$ & $ 1.04400\pm 0.00041$ & $ 1.04443\pm 0.00039$ & $ 1.0615\pm 0.0052$ & $ 1.04463\pm 0.00038$ \\
$z_\mathrm{re} $ & $ 8.32^{+0.11}_{-2.1}$ & $ 7.73\pm 0.54$ & $ 8.11\pm 0.53$ & $ 7.32\pm 0.22$ & $ 8.28\pm 0.50$ \\
$\sigma_8 (\Omega_\mathrm{m}/0.3)^{0.5} $ & $ 0.78^{+0.16}_{-0.36}$ & $ 0.8367\pm 0.0089$ & $ 0.8192\pm 0.0064$ & $ 0.80^{+0.21}_{-0.37}$ & $ 0.8138\pm 0.0060$ \\
\hline
$\chi^2$ & 1416.17 & 604.17 & 2024.51 & 1465.35 & 2103.85 

%% file: Tables_CCG/summarytable_Model-I.txt
Model I parameters  & PP+DESI & CMB-SPA & CMB-SPA+PP+DESI & PPS+DESI & CMB-SPA+PPS+DESI \\
\hline
$\Omega_\mathrm{b} h^2 $ & $ 0.0272^{+0.012}_{-0.0049}$ & $ 0.022397\pm 0.000097$ & $ 0.022356\pm 0.000094$ & $ 0.0238\pm 0.0013$ & $ 0.022386\pm 0.000093$ \\
$\Omega_\mathrm{c} h^2 $ & $ 0.161^{+0.039}_{-0.024}$ & $ 0.12035\pm 0.00093$ & $ 0.12089\pm 0.00058$ & $ 0.1485\pm 0.0055$ & $ 0.12057\pm 0.00058$ \\
$H_0 $ & $ 77^{+10}_{-5}$ & $ 71.97\pm 0.45$ & $ 71.70\pm 0.28$ & $ 74.3\pm 1.0$ & $ 71.88\pm 0.27$ \\
$\tau_\mathrm{reio} $ & $--$ & $ 0.0547\pm 0.0055$ & $ 0.0533\pm 0.0054$ & $--$ & $ 0.0540\pm 0.0053$ \\
$\log(10^{10} A_\mathrm{s}) $ & $ ---$ & $ 3.050\pm 0.010$ & $ 3.0463\pm 0.0097$ & $ ---$ & $ 3.0487\pm 0.0094$ \\
$n_\mathrm{s} $ & $ ---$ & $ 0.9692\pm 0.0032$ & $ 0.9683\pm 0.0029$ & $ ---$ & $ 0.9689\pm 0.0029$ \\
\hline
$\Omega_\mathrm{m} $ & $ 0.3128\pm 0.0079$ & $ 0.2769\pm 0.0051$ & $ 0.2799\pm 0.0031$ & $ 0.3133\pm 0.0080$ & $ 0.2780\pm 0.0031$ \\
$\sigma_8 $ & $ 0.93^{+0.20}_{-0.43}$ & $ 0.8546\pm 0.0040$ & $ 0.8548\pm 0.0041$ & $ 0.91^{+0.19}_{-0.43}$ & $ 0.8549\pm 0.0040$ \\
$z_\star $ & $ 1086.01^{+0.34}_{-11}$ & $ 1085.12\pm 0.57$ & $ 1085.35\pm 0.52$ & $ 1085.3^{+1.3}_{-1.5}$ & $ 1085.21\pm 0.51$ \\
$r_\star $ & $ 133.3^{+6.1}_{-16}$ & $ 144.68\pm 0.25$ & $ 144.56\pm 0.17$ & $ 137.1\pm 1.7$ & $ 144.62\pm 0.17$ \\
$100\theta_\star $ & $ 1.074^{+0.026}_{-0.0063}$ & $ 1.04398\pm 0.00041$ & $ 1.04390\pm 0.00038$ & $ 1.0734\pm 0.0049$ & $ 1.04398\pm 0.00038$ \\
$z_\mathrm{re} $ & $ 8.50^{+0.14}_{-2.0}$ & $ 7.69\pm 0.56$ & $ 7.57\pm 0.55$ & $ 8.44^{+0.25}_{-0.28}$ & $ 7.63\pm 0.54$ \\
$\sigma_8 (\Omega_\mathrm{m}/0.3)^{0.5} $ & $ 0.95^{+0.21}_{-0.44}$ & $ 0.8210\pm 0.0090$ & $ 0.8256\pm 0.0062$ & $ 0.93^{+0.20}_{-0.43}$ & $ 0.8229\pm 0.0061$ \\
\hline
$\chi^2$ & 1448.08 & 605.81 & 2080.22 & 1497.43 & 2137.27 \\
$\Delta\chi^2$ & 31.91 & 1.64 & 55.71 & 32.07 & 33.43 

%% file: Tables_CCG/summarytable_Model-II.txt
Model II parameters  & PP+DESI & CMB-SPA & CMB-SPA+PP+DESI & PPS+DESI & CMB-SPA+PPS+DESI \\
\hline
$\Omega_\mathrm{b} h^2 $ & $ 0.0237^{+0.011}_{-0.0045}$ & $ 0.022259\pm 0.000096$ & $ 0.021681\pm 0.000090$ & $ 0.01346\pm 0.00098$ & $ 0.021648\pm 0.000088$ \\
$\Omega_\mathrm{c} h^2 $ & $ 0.265^{+0.077}_{-0.039}$ & $ 0.12255\pm 0.00098$ & $ 0.13785\pm 0.00067$ & $ 0.1948\pm 0.0068$ & $ 0.13825\pm 0.00063$ \\
$H_0 $ & $ > 82.3$ & $ 83.27\pm 0.36$ & $ 78.00\pm 0.22$ & $ 73.8\pm 1.0$ & $ 77.84\pm 0.21$ \\
$\tau_\mathrm{reio} $ & $--$ & $ 0.0609\pm 0.0054$ & $ 0.0257^{+0.0060}_{-0.0053}$ & $--$ & $ 0.0247^{+0.0059}_{-0.0051}$ \\
$\log(10^{10} A_\mathrm{s}) $ & $ ---$ & $ 3.084\pm 0.010$ & $ 3.0072\pm 0.0097$ & $ ---$ & $ 3.0048\pm 0.0095$ \\
$n_\mathrm{s} $ & $ ---$ & $ 0.9657\pm 0.0032$ & $ 0.9375\pm 0.0029$ & $ ---$ & $ 0.9369\pm 0.0028$ \\
\hline
$\Omega_\mathrm{m} $ & $ 0.3834\pm 0.0080$ & $ 0.2098\pm 0.0031$ & $ 0.2633\pm 0.0025$ & $ 0.3834\pm 0.0080$ & $ 0.2650\pm 0.0023$ \\
$\sigma_8 $ & $ 1.39^{+0.31}_{-0.63}$ & $ 0.9565\pm 0.0051$ & $ 0.9828\pm 0.0046$ & $ 1.22^{+0.24}_{-0.56}$ & $ 0.9832\pm 0.0046$ \\
$z_\star $ & $ 1098.198^{-0.027}_{-14}$ & $ 1084.94\pm 0.57$ & $ 1090.22\pm 0.55$ & $ 1112.2^{+3.1}_{-3.8}$ & $ 1090.39\pm 0.53$ \\
$r_\star $ & $ 118.9^{+5.1}_{-15}$ & $ 144.27\pm 0.25$ & $ 140.68\pm 0.17$ & $ 134.1\pm 1.6$ & $ 140.60\pm 0.17$ \\
$100\theta_\star $ & $ 1.116^{+0.028}_{-0.0051}$ & $ 1.04416\pm 0.00041$ & $ 1.04092\pm 0.00039$ & $ 1.0893^{+0.0055}_{-0.0049}$ & $ 1.04079\pm 0.00039$ \\
$z_\mathrm{re} $ & $ 10.735^{+0.065}_{-2.6}$ & $ 8.15\pm 0.53$ & $ 4.44^{+0.84}_{-0.68}$ & $ 13.28^{+0.57}_{-0.67}$ & $ 4.32^{+0.84}_{-0.65}$ \\
$\sigma_8 (\Omega_\mathrm{m}/0.3)^{0.5} $ & $ 1.57^{+0.35}_{-0.72}$ & $ 0.7998\pm 0.0087$ & $ 0.9207\pm 0.0065$ & $ 1.38^{+0.28}_{-0.63}$ & $ 0.9240\pm 0.0063$ \\
\hline
$\chi^2$ & 1454.23 & 681.56 & 2656.34 & 1503.64 & 2716.8 \\
$\Delta\chi^2$ & 38.06 & 77.39 & 631.84 & 38.29 & 612.95

%% file: biblio.bib
@ARTICLE{LIGOScientific:2016aoc,
  author       = {Abbott, B. P. and others},
  collaboration= {LIGO Scientific, Virgo},
  title        = {{Observation of Gravitational Waves from a Binary Black Hole Merger}},
  eprint       = {1602.03837},
  archiveprefix= {arXiv},
  primaryclass = {gr-qc},
  reportnumber = {LIGO-P150914},
  doi          = {10.1103/PhysRevLett.116.061102},
  journal      = {Phys. Rev. Lett.},
  volume       = 116,
  number       = 6,
  pages        = 061102,
  year         = 2016,
}

@ARTICLE{Abdalla:2022yfr,
  author       = {Abdalla, Elcio and others},
  title        = {{Cosmology intertwined: A review of the particle physics, astrophysics, and cosmology associated with the cosmological tensions and anomalies}},
  eprint       = {2203.06142},
  archiveprefix= {arXiv},
  primaryclass = {astro-ph.CO},
  reportnumber = {FERMILAB-CONF-22-192-SCD},
  doi          = {10.1016/j.jheap.2022.04.002},
  journal      = {JHEAp},
  volume       = 34,
  pages        = {49--211},
  year         = 2022,
}

@ARTICLE{DESI:2025zgx,
  author       = {Abdul Karim, M. and others},
  collaboration= {DESI},
  title        = {{DESI DR2 results. II. Measurements of baryon acoustic oscillations and cosmological constraints}},
  eprint       = {2503.14738},
  archiveprefix= {arXiv},
  primaryclass = {astro-ph.CO},
  reportnumber = {FERMILAB-PUB-25-0169-PPD},
  doi          = {10.1103/tr6y-kpc6},
  journal      = {Phys. Rev. D},
  volume       = 112,
  number       = 8,
  pages        = 083515,
  year         = 2025,
}

@ARTICLE{DESI:2024mwx,
  author       = {Adame, A. G. and others},
  collaboration= {DESI},
  title        = {{DESI 2024 VI: cosmological constraints from the measurements of baryon acoustic oscillations}},
  eprint       = {2404.03002},
  archiveprefix= {arXiv},
  primaryclass = {astro-ph.CO},
  reportnumber = {FERMILAB-PUB-24-0154-PPD},
  doi          = {10.1088/1475-7516/2025/02/021},
  journal      = {JCAP},
  volume       = 02,
  pages        = 021,
  year         = 2025,
}

@ARTICLE{Adams_2022,
  title        = {Discovery and properties of ultra-high redshift galaxies ($9 < z < 12$) in the JWST ERO SMACS 0723 Field},
  volume       = 518,
  issn         = {1365-2966},
  url          = {http://dx.doi.org/10.1093/mnras/stac3347},
  doi          = {10.1093/mnras/stac3347},
  number       = 3,
  journal      = {Monthly Notices of the Royal Astronomical Society},
  publisher    = {Oxford University Press (OUP)},
  author       = {Adams, N J and Conselice, C J and Ferreira, L and Austin, D and Trussler, J A A and Juodžbalis, I and Wilkins, S M and Caruana, J and Dayal, P and Verma, A and Vijayan, A P},
  year         = 2022,
  month        = nov, pages={4755–4766},
}

@ARTICLE{COSINE-100:2021xqn,
  author       = {Adhikari, Govinda and others},
  collaboration= {COSINE-100},
  title        = {{Strong constraints from COSINE-100 on the DAMA dark matter results using the same sodium iodide target}},
  eprint       = {2104.03537},
  archiveprefix= {arXiv},
  primaryclass = {hep-ex},
  doi          = {10.1126/sciadv.abk2699},
  journal      = {Sci. Adv.},
  volume       = 7,
  number       = 46,
  pages        = {abk2699},
  year         = 2021,
}

@ARTICLE{Planck:2018vyg,
  author       = {Aghanim, N. and others},
  collaboration= {Planck},
  title        = {{Planck 2018 results. VI. Cosmological parameters}},
  eprint       = {1807.06209},
  archiveprefix= {arXiv},
  primaryclass = {astro-ph.CO},
  doi          = {10.1051/0004-6361/201833910},
  journal      = {Astron. Astrophys.},
  volume       = 641,
  pages        = A6,
  year         = 2020,
  note         = {[Erratum: Astron.Astrophys. 652, C4 (2021)]},
}

@ARTICLE{Planck:2018nkj,
  author       = {Aghanim, N. and others},
  collaboration= {Planck},
  title        = {{Planck 2018 results. I. Overview and the cosmological legacy of Planck}},
  eprint       = {1807.06205},
  archiveprefix= {arXiv},
  primaryclass = {astro-ph.CO},
  doi          = {10.1051/0004-6361/201833880},
  journal      = {Astron. Astrophys.},
  volume       = 641,
  pages        = A1,
  year         = 2020,
}

@ARTICLE{Planck:2019nip,
  author       = {Aghanim, N. and others},
  collaboration= {Planck},
  title        = {{Planck 2018 results. V. CMB power spectra and likelihoods}},
  eprint       = {1907.12875},
  archiveprefix= {arXiv},
  primaryclass = {astro-ph.CO},
  doi          = {10.1051/0004-6361/201936386},
  journal      = {Astron. Astrophys.},
  volume       = 641,
  pages        = A5,
  year         = 2020,
}

@ARTICLE{Agrawal:2025tuv,
  author       = {Agrawal, Aadya and others},
  title        = {{Testing Lens Models of PLCK G165.7+67.0 Using Lensed Supernova H0pe}},
  eprint       = {2510.07637},
  archiveprefix= {arXiv},
  primaryclass = {astro-ph.CO},
  doi          = {10.3847/1538-4357/ae5f65},
  journal      = {Astrophys. J.},
  volume       = 1002,
  number       = 2,
  pages        = 187,
  year         = 2026,
}

@ARTICLE{Akarsu:2024nas,
  author       = {Akarsu, Ozgur and Bulduk, Bilal and De Felice, Antonio and Kat{\i}rc{\i}, Nihan and Uzun, N. Merve},
  title        = {{Unexplored regions in teleparallel f(T) gravity: Sign-changing dark energy density}},
  eprint       = {2410.23068},
  archiveprefix= {arXiv},
  primaryclass = {gr-qc},
  reportnumber = {YITP-24-119},
  doi          = {10.1103/1xd4-k91h},
  journal      = {Phys. Rev. D},
  volume       = 112,
  number       = 8,
  pages        = 083532,
  year         = 2025,
}

@ARTICLE{Akarsu:2025ijk,
  author       = {Akarsu, {\"O}zg{\"u}r and {\c{C}}am, Arman and Paraskevas, Evangelos A. and Perivolaropoulos, Leandros},
  title        = {{Linear matter density perturbations in the {\ensuremath{\Lambda}}$_{s}$CDM model: Examining growth dynamics and addressing the S $_{8}$ tension}},
  eprint       = {2502.20384},
  archiveprefix= {arXiv},
  primaryclass = {astro-ph.CO},
  doi          = {10.1088/1475-7516/2025/08/089},
  journal      = {JCAP},
  volume       = 08,
  pages        = 089,
  year         = 2025,
}

@ARTICLE{Akarsu:2024eoo,
  author       = {Akarsu, {\"O}zg{\"u}r and De Felice, Antonio and Di Valentino, Eleonora and Kumar, Suresh and Nunes, Rafael C. and {\"O}z{\"u}lker, Emre and Vazquez, J. Alberto and Yadav, Anita},
  title        = {{Cosmological constraints on {\ensuremath{\Lambda}}sCDM scenario in a type II minimally modified gravity}},
  eprint       = {2406.07526},
  archiveprefix= {arXiv},
  primaryclass = {astro-ph.CO},
  reportnumber = {YITP-24-57},
  doi          = {10.1103/PhysRevD.110.103527},
  journal      = {Phys. Rev. D},
  volume       = 110,
  number       = 10,
  pages        = 103527,
  year         = 2024,
}

@ARTICLE{Akarsu:2024qsi,
  author       = {Akarsu, {\"O}zg{\"u}r and De Felice, Antonio and Di Valentino, Eleonora and Kumar, Suresh and Nunes, Rafael C. and {\"O}z{\"u}lker, Emre and Vazquez, J. Alberto and Yadav, Anita},
  title        = {{{\ensuremath{\Lambda}}sCDM cosmology from a type-II minimally modified gravity}},
  eprint       = {2402.07716},
  archiveprefix= {arXiv},
  primaryclass = {astro-ph.CO},
  reportnumber = {YITP-24-18},
  doi          = {10.1093/mnras/staf2276},
  journal      = {Mon. Not. Roy. Astron. Soc.},
  volume       = 546,
  number       = 1,
  pages        = {staf2276},
  year         = 2026,
}

@ARTICLE{Akarsu:2021fol,
  author       = {Akarsu, {\"O}zg{\"u}r and Kumar, Suresh and {\"O}z{\"u}lker, Emre and Vazquez, J. Alberto},
  title        = {{Relaxing cosmological tensions with a sign switching cosmological constant}},
  eprint       = {2108.09239},
  archiveprefix= {arXiv},
  primaryclass = {astro-ph.CO},
  doi          = {10.1103/PhysRevD.104.123512},
  journal      = {Phys. Rev. D},
  volume       = 104,
  number       = 12,
  pages        = 123512,
  year         = 2021,
}

@ARTICLE{Akarsu:2022typ,
  author       = {Akarsu, Ozgur and Kumar, Suresh and {\"O}z{\"u}lker, Emre and Vazquez, J. Alberto and Yadav, Anita},
  title        = {{Relaxing cosmological tensions with a sign switching cosmological constant: Improved results with Planck, BAO, and Pantheon data}},
  eprint       = {2211.05742},
  archiveprefix= {arXiv},
  primaryclass = {astro-ph.CO},
  doi          = {10.1103/PhysRevD.108.023513},
  journal      = {Phys. Rev. D},
  volume       = 108,
  number       = 2,
  pages        = 023513,
  year         = 2023,
}

@ARTICLE{Akarsu:2025gwi,
  author       = {Akarsu, {\"O}zg{\"u}r and Perivolaropoulos, Leandros and Tsikoundoura, Anna and Y{\"u}kselci, A. Emrah and Zhuk, Alexander},
  title        = {{Dynamical dark energy with AdS-to-dS and dS-to-dS transitions: Implications for the $H_0$ tension}},
  eprint       = {2502.14667},
  archiveprefix= {arXiv},
  primaryclass = {astro-ph.CO},
  month        = 2,
  year         = 2025,
}

@ARTICLE{EventHorizonTelescope:2019dse,
  author       = {Akiyama, Kazunori and others},
  collaboration= {Event Horizon Telescope},
  title        = {{First M87 Event Horizon Telescope Results. I. The Shadow of the Supermassive Black Hole}},
  eprint       = {1906.11238},
  archiveprefix= {arXiv},
  primaryclass = {astro-ph.GA},
  doi          = {10.3847/2041-8213/ab0ec7},
  journal      = {Astrophys. J. Lett.},
  volume       = 875,
  pages        = L1,
  year         = 2019,
}

@ARTICLE{EventHorizonTelescope:2019pgp,
  author       = {Akiyama, Kazunori and others},
  collaboration= {Event Horizon Telescope},
  title        = {{First M87 Event Horizon Telescope Results. V. Physical Origin of the Asymmetric Ring}},
  eprint       = {1906.11242},
  archiveprefix= {arXiv},
  primaryclass = {astro-ph.GA},
  doi          = {10.3847/2041-8213/ab0f43},
  journal      = {Astrophys. J. Lett.},
  volume       = 875,
  number       = 1,
  pages        = L5,
  year         = 2019,
}

@ARTICLE{Anand:2021sum,
  author       = {Anand, Gagandeep S. and Tully, R. Brent and Rizzi, Luca and Riess, Adam G. and Yuan, Wenlong},
  title        = {{Comparing Tip of the Red Giant Branch Distance Scales: An Independent Reduction of the Carnegie-Chicago Hubble Program and the Value of the Hubble Constant}},
  eprint       = {2108.00007},
  archiveprefix= {arXiv},
  primaryclass = {astro-ph.CO},
  doi          = {10.3847/1538-4357/ac68df},
  journal      = {Astrophys. J.},
  volume       = 932,
  number       = 1,
  pages        = 15,
  year         = 2022,
}

@ARTICLE{Anchordoqui:2024gfa,
  author       = {Anchordoqui, Luis A. and Antoniadis, Ignatios and Lust, Dieter and Noble, Neena T. and Soriano, Jorge F.},
  title        = {{From infinite to infinitesimal: Using the universe as a dataset to probe Casimir corrections to the vacuum energy from fields inhabiting the dark dimension}},
  eprint       = {2404.17334},
  archiveprefix= {arXiv},
  primaryclass = {astro-ph.CO},
  reportnumber = {MPP-2024-91; LMU-ASC 05/24},
  doi          = {10.1016/j.dark.2024.101715},
  journal      = {Phys. Dark Univ.},
  volume       = 46,
  pages        = 101715,
  year         = 2024,
}

@ARTICLE{Anderson:2023aga,
  author       = {Anderson, Richard I. and Koblischke, Nolan W. and Eyer, Laurent},
  title        = {{Small-amplitude Red Giants Elucidate the Nature of the Tip of the Red Giant Branch as a Standard Candle}},
  eprint       = {2303.04790},
  archiveprefix= {arXiv},
  primaryclass = {astro-ph.CO},
  doi          = {10.3847/2041-8213/ad284d},
  journal      = {Astrophys. J. Lett.},
  volume       = 963,
  number       = 2,
  pages        = L43,
  year         = 2024,
}

@ARTICLE{Awad:2017yod,
  author       = {Awad, A. and El Hanafy, W. and Nashed, G. G. L. and Saridakis, Emmanuel N.},
  title        = {{Phase Portraits of general f(T) Cosmology}},
  eprint       = {1710.10194},
  archiveprefix= {arXiv},
  primaryclass = {gr-qc},
  doi          = {10.1088/1475-7516/2018/02/052},
  journal      = {JCAP},
  volume       = 02,
  pages        = 052,
  year         = 2018,
}

@ARTICLE{Awad:2013tha,
  author       = {Awad, Adel},
  title        = {{Fixed points and FLRW cosmologies: Flat case}},
  eprint       = {1303.2014},
  archiveprefix= {arXiv},
  primaryclass = {gr-qc},
  doi          = {10.1103/PhysRevD.87.103001},
  journal      = {Phys. Rev. D},
  volume       = 87,
  number       = 10,
  pages        = 103001,
  year         = 2013,
  note         = {[Erratum: Phys.Rev.D 87, 109902 (2013)]},
}

@ARTICLE{Bahamonde:2021gfp,
  author       = {Bahamonde, Sebastian and Dialektopoulos, Konstantinos F. and Escamilla-Rivera, Celia and Farrugia, Gabriel and Gakis, Viktor and Hendry, Martin and Hohmann, Manuel and Levi Said, Jackson and Mifsud, Jurgen and Di Valentino, Eleonora},
  title        = {{Teleparallel gravity: from theory to cosmology}},
  eprint       = {2106.13793},
  archiveprefix= {arXiv},
  primaryclass = {gr-qc},
  doi          = {10.1088/1361-6633/ac9cef},
  journal      = {Rept. Prog. Phys.},
  volume       = 86,
  number       = 2,
  pages        = 026901,
  year         = 2023,
}

@ARTICLE{2023MNRAS.519.5076B,
  author       = {{Bakx}, Tom J.~L.~C. and {Zavala}, Jorge A. and {Mitsuhashi}, Ikki and {Treu}, Tommaso and {Fontana}, Adriano and {Tadaki}, Ken-ichi and {Casey}, Caitlin M. and {Castellano}, Marco and {Glazebrook}, Karl and {Hagimoto}, Masato and {Ikeda}, Ryota and {Jones}, Tucker and {Leethochawalit}, Nicha and {Mason}, Charlotte and {Morishita}, Takahiro and {Nanayakkara}, Themiya and {Pentericci}, Laura and {Roberts-Borsani}, Guido and {Santini}, Paola and {Serjeant}, Stephen and {Tamura}, Yoichi and {Trenti}, Michele and {Vanzella}, Eros},
  title        = {{Deep ALMA redshift search of a z {\ensuremath{\sim}} 12 GLASS-JWST galaxy candidate}},
  journal      = {Mon. Not. Roy. Astron. Soc.},
  year         = 2023,
  month        = mar,
  volume       = 519,
  number       = 4,
  pages        = {5076-5085},
  doi          = {10.1093/mnras/stac3723},
  archiveprefix= {arXiv},
  eprint       = {2208.13642},
  primaryclass = {astro-ph.GA},
  adsurl       = {https://ui.adsabs.harvard.edu/abs/2023MNRAS.519.5076B},
}

@ARTICLE{Bean:2010zq,
  author       = {Bean, Rachel and Tangmatitham, Matipon},
  title        = {{Current constraints on the cosmic growth history}},
  eprint       = {1002.4197},
  archiveprefix= {arXiv},
  primaryclass = {astro-ph.CO},
  doi          = {10.1103/PhysRevD.81.083534},
  journal      = {Phys. Rev. D},
  volume       = 81,
  pages        = 083534,
  year         = 2010,
}

@ARTICLE{Bella:2026zuk,
  author       = {Bella, Marco and Poulin, Vivian and Vagnozzi, Sunny and Knox, Lloyd},
  title        = {{Double the axions, half the tension: multi-field early dark energy eases the Hubble tension}},
  eprint       = {2604.13535},
  archiveprefix= {arXiv},
  primaryclass = {astro-ph.CO},
  month        = 4,
  year         = 2026,
}

@ARTICLE{Benisty:2024lmj,
  author       = {Benisty, David and Pan, Supriya and Staicova, Denitsa and Di Valentino, Eleonora and Nunes, Rafael C.},
  title        = {{Late-time constraints on interacting dark energy: Analysis independent of H0, rd, and MB}},
  eprint       = {2403.00056},
  archiveprefix= {arXiv},
  primaryclass = {astro-ph.CO},
  doi          = {10.1051/0004-6361/202449883},
  journal      = {Astron. Astrophys.},
  volume       = 688,
  pages        = A156,
  year         = 2024,
}

@ARTICLE{Benisty:2025tct,
  author       = {Benisty, David and Wagner, Jenny and Haridasu, Sandeep and Salucci, Paolo},
  title        = {{Unveiling the Coma cluster structure: from the core to the Hubble flow}},
  eprint       = {2504.04135},
  archiveprefix= {arXiv},
  primaryclass = {astro-ph.CO},
  doi          = {10.1088/1475-7516/2026/06/005},
  journal      = {JCAP},
  volume       = 06,
  pages        = 005,
  year         = 2026,
}

@ARTICLE{Bertschinger:2008zb,
  author       = {Bertschinger, Edmund and Zukin, Phillip},
  title        = {{Distinguishing Modified Gravity from Dark Energy}},
  eprint       = {0801.2431},
  archiveprefix= {arXiv},
  primaryclass = {astro-ph},
  doi          = {10.1103/PhysRevD.78.024015},
  journal      = {Phys. Rev. D},
  volume       = 78,
  pages        = 024015,
  year         = 2008,
}

@ARTICLE{Bhardwaj:2025kbw,
  author       = {Bhardwaj, Anupam and Matsunaga, Noriyuki and Huang, Caroline D. and Riess, Adam G. and Rejkuba, Marina},
  title        = {{Absolute Calibration of Cluster Mira Variables to Provide a New Anchor for the Hubble Constant Determination}},
  eprint       = {2507.10658},
  archiveprefix= {arXiv},
  primaryclass = {astro-ph.GA},
  doi          = {10.3847/1538-4357/adf20b},
  journal      = {Astrophys. J.},
  volume       = 990,
  number       = 1,
  pages        = 63,
  year         = 2025,
}

@ARTICLE{Bhattacharya:2024hep,
  author       = {Bhattacharya, Sukannya and Borghetto, Giulia and Malhotra, Ameek and Parameswaran, Susha and Tasinato, Gianmassimo and Zavala, Ivonne},
  title        = {{Cosmological constraints on curved quintessence}},
  eprint       = {2405.17396},
  archiveprefix= {arXiv},
  primaryclass = {astro-ph.CO},
  doi          = {10.1088/1475-7516/2024/09/073},
  journal      = {JCAP},
  volume       = 09,
  pages        = 073,
  year         = 2024,
}

@ARTICLE{Birrer:2020tax,
  author       = {Birrer, S. and others},
  title        = {{TDCOSMO - IV. Hierarchical time-delay cosmography {\textendash} joint inference of the Hubble constant and galaxy density profiles}},
  eprint       = {2007.02941},
  archiveprefix= {arXiv},
  primaryclass = {astro-ph.CO},
  doi          = {10.1051/0004-6361/202038861},
  journal      = {Astron. Astrophys.},
  volume       = 643,
  pages        = A165,
  year         = 2020,
}

@ARTICLE{Blakeslee:2021rqi,
  author       = {Blakeslee, John P. and Jensen, Joseph B. and Ma, Chung-Pei and Milne, Peter A. and Greene, Jenny E.},
  title        = {{The Hubble Constant from Infrared Surface Brightness Fluctuation Distances}},
  eprint       = {2101.02221},
  archiveprefix= {arXiv},
  primaryclass = {astro-ph.CO},
  doi          = {10.3847/1538-4357/abe86a},
  journal      = {Astrophys. J.},
  volume       = 911,
  number       = 1,
  pages        = 65,
  year         = 2021,
}

@ARTICLE{Boubel:2024cqw,
  author       = {Boubel, Paula and Colless, Matthew and Said, Khaled and Staveley-Smith, Lister},
  title        = {{An improved Tully{\textendash}Fisher estimate of H0}},
  eprint       = {2408.03660},
  archiveprefix= {arXiv},
  primaryclass = {astro-ph.CO},
  doi          = {10.1093/mnras/stae1925},
  journal      = {Mon. Not. Roy. Astron. Soc.},
  volume       = 533,
  number       = 2,
  pages        = {1550--1559},
  year         = 2024,
}

@ARTICLE{Bouhmadi-Lopez:2025ggl,
  author       = {Bouhmadi-L{\'o}pez, Mariam and Ibarra-Uriondo, Be{\~n}at},
  title        = {{Cosmographic analysis of sign-switching dark energy}},
  eprint       = {2506.12139},
  archiveprefix= {arXiv},
  primaryclass = {gr-qc},
  doi          = {10.1103/v1cl-pr54},
  journal      = {Phys. Rev. D},
  volume       = 112,
  number       = 6,
  pages        = 063559,
  year         = 2025,
}

@ARTICLE{Bousis:2024rnb,
  author       = {Bousis, Dimitrios and Perivolaropoulos, Leandros},
  title        = {{Hubble tension tomography: BAO vs SN Ia distance tension}},
  eprint       = {2405.07039},
  archiveprefix= {arXiv},
  primaryclass = {astro-ph.CO},
  doi          = {10.1103/PhysRevD.110.103546},
  journal      = {Phys. Rev. D},
  volume       = 110,
  number       = 10,
  pages        = 103546,
  year         = 2024,
}

@ARTICLE{Breuval:2024lsv,
  author       = {Breuval, Louise and Riess, Adam G. and Casertano, Stefano and Yuan, Wenlong and Macri, Lucas M. and Romaniello, Martino and Murakami, Yukei S. and Scolnic, Daniel and Anand, Gagandeep S. and Soszy{\'n}ski, Igor},
  title        = {{Small Magellanic Cloud Cepheids Observed with the Hubble Space Telescope Provide a New Anchor for the SH0ES Distance Ladder}},
  eprint       = {2404.08038},
  archiveprefix= {arXiv},
  primaryclass = {astro-ph.CO},
  doi          = {10.3847/1538-4357/ad630e},
  journal      = {Astrophys. J.},
  volume       = 973,
  number       = 1,
  pages        = 30,
  year         = 2024,
}

@ARTICLE{Brout:2022vxf,
  author       = {Brout, Dillon and others},
  title        = {{The Pantheon+ Analysis: Cosmological Constraints}},
  eprint       = {2202.04077},
  archiveprefix= {arXiv},
  primaryclass = {astro-ph.CO},
  doi          = {10.3847/1538-4357/ac8e04},
  journal      = {Astrophys. J.},
  volume       = 938,
  number       = 2,
  pages        = 110,
  year         = 2022,
}

@ARTICLE{SPT-3G:2025bzu,
  author       = {Camphuis, E. and others},
  collaboration= {SPT-3G},
  title        = {{SPT-3G D1: CMB temperature and polarization power spectra and cosmology from 2019 and 2020 observations of the SPT-3G main field}},
  eprint       = {2506.20707},
  archiveprefix= {arXiv},
  primaryclass = {astro-ph.CO},
  reportnumber = {FERMILAB-PUB-25-0144-PPD},
  doi          = {10.1103/7wt3-9v2y},
  journal      = {Phys. Rev. D},
  volume       = 113,
  number       = 8,
  pages        = 083504,
  year         = 2026,
}

@ARTICLE{Carloni:2026yut,
  author       = {Carloni, Youri and Luongo, Orlando},
  title        = {{A barotropic alternative to Early Dark Energy for alleviating the $H_0$ tension}},
  eprint       = {2604.18053},
  archiveprefix= {arXiv},
  primaryclass = {astro-ph.CO},
  month        = 4,
  year         = 2026,
}

@ARTICLE{Carloni:2024zpl,
  author       = {Carloni, Youri and Luongo, Orlando and Muccino, Marco},
  title        = {{Does dark energy really revive using DESI 2024 data?}},
  eprint       = {2404.12068},
  archiveprefix= {arXiv},
  primaryclass = {astro-ph.CO},
  doi          = {10.1103/PhysRevD.111.023512},
  journal      = {Phys. Rev. D},
  volume       = 111,
  number       = 2,
  pages        = 023512,
  year         = 2025,
}

@ARTICLE{Carron:2022eyg,
  author       = {Carron, Julien and Mirmelstein, Mark and Lewis, Antony},
  title        = {{CMB lensing from Planck PR4~maps}},
  eprint       = {2206.07773},
  archiveprefix= {arXiv},
  primaryclass = {astro-ph.CO},
  doi          = {10.1088/1475-7516/2022/09/039},
  journal      = {JCAP},
  volume       = 09,
  pages        = 039,
  year         = 2022,
}

@ARTICLE{H0DN:2025lyy,
  author       = {Casertano, Stefano and others},
  collaboration= {H0DN},
  title        = {{The Local Distance Network: A community consensus report on the measurement of the Hubble constant at {\ensuremath{\sim}}1{\%} precision}},
  eprint       = {2510.23823},
  archiveprefix= {arXiv},
  primaryclass = {astro-ph.CO},
  doi          = {10.1051/0004-6361/202557993},
  journal      = {Astron. Astrophys.},
  volume       = 708,
  pages        = A166,
  year         = 2026,
}

@ARTICLE{Cepa:2004bc,
  author       = {Cepa, Jordi},
  title        = {{Constraints on the cosmic equation of state: Age conflict versus phantom energy. Age - redshift relations in an accelerated universe}},
  eprint       = {astro-ph/0403616},
  archiveprefix= {arXiv},
  reportnumber = {PP-12-2004},
  doi          = {10.1051/0004-6361:20035734},
  journal      = {Astron. Astrophys.},
  volume       = 422,
  pages        = {831--839},
  year         = 2004,
}

@ARTICLE{Cheng:2025lod,
  author       = {Cheng, Hanyu and Di Valentino, Eleonora and Escamilla, Luis A. and Sen, Anjan A. and Visinelli, Luca},
  title        = {{Pressure parametrization of dark energy: first and second-order constraints with latest cosmological data}},
  eprint       = {2505.02932},
  archiveprefix= {arXiv},
  primaryclass = {astro-ph.CO},
  reportnumber = {CA21106; CA21136},
  doi          = {10.1088/1475-7516/2025/09/031},
  journal      = {JCAP},
  volume       = 09,
  pages        = 031,
  year         = 2025,
}

@ARTICLE{Cheng:2025hug,
  author       = {Cheng, Hanyu and Di Valentino, Eleonora and Visinelli, Luca},
  title        = {{Cosmic strings as dynamical dark energy: Novel constraints}},
  eprint       = {2505.22066},
  archiveprefix= {arXiv},
  primaryclass = {astro-ph.CO},
  doi          = {10.1016/j.jheap.2026.100610},
  journal      = {JHEAp},
  volume       = 53,
  pages        = 100610,
  year         = 2026,
}

@ARTICLE{Cheng:2025yue,
  author       = {Cheng, Hanyu and Pan, Supriya and Di Valentino, Eleonora},
  title        = {{Beyond Two Parameters: Revisiting Dark Energy with the Latest Cosmic Probes}},
  eprint       = {2512.09866},
  archiveprefix= {arXiv},
  primaryclass = {astro-ph.CO},
  doi          = {10.3847/1538-4357/ae3a8f},
  journal      = {Astrophys. J.},
  volume       = 999,
  number       = 2,
  pages        = 190,
  year         = 2026,
}

@ARTICLE{Chluba:2023xqj,
  author       = {Chluba, Jens and Hart, Luke},
  title        = {{Varying fundamental constants meet Hubble}},
  eprint       = {2309.12083},
  archiveprefix= {arXiv},
  primaryclass = {astro-ph.CO},
  month        = 9,
  year         = 2023,
}

@ARTICLE{Cooke:2017cwo,
  author       = {Cooke, Ryan J. and Pettini, Max and Steidel, Charles C.},
  title        = {{One Percent Determination of the Primordial Deuterium Abundance}},
  eprint       = {1710.11129},
  archiveprefix= {arXiv},
  primaryclass = {astro-ph.CO},
  doi          = {10.3847/1538-4357/aaab53},
  journal      = {Astrophys. J.},
  volume       = 855,
  number       = 2,
  pages        = 102,
  year         = 2018,
}

@ARTICLE{Cortes:2024lgw,
  author       = {Cort{\^e}s, Marina and Liddle, Andrew R.},
  title        = {{Interpreting DESI's evidence for evolving dark energy}},
  eprint       = {2404.08056},
  archiveprefix= {arXiv},
  primaryclass = {astro-ph.CO},
  doi          = {10.1088/1475-7516/2024/12/007},
  journal      = {JCAP},
  volume       = 12,
  pages        = 007,
  year         = 2024,
}

@INPROCEEDINGS{Craig:2013cxa,
  author       = {Craig, Nathaniel},
  title        = {{The State of Supersymmetry after Run I of the LHC}},
  booktitle    = {{Beyond the Standard Model after the first run of the LHC}},
  eprint       = {1309.0528},
  archiveprefix= {arXiv},
  primaryclass = {hep-ph},
  month        = 9,
  year         = 2013,
}

@ARTICLE{Craig:2024tky,
  author       = {Craig, Nathaniel and Green, Daniel and Meyers, Joel and Rajendran, Surjeet},
  title        = {{No {\ensuremath{\nu}}s is Good News}},
  eprint       = {2405.00836},
  archiveprefix= {arXiv},
  primaryclass = {astro-ph.CO},
  reportnumber = {FERMILAB-PUB-24-0492-SQMS-V},
  doi          = {10.1007/JHEP09(2024)097},
  journal      = {JHEP},
  volume       = 09,
  pages        = 097,
  year         = 2024,
}

@ARTICLE{deJaeger:2022lit,
  author       = {de Jaeger, T. and Galbany, L. and Riess, A. G. and Stahl, B. E. and Shappee, B. J. and Filippenko, A. V. and Zheng, W.},
  title        = {{A 5~per{\,}cent measurement of the Hubble{\textendash}Lema{\^\i}tre constant from Type II supernovae}},
  eprint       = {2203.08974},
  archiveprefix= {arXiv},
  primaryclass = {astro-ph.CO},
  doi          = {10.1093/mnras/stac1661},
  journal      = {Mon. Not. Roy. Astron. Soc.},
  volume       = 514,
  number       = 3,
  pages        = {4620--4628},
  year         = 2022,
}

@ARTICLE{DiValentino:2020vnx,
  author       = {Di Valentino, Eleonora},
  title        = {{A combined analysis of the $H_0$ late time direct measurements and the impact on the Dark Energy sector}},
  eprint       = {2011.00246},
  archiveprefix= {arXiv},
  primaryclass = {astro-ph.CO},
  reportnumber = {IPPP/20/72},
  doi          = {10.1093/mnras/stab187},
  journal      = {Mon. Not. Roy. Astron. Soc.},
  volume       = 502,
  number       = 2,
  pages        = {2065--2073},
  year         = 2021,
}

@ARTICLE{DiValentino:2022fjm,
  author       = {Di Valentino, Eleonora},
  title        = {{Challenges of the Standard Cosmological Model}},
  doi          = {10.3390/universe8080399},
  journal      = {Universe},
  volume       = 8,
  number       = 8,
  pages        = 399,
  year         = 2022,
}

@BOOK{DiValentino:2024yew,
  editor       = {Di Valentino, Eleonora and Brout, Dillon},
  title        = {{The Hubble Constant Tension}},
  doi          = {10.1007/978-981-99-0177-7},
  isbn         = {978-981-99-0176-0, 978-981-99-0179-1, 978-981-99-0177-7},
  publisher    = {Springer},
  series       = {Springer Series in Astrophysics and Cosmology},
  year         = 2024,
}

@ARTICLE{DiValentino:2017iww,
  author       = {Di Valentino, Eleonora and Melchiorri, Alessandro and Mena, Olga},
  title        = {{Can interacting dark energy solve the $H_0$ tension?}},
  eprint       = {1704.08342},
  archiveprefix= {arXiv},
  primaryclass = {astro-ph.CO},
  doi          = {10.1103/PhysRevD.96.043503},
  journal      = {Phys. Rev. D},
  volume       = 96,
  number       = 4,
  pages        = 043503,
  year         = 2017,
}

@ARTICLE{DiValentino:2019ffd,
  author       = {Di Valentino, Eleonora and Melchiorri, Alessandro and Mena, Olga and Vagnozzi, Sunny},
  title        = {{Interacting dark energy in the early 2020s: A promising solution to the $H_0$ and cosmic shear tensions}},
  eprint       = {1908.04281},
  archiveprefix= {arXiv},
  primaryclass = {astro-ph.CO},
  doi          = {10.1016/j.dark.2020.100666},
  journal      = {Phys. Dark Univ.},
  volume       = 30,
  pages        = 100666,
  year         = 2020,
}

@ARTICLE{DiValentino:2016hlg,
  author       = {Di Valentino, Eleonora and Melchiorri, Alessandro and Silk, Joseph},
  title        = {{Reconciling Planck with the local value of $H_0$ in extended parameter space}},
  eprint       = {1606.00634},
  archiveprefix= {arXiv},
  primaryclass = {astro-ph.CO},
  doi          = {10.1016/j.physletb.2016.08.043},
  journal      = {Phys. Lett. B},
  volume       = 761,
  pages        = {242--246},
  year         = 2016,
}

@ARTICLE{DiValentino:2021izs,
  author       = {Di Valentino, Eleonora and Mena, Olga and Pan, Supriya and Visinelli, Luca and Yang, Weiqiang and Melchiorri, Alessandro and Mota, David F. and Riess, Adam G. and Silk, Joseph},
  title        = {{In the realm of the Hubble tension{\textemdash}a review of solutions}},
  eprint       = {2103.01183},
  archiveprefix= {arXiv},
  primaryclass = {astro-ph.CO},
  reportnumber = {IPPP/20/108},
  doi          = {10.1088/1361-6382/ac086d},
  journal      = {Class. Quant. Grav.},
  volume       = 38,
  number       = 15,
  pages        = 153001,
  year         = 2021,
}

@ARTICLE{DiValentino:2020naf,
  author       = {Di Valentino, Eleonora and Mukherjee, Ankan and Sen, Anjan A.},
  title        = {{Dark Energy with Phantom Crossing and the $H_0$ Tension}},
  eprint       = {2005.12587},
  archiveprefix= {arXiv},
  primaryclass = {astro-ph.CO},
  reportnumber = {IPPP/20/89},
  doi          = {10.3390/e23040404},
  journal      = {Entropy},
  volume       = 23,
  number       = 4,
  pages        = 404,
  year         = 2021,
}

@ARTICLE{DiValentino:2020zio,
  author       = {Di Valentino, Eleonora and others},
  title        = {{Snowmass2021 - Letter of interest cosmology intertwined II: The hubble constant tension}},
  eprint       = {2008.11284},
  archiveprefix= {arXiv},
  primaryclass = {astro-ph.CO},
  reportnumber = {FERMILAB-PUB-21-590-PPD},
  doi          = {10.1016/j.astropartphys.2021.102605},
  journal      = {Astropart. Phys.},
  volume       = 131,
  pages        = 102605,
  year         = 2021,
}

@ARTICLE{CosmoVerseNetwork:2025alb,
  author       = {Di Valentino, Eleonora and others},
  collaboration= {CosmoVerse Network},
  title        = {{The CosmoVerse White Paper: Addressing observational tensions in cosmology with systematics and fundamental physics}},
  eprint       = {2504.01669},
  archiveprefix= {arXiv},
  primaryclass = {astro-ph.CO},
  doi          = {10.1016/j.dark.2025.101965},
  journal      = {Phys. Dark Univ.},
  volume       = 49,
  pages        = 101965,
  year         = 2025,
}

@ARTICLE{Dossett:2014oia,
  author       = {Dossett, Jason and Hu, Bin and Parkinson, David},
  title        = {{Constraining models of f(R) gravity with Planck and WiggleZ power spectrum data}},
  eprint       = {1401.3980},
  archiveprefix= {arXiv},
  primaryclass = {astro-ph.CO},
  doi          = {10.1088/1475-7516/2014/03/046},
  journal      = {JCAP},
  volume       = 03,
  pages        = 046,
  year         = 2014,
}

@ARTICLE{Dossett:2011tn,
  author       = {Dossett, Jason N. and Ishak, Mustapha and Moldenhauer, Jacob},
  title        = {{Testing General Relativity at Cosmological Scales: Implementation and Parameter Correlations}},
  eprint       = {1109.4583},
  archiveprefix= {arXiv},
  primaryclass = {astro-ph.CO},
  doi          = {10.1103/PhysRevD.84.123001},
  journal      = {Phys. Rev. D},
  volume       = 84,
  pages        = 123001,
  year         = 2011,
}

@ARTICLE{Dossett:2015nda,
  author       = {Dossett, Jason N. and Ishak, Mustapha and Parkinson, David and Davis, Tamara},
  title        = {{Constraints and tensions in testing general relativity from Planck and CFHTLenS data including intrinsic alignment systematics}},
  eprint       = {1501.03119},
  archiveprefix= {arXiv},
  primaryclass = {astro-ph.CO},
  doi          = {10.1103/PhysRevD.92.023003},
  journal      = {Phys. Rev. D},
  volume       = 92,
  number       = 2,
  pages        = 023003,
  year         = 2015,
}

@ARTICLE{Driver:2021cde,
  author       = {Driver, Simon P.},
  title        = {{The challenge of measuring and mapping the missing baryons}},
  eprint       = {2203.08541},
  archiveprefix= {arXiv},
  primaryclass = {astro-ph.CO},
  doi          = {10.1038/s41550-021-01441-w},
  journal      = {Nature Astron.},
  volume       = 5,
  number       = 9,
  pages        = {852--854},
  year         = 2021,
}

@ARTICLE{Efstathiou:2023fbn,
  author       = {Efstathiou, George and Rosenberg, Erik and Poulin, Vivian},
  title        = {{Improved Planck Constraints on Axionlike Early Dark Energy as a Resolution of the Hubble Tension}},
  eprint       = {2311.00524},
  archiveprefix= {arXiv},
  primaryclass = {astro-ph.CO},
  doi          = {10.1103/PhysRevLett.132.221002},
  journal      = {Phys. Rev. Lett.},
  volume       = 132,
  number       = 22,
  pages        = 221002,
  year         = 2024,
}

@ARTICLE{El-Zant:2018bsc,
  author       = {El-Zant, Amr and El Hanafy, Waleed and Elgammal, Sherif},
  title        = {{$H_0$ Tension and the Phantom Regime: A Case Study in Terms of an Infrared $f(T)$ Gravity}},
  eprint       = {1809.09390},
  archiveprefix= {arXiv},
  primaryclass = {gr-qc},
  doi          = {10.3847/1538-4357/aafa12},
  journal      = {Astrophys. J.},
  volume       = 871,
  number       = 2,
  pages        = 210,
  year         = 2019,
}

@ARTICLE{Elbers:2025vlz,
  author       = {Elbers, W. and others},
  title        = {{Constraints on neutrino physics from DESI DR2 BAO and DR1 full shape}},
  eprint       = {2503.14744},
  archiveprefix= {arXiv},
  primaryclass = {astro-ph.CO},
  reportnumber = {FERMILAB-PUB-25-0168-PPD},
  doi          = {10.1103/w9pk-xsk7},
  journal      = {Phys. Rev. D},
  volume       = 112,
  number       = 8,
  pages        = 083513,
  year         = 2025,
}

@ARTICLE{Elbers:2024sha,
  author       = {Elbers, Willem and Frenk, Carlos S. and Jenkins, Adrian and Li, Baojiu and Pascoli, Silvia},
  title        = {{Negative neutrino masses as a mirage of dark energy}},
  eprint       = {2407.10965},
  archiveprefix= {arXiv},
  primaryclass = {astro-ph.CO},
  doi          = {10.1103/PhysRevD.111.063534},
  journal      = {Phys. Rev. D},
  volume       = 111,
  number       = 6,
  pages        = 063534,
  year         = 2025,
}

@ARTICLE{Escamilla:2025imi,
  author       = {Escamilla, Luis A. and Akarsu, {\"O}zg{\"u}r and Di Valentino, Eleonora and {\"O}z{\"u}lker, Emre and Vazquez, J. Alberto},
  title        = {{Exploring the Growth-Index ($\gamma$) Tension with $\Lambda_{\rm s}$CDM}},
  eprint       = {2503.12945},
  archiveprefix= {arXiv},
  primaryclass = {astro-ph.CO},
  month        = 3,
  year         = 2025,
}

@ARTICLE{Fazzari:2025lzd,
  author       = {Fazzari, Elisa and Giar{\`e}, William and Di Valentino, Eleonora},
  title        = {{Cosmographic Footprints of Dynamical Dark Energy}},
  eprint       = {2509.16196},
  archiveprefix= {arXiv},
  primaryclass = {astro-ph.CO},
  doi          = {10.3847/2041-8213/ae2917},
  journal      = {Astrophys. J. Lett.},
  volume       = 996,
  number       = 1,
  pages        = L5,
  year         = 2026,
}

@ARTICLE{Ferreira:2025lrd,
  author       = {Ferreira, Elisa G. M. and McDonough, Evan and Balkenhol, Lennart and Kallosh, Renata and Knox, Lloyd and Linde, Andrei},
  title        = {{BAO-CMB tension and implications for inflation}},
  eprint       = {2507.12459},
  archiveprefix= {arXiv},
  primaryclass = {astro-ph.CO},
  doi          = {10.1103/lq71-b84v},
  journal      = {Phys. Rev. D},
  volume       = 113,
  number       = 4,
  pages        = 043524,
  year         = 2026,
}

@ARTICLE{Freedman:2021ahq,
  author       = {Freedman, Wendy L.},
  title        = {{Measurements of the Hubble Constant: Tensions in Perspective}},
  eprint       = {2106.15656},
  archiveprefix= {arXiv},
  primaryclass = {astro-ph.CO},
  doi          = {10.3847/1538-4357/ac0e95},
  journal      = {Astrophys. J.},
  volume       = 919,
  number       = 1,
  pages        = 16,
  year         = 2021,
}

@ARTICLE{Freedman:2024eph,
  author       = {Freedman, Wendy L. and Madore, Barry F. and Hoyt, Taylor J. and Jang, In Sung and Lee, Abigail J. and Owens, Kayla A.},
  title        = {{Status Report on the Chicago-Carnegie Hubble Program (CCHP): Measurement of the Hubble Constant Using the Hubble and James Webb Space Telescopes}},
  eprint       = {2408.06153},
  archiveprefix= {arXiv},
  primaryclass = {astro-ph.CO},
  doi          = {10.3847/1538-4357/adce78},
  journal      = {Astrophys. J.},
  volume       = 985,
  number       = 2,
  pages        = 203,
  year         = 2025,
  note         = {[Erratum: Astrophys.J. 993, 252 (2025)]},
}

@ARTICLE{Freedman:2020dne,
  author       = {Freedman, Wendy L. and Madore, Barry F. and Hoyt, Taylor and Jang, In Sung and Beaton, Rachael and Lee, Myung Gyoon and Monson, Andrew and Neeley, Jill and Rich, Jeffrey},
  title        = {{Calibration of the Tip of the Red Giant Branch}},
  eprint       = {2002.01550},
  archiveprefix= {arXiv},
  primaryclass = {astro-ph.GA},
  doi          = {10.3847/1538-4357/ab7339},
  journal      = {Astrophys. J.},
  volume       = 891,
  number       = 1,
  pages        = 57,
  year         = 2020,
}

@ARTICLE{GarciaEscudero:2025lef,
  author       = {Garc{\'\i}a Escudero, Helena and Mirpoorian, Seyed Hamidreza and Pogosian, Levon},
  title        = {{Sound-horizon-agnostic Inference of the Hubble Constant and Neutrino Masses from Baryon Acoustic Oscillations, Cosmic Microwave Background Lensing, and Galaxy Weak Lensing and Clustering}},
  eprint       = {2509.16202},
  archiveprefix= {arXiv},
  primaryclass = {astro-ph.CO},
  doi          = {10.3847/2041-8213/ae40bf},
  journal      = {Astrophys. J. Lett.},
  volume       = 998,
  number       = 2,
  pages        = L41,
  year         = 2026,
}

@ARTICLE{Garramone:2026evc,
  author       = {Garramone, Michela and Fornengo, Nicolao and Gariazzo, Stefano},
  title        = {{Early-Universe constraints on the electron mass}},
  eprint       = {2602.05720},
  archiveprefix= {arXiv},
  primaryclass = {hep-ph},
  doi          = {10.1103/s3c6-h22g},
  journal      = {Phys. Rev. D},
  volume       = 113,
  number       = 12,
  pages        = 123008,
  year         = 2026,
}

@ARTICLE{SPT-3G:2024atg,
  author       = {Ge, F. and others},
  collaboration= {SPT-3G},
  title        = {{Cosmology from CMB lensing and delensed EE power spectra using 2019{\textendash}2020 SPT-3G polarization data}},
  eprint       = {2411.06000},
  archiveprefix= {arXiv},
  primaryclass = {astro-ph.CO},
  reportnumber = {FERMILAB-PUB-24-0840-PPD},
  doi          = {10.1103/PhysRevD.111.083534},
  journal      = {Phys. Rev. D},
  volume       = 111,
  number       = 8,
  pages        = 083534,
  year         = 2025,
}

@ARTICLE{Ghafari:2025eql,
  author       = {Ghafari, Payam and Najafi, Mahdi and Ghodsi Yengejeh, Mina and {\"O}z{\"u}lker, Emre and Di Valentino, Eleonora and Firouzjaee, Javad T.},
  title        = {{A Multi-Probe ISW Study of Dark Energy Models with Negative Energy Density: Galaxy Correlations, Lensing Bispectrum, and Planck ISW-Lensing Likelihood}},
  eprint       = {2512.07060},
  archiveprefix= {arXiv},
  primaryclass = {astro-ph.CO},
  month        = 12,
  year         = 2025,
}

@ARTICLE{Gialamas:2024lyw,
  author       = {Gialamas, Ioannis D. and H{\"u}tsi, Gert and Kannike, Kristjan and Racioppi, Antonio and Raidal, Martti and Vasar, Martin and Veerm{\"a}e, Hardi},
  title        = {{Interpreting DESI 2024 BAO: Late-time dynamical dark energy or a local effect?}},
  eprint       = {2406.07533},
  archiveprefix= {arXiv},
  primaryclass = {astro-ph.CO},
  doi          = {10.1103/PhysRevD.111.043540},
  journal      = {Phys. Rev. D},
  volume       = 111,
  number       = 4,
  pages        = 043540,
  year         = 2025,
}

@ARTICLE{Giare:2024akf,
  author       = {Giar{\`e}, William},
  title        = {{Inflation, the Hubble tension, and early dark energy: An alternative overview}},
  eprint       = {2404.12779},
  archiveprefix= {arXiv},
  primaryclass = {astro-ph.CO},
  doi          = {10.1103/PhysRevD.109.123545},
  journal      = {Phys. Rev. D},
  volume       = 109,
  number       = 12,
  pages        = 123545,
  year         = 2024,
}

@ARTICLE{Giare:2024ocw,
  author       = {Giar{\`e}, William},
  title        = {{Dynamical dark energy beyond Planck? Constraints from multiple CMB probes, DESI BAO, and type-Ia supernovae}},
  eprint       = {2409.17074},
  archiveprefix= {arXiv},
  primaryclass = {astro-ph.CO},
  doi          = {10.1103/ss37-cxhn},
  journal      = {Phys. Rev. D},
  volume       = 112,
  number       = 2,
  pages        = 023508,
  year         = 2025,
}

@ARTICLE{Giare:2024oil,
  author       = {Giar{\`e}, William},
  title        = {{Dynamical dark energy beyond Planck? Constraints from multiple CMB probes, DESI BAO, and type-Ia supernovae}},
  eprint       = {2409.17074},
  archiveprefix= {arXiv},
  primaryclass = {astro-ph.CO},
  doi          = {10.1103/ss37-cxhn},
  journal      = {Phys. Rev. D},
  volume       = 112,
  number       = 2,
  pages        = 023508,
  year         = 2025,
}

@ARTICLE{Giare:2024syw,
  author       = {Giar{\`e}, William and Betts, Jonathan and van de Bruck, Carsten and Di Valentino, Eleonora},
  title        = {{Model-Independent Test of Prerecombination New Physics: Measuring the Sound Horizon with Gravitational Wave Standard Sirens and the Baryon Acoustic Oscillation Angular Scale}},
  eprint       = {2406.07493},
  archiveprefix= {arXiv},
  primaryclass = {astro-ph.CO},
  doi          = {10.1103/k6mg-g23d},
  journal      = {Phys. Rev. Lett.},
  volume       = 135,
  number       = 7,
  pages        = 071003,
  year         = 2025,
}

@ARTICLE{Giare:2023ejv,
  author       = {Giar{\`e}, William and Di Valentino, Eleonora and Melchiorri, Alessandro},
  title        = {{Measuring the reionization optical depth without large-scale CMB polarization}},
  eprint       = {2312.06482},
  archiveprefix= {arXiv},
  primaryclass = {astro-ph.CO},
  doi          = {10.1103/PhysRevD.109.103519},
  journal      = {Phys. Rev. D},
  volume       = 109,
  number       = 10,
  pages        = 103519,
  year         = 2024,
}

@ARTICLE{Giare:2025pzu,
  author       = {Giar{\`e}, William and Mahassen, Tariq and Di Valentino, Eleonora and Pan, Supriya},
  title        = {{An overview of what current data can (and cannot yet) say about evolving dark energy}},
  eprint       = {2502.10264},
  archiveprefix= {arXiv},
  primaryclass = {astro-ph.CO},
  doi          = {10.1016/j.dark.2025.101906},
  journal      = {Phys. Dark Univ.},
  volume       = 48,
  pages        = 101906,
  year         = 2025,
}

@ARTICLE{Giare:2024gpk,
  author       = {Giar{\`e}, William and Najafi, Mahdi and Pan, Supriya and Di Valentino, Eleonora and Firouzjaee, Javad T.},
  title        = {{Robust preference for Dynamical Dark Energy in DESI BAO and SN measurements}},
  eprint       = {2407.16689},
  archiveprefix= {arXiv},
  primaryclass = {astro-ph.CO},
  doi          = {10.1088/1475-7516/2024/10/035},
  journal      = {JCAP},
  volume       = 10,
  pages        = 035,
  year         = 2024,
}

@ARTICLE{Giare:2024ytc,
  author       = {Giar{\`e}, William and Zhai, Yuejia and Pan, Supriya and Di Valentino, Eleonora and Nunes, Rafael C. and van de Bruck, Carsten},
  title        = {{Tightening the reins on nonminimal dark sector physics: Interacting dark energy with dynamical and nondynamical equation of state}},
  eprint       = {2404.02110},
  archiveprefix= {arXiv},
  primaryclass = {astro-ph.CO},
  doi          = {10.1103/PhysRevD.110.063527},
  journal      = {Phys. Rev. D},
  volume       = 110,
  number       = 6,
  pages        = 063527,
  year         = 2024,
}

@ARTICLE{Gokcen:2026pkq,
  author       = {G{\"o}k{\c{c}}en, Mine and Akarsu, {\"O}zg{\"u}r and Di Valentino, Eleonora},
  title        = {{Revisiting CPL with sign-switching density: To cross or not to cross the NECB}},
  eprint       = {2602.21169},
  archiveprefix= {arXiv},
  primaryclass = {astro-ph.CO},
  doi          = {10.1016/j.dark.2026.102273},
  journal      = {Phys. Dark Univ.},
  volume       = 52,
  pages        = 102273,
  year         = 2026,
}

@ARTICLE{Golovnev:2018wbh,
  author       = {Golovnev, Alexey and Koivisto, Tomi},
  title        = {{Cosmological perturbations in modified teleparallel gravity models}},
  eprint       = {1808.05565},
  archiveprefix= {arXiv},
  primaryclass = {gr-qc},
  reportnumber = {NORDITA 2018-073},
  doi          = {10.1088/1475-7516/2018/11/012},
  journal      = {JCAP},
  volume       = 11,
  pages        = 012,
  year         = 2018,
}

@ARTICLE{Gomez-Valent:2024tdb,
  author       = {Gomez-Valent, Adria and Sol{\`a} Peracaula, Joan},
  title        = {{Phantom Matter: A Challenging Solution to the Cosmological Tensions}},
  eprint       = {2404.18845},
  archiveprefix= {arXiv},
  primaryclass = {astro-ph.CO},
  doi          = {10.3847/1538-4357/ad7a62},
  journal      = {Astrophys. J.},
  volume       = 975,
  number       = 1,
  pages        = 64,
  year         = 2024,
}

@ARTICLE{Gomez-Valent:2024ejh,
  author       = {G{\'o}mez-Valent, Adria and Sol{\`a} Peracaula, Joan},
  title        = {{Composite dark energy and the cosmological tensions}},
  eprint       = {2412.15124},
  archiveprefix= {arXiv},
  primaryclass = {astro-ph.CO},
  doi          = {10.1016/j.physletb.2025.139391},
  journal      = {Phys. Lett. B},
  volume       = 864,
  pages        = 139391,
  year         = 2025,
}

@ARTICLE{Graham:2025dqn,
  author       = {Graham, Peter W. and Green, Daniel and Meyers, Joel},
  title        = {{New interpretations of the cosmological preference for a negative neutrino mass}},
  eprint       = {2508.20999},
  archiveprefix= {arXiv},
  primaryclass = {astro-ph.CO},
  doi          = {10.1103/1bqb-qlrj},
  journal      = {Phys. Rev. D},
  volume       = 113,
  number       = 4,
  pages        = 043514,
  year         = 2026,
}

@ARTICLE{Green:2024xbb,
  author       = {Green, Daniel and Meyers, Joel},
  title        = {{Cosmological preference for a negative neutrino mass}},
  eprint       = {2407.07878},
  archiveprefix= {arXiv},
  primaryclass = {astro-ph.CO},
  doi          = {10.1103/PhysRevD.111.083507},
  journal      = {Phys. Rev. D},
  volume       = 111,
  number       = 8,
  pages        = 083507,
  year         = 2025,
}

@ARTICLE{Greene:2024qis,
  author       = {Greene, Kylar and Cyr-Racine, Francis-Yan},
  title        = {{Ratio-preserving approach to cosmological concordance}},
  eprint       = {2403.05619},
  archiveprefix= {arXiv},
  primaryclass = {astro-ph.CO},
  reportnumber = {FERMILAB-PUB-24-0721-V},
  doi          = {10.1103/PhysRevD.110.043524},
  journal      = {Phys. Rev. D},
  volume       = 110,
  number       = 4,
  pages        = 043524,
  year         = 2024,
}

@ARTICLE{Greene:2023cro,
  author       = {Greene, Kylar L. and Cyr-Racine, Francis-Yan},
  title        = {{Thomson scattering: one rate to rule them all}},
  eprint       = {2306.06165},
  archiveprefix= {arXiv},
  primaryclass = {astro-ph.CO},
  doi          = {10.1088/1475-7516/2023/10/065},
  journal      = {JCAP},
  volume       = 10,
  pages        = 065,
  year         = 2023,
}

@ARTICLE{Halder:2024uao,
  author       = {Halder, Sudip and de Haro, Jaume and Saha, Tapan and Pan, Supriya},
  title        = {{Phase space analysis of sign-shifting interacting dark energy models}},
  eprint       = {2403.01397},
  archiveprefix= {arXiv},
  primaryclass = {gr-qc},
  doi          = {10.1103/PhysRevD.109.083522},
  journal      = {Phys. Rev. D},
  volume       = 109,
  number       = 8,
  pages        = 083522,
  year         = 2024,
}

@ARTICLE{Hart:2017ndk,
  author       = {Hart, Luke and Chluba, Jens},
  title        = {{New constraints on time-dependent variations of fundamental constants using Planck data}},
  eprint       = {1705.03925},
  archiveprefix= {arXiv},
  primaryclass = {astro-ph.CO},
  doi          = {10.1093/mnras/stx2783},
  journal      = {Mon. Not. Roy. Astron. Soc.},
  volume       = 474,
  number       = 2,
  pages        = {1850--1861},
  year         = 2018,
}

@ARTICLE{Hart:2019dxi,
  author       = {Hart, Luke and Chluba, Jens},
  title        = {{Updated fundamental constant constraints from Planck 2018 data and possible relations to the Hubble tension}},
  eprint       = {1912.03986},
  archiveprefix= {arXiv},
  primaryclass = {astro-ph.CO},
  doi          = {10.1093/mnras/staa412},
  journal      = {Mon. Not. Roy. Astron. Soc.},
  volume       = 493,
  number       = 3,
  pages        = {3255--3263},
  year         = 2020,
}

@ARTICLE{Hart:2021kad,
  author       = {Hart, Luke and Chluba, Jens},
  title        = {{Varying fundamental constants principal component analysis: additional hints about the Hubble tension}},
  eprint       = {2107.12465},
  archiveprefix= {arXiv},
  primaryclass = {astro-ph.CO},
  doi          = {10.1093/mnras/stab2777},
  journal      = {Mon. Not. Roy. Astron. Soc.},
  volume       = 510,
  number       = 2,
  pages        = {2206--2227},
  year         = 2022,
}

@ARTICLE{Hashim:2020sez,
  author       = {Hashim, Mahmoud and El Hanafy, Waleed and Golovnev, Alexey and El-Zant, Amr A.},
  title        = {{Toward a concordance teleparallel cosmology. Part I. Background dynamics}},
  eprint       = {2010.14964},
  archiveprefix= {arXiv},
  primaryclass = {astro-ph.CO},
  doi          = {10.1088/1475-7516/2021/07/052},
  journal      = {JCAP},
  volume       = 07,
  pages        = 052,
  year         = 2021,
}

@ARTICLE{Hashim:2021pkq,
  author       = {Hashim, Mahmoud and El-Zant, Amr A. and El Hanafy, Waleed and Golovnev, Alexey},
  title        = {{Toward a concordance teleparallel cosmology. Part~II. Linear perturbation}},
  eprint       = {2104.08311},
  archiveprefix= {arXiv},
  primaryclass = {astro-ph.CO},
  doi          = {10.1088/1475-7516/2021/07/053},
  journal      = {JCAP},
  volume       = 07,
  pages        = 053,
  year         = 2021,
}

@ARTICLE{Herold:2025hkb,
  author       = {Herold, Laura and Karwal, Tanvi},
  title        = {{Bayesian and frequentist perspectives agree on dynamical dark energy}},
  eprint       = {2506.12004},
  archiveprefix= {arXiv},
  primaryclass = {astro-ph.CO},
  month        = 6,
  year         = 2025,
}

@ARTICLE{Hoyt:2026fve,
  author       = {Hoyt, Taylor J. and Rubin, David and Aldering, Greg and Perlmutter, Saul and Cuceu, Andrei and Gupta, Ravi},
  title        = {{Union3.1: Self-consistent Measurements of Host Galaxy Properties for 2000 Type Ia Supernovae}},
  eprint       = {2601.19424},
  archiveprefix= {arXiv},
  primaryclass = {astro-ph.CO},
  month        = 1,
  year         = 2026,
}

@ARTICLE{Hu:2023jqc,
  author       = {Hu, Jian-Ping and Wang, Fa-Yin},
  title        = {{Hubble Tension: The Evidence of New Physics}},
  eprint       = {2302.05709},
  archiveprefix= {arXiv},
  primaryclass = {astro-ph.CO},
  doi          = {10.3390/universe9020094},
  journal      = {Universe},
  volume       = 9,
  number       = 2,
  pages        = 94,
  year         = 2023,
}

@ARTICLE{Huang:2023frr,
  author       = {Huang, Caroline D. and others},
  title        = {{The Mira Distance to M101 and a 4{\%} Measurement of H $_{0}$}},
  eprint       = {2312.08423},
  archiveprefix= {arXiv},
  primaryclass = {astro-ph.CO},
  doi          = {10.3847/1538-4357/ad1ff8},
  journal      = {Astrophys. J.},
  volume       = 963,
  number       = 2,
  pages        = 83,
  year         = 2024,
}

@ARTICLE{Huang:2025xyf,
  author       = {Huang, Zhiqi},
  title        = {{Reionization optical depth and CMB{\textendash}BAO tension in punctuated inflation}},
  eprint       = {2509.09086},
  archiveprefix= {arXiv},
  primaryclass = {astro-ph.CO},
  reportnumber = {SYSU-SPA-2025},
  doi          = {10.1093/mnras/staf1892},
  journal      = {Mon. Not. Roy. Astron. Soc.},
  volume       = 544,
  number       = 2,
  pages        = {2193--2199},
  year         = 2025,
}

@ARTICLE{Hulse:1974eb,
  author       = {Hulse, R. A. and Taylor, J. H.},
  title        = {{Discovery of a pulsar in a binary system}},
  doi          = {10.1086/181708},
  journal      = {Astrophys. J. Lett.},
  volume       = 195,
  pages        = {L51--L53},
  year         = 1975,
}

@ARTICLE{Ishak:2025cay,
  author       = {Ishak, Mustapha and Medina-Varela, Leonel},
  title        = {{Persistent and serious challenge to the $Λ$CDM throne: Evidence for dynamical dark energy rising from combinations of different types of datasets}},
  eprint       = {2507.22856},
  archiveprefix= {arXiv},
  primaryclass = {astro-ph.CO},
  month        = 7,
  year         = 2025,
}

@ARTICLE{Jensen:2025aai,
  author       = {Jensen, Joseph B. and Blakeslee, John P. and Cantiello, Michele and Cowles, Mikaela and Anand, Gagandeep S. and Tully, R. Brent and Kourkchi, Ehsan and Raimondo, Gabriella},
  title        = {{The TRGB-SBF Project. III. Refining the HST Surface Brightness Fluctuation Distance Scale Calibration with JWST}},
  eprint       = {2502.15935},
  archiveprefix= {arXiv},
  primaryclass = {astro-ph.CO},
  doi          = {10.3847/1538-4357/addfd6},
  month        = 6,
  year         = 2025,
}

@ARTICLE{Jhaveri:2026bla,
  author       = {Jhaveri, Tanisha and Karwal, Tanvi and Crawford, Thomas and Hu, Wayne and Khalife, Ali Rida and Balkenhol, Lennart and Ge, Fei},
  title        = {{Disentangling cosmic distance tensions with early and late dark energy}},
  eprint       = {2604.08530},
  archiveprefix= {arXiv},
  primaryclass = {astro-ph.CO},
  month        = 4,
  year         = 2026,
}

@ARTICLE{Jhaveri:2025neg,
  author       = {Jhaveri, Tanisha and Karwal, Tanvi and Hu, Wayne},
  title        = {{Turning a negative neutrino mass into a positive optical depth}},
  eprint       = {2504.21813},
  archiveprefix= {arXiv},
  primaryclass = {astro-ph.CO},
  doi          = {10.1103/6vd2-rbfn},
  journal      = {Phys. Rev. D},
  volume       = 112,
  number       = 4,
  pages        = 043541,
  year         = 2025,
}

@ARTICLE{Jiang:2024xnu,
  author       = {Jiang, Jun-Qian and Pedrotti, Davide and da Costa, Simony Santos and Vagnozzi, Sunny},
  title        = {{Nonparametric late-time expansion history reconstruction and implications for the Hubble tension in light of recent DESI and type Ia supernovae data}},
  eprint       = {2408.02365},
  archiveprefix= {arXiv},
  primaryclass = {astro-ph.CO},
  doi          = {10.1103/PhysRevD.110.123519},
  journal      = {Phys. Rev. D},
  volume       = 110,
  number       = 12,
  pages        = 123519,
  year         = 2024,
}

@ARTICLE{Jones:2022mvo,
  author       = {Jones, D. O. and others},
  title        = {{Cosmological Results from the RAISIN Survey: Using Type Ia Supernovae in the Near Infrared as a Novel Path to Measure the Dark Energy Equation of State}},
  eprint       = {2201.07801},
  archiveprefix= {arXiv},
  primaryclass = {astro-ph.CO},
  doi          = {10.3847/1538-4357/ac755b},
  journal      = {Astrophys. J.},
  volume       = 933,
  number       = 2,
  pages        = 172,
  year         = 2022,
}

@ARTICLE{Jungman:1995df,
  author       = {Jungman, Gerard and Kamionkowski, Marc and Griest, Kim},
  title        = {{Supersymmetric dark matter}},
  eprint       = {hep-ph/9506380},
  archiveprefix= {arXiv},
  reportnumber = {SU-4240-605, UCSD-PTH-95-02, IASSNS-HEP-95-14, CU-TP-677},
  doi          = {10.1016/0370-1573(95)00058-5},
  journal      = {Phys. Rept.},
  volume       = 267,
  pages        = {195--373},
  year         = 1996,
}

@ARTICLE{Kamionkowski:2022pkx,
  author       = {Kamionkowski, Marc and Riess, Adam G.},
  title        = {{The Hubble Tension and Early Dark Energy}},
  eprint       = {2211.04492},
  archiveprefix= {arXiv},
  primaryclass = {astro-ph.CO},
  doi          = {10.1146/annurev-nucl-111422-024107},
  journal      = {Ann. Rev. Nucl. Part. Sci.},
  volume       = 73,
  pages        = {153--180},
  year         = 2023,
}

@ARTICLE{Kessler:2026dbi,
  author       = {Kessler, Daniel A. and Di Valentino, Eleonora and Escamilla, Luis A. and Huterer, Dragan},
  title        = {{Reconstructing dark energy with fewer assumptions}},
  eprint       = {2606.05853},
  archiveprefix= {arXiv},
  primaryclass = {astro-ph.CO},
  month        = 6,
  year         = 2026,
}

@ARTICLE{Kessler:2025kju,
  author       = {Kessler, Daniel A. and Escamilla, Luis A. and Pan, Supriya and Di Valentino, Eleonora},
  title        = {{One-parameter dynamical dark energy: Hints for oscillations}},
  eprint       = {2504.00776},
  archiveprefix= {arXiv},
  primaryclass = {astro-ph.CO},
  month        = 4,
  year         = 2025,
}

@ARTICLE{Kibris:2026cqq,
  author       = {K{\i}br{\i}s, Cihad and Elbers, Willem and Akarsu, {\"O}zg{\"u}r and Di Valentino, Eleonora},
  title        = {{Negative neutrino mass or negative dark energy?}},
  eprint       = {2605.21456},
  archiveprefix= {arXiv},
  primaryclass = {astro-ph.CO},
  month        = 5,
  year         = 2026,
}

@ARTICLE{Kochappan:2024jyf,
  author       = {Kochappan, Joby and Yin, Lu and Lee, Bum-Hoon and Ghosh, Tuhin},
  title        = {{Observational evidence for early dark energy as a unified explanation for cosmic birefringence and the Hubble tension}},
  eprint       = {2408.09521},
  archiveprefix= {arXiv},
  primaryclass = {astro-ph.CO},
  doi          = {10.1103/x1qj-t4jz},
  journal      = {Phys. Rev. D},
  volume       = 112,
  number       = 6,
  pages        = 063562,
  year         = 2025,
}

@ARTICLE{Kourkchi:2020iyz,
  author       = {Kourkchi, Ehsan and Tully, R. Brent and Anand, Gagandeep S. and Courtois, Helene M. and Dupuy, Alexandra and Neill, James D. and Rizzi, Luca and Seibert, Mark},
  title        = {{Cosmicflows-4: The Calibration of Optical and Infrared Tully{\textendash}Fisher Relations}},
  eprint       = {2004.14499},
  archiveprefix= {arXiv},
  primaryclass = {astro-ph.GA},
  doi          = {10.3847/1538-4357/ab901c},
  journal      = {Astrophys. J.},
  volume       = 896,
  number       = 1,
  pages        = 3,
  year         = 2020,
}

@ARTICLE{Krishnan:2020obg,
  author       = {Krishnan, C. and Colg{\'a}in, Eoin {\'O}. and Ruchika and Sen, Anjan A. and Sheikh-Jabbari, M. M. and Yang, Tao},
  title        = {{Is there an early Universe solution to Hubble tension?}},
  eprint       = {2002.06044},
  archiveprefix= {arXiv},
  primaryclass = {astro-ph.CO},
  doi          = {10.1103/PhysRevD.102.103525},
  journal      = {Phys. Rev. D},
  volume       = 102,
  number       = 10,
  pages        = 103525,
  year         = 2020,
}

@ARTICLE{Krolewski:2024jwj,
  author       = {Krolewski, Alex and Percival, Will J. and Woodfinden, Alex},
  title        = {{New Method to Determine the Hubble Parameter from Cosmological Energy-Density Measurements}},
  eprint       = {2403.19227},
  archiveprefix= {arXiv},
  primaryclass = {astro-ph.CO},
  doi          = {10.1103/PhysRevLett.134.101002},
  journal      = {Phys. Rev. Lett.},
  volume       = 134,
  number       = 10,
  pages        = 101002,
  year         = 2025,
}

@ARTICLE{Lee:2026yzs,
  author       = {Lee, Dong Ha and van de Bruck, Carsten and Di Valentino, Eleonora and Van Waerbeke, Ludovic and Zhitnitsky, Ariel},
  title        = {{Evolving Dark Energy Is Vacuum Energy After All}},
  eprint       = {2606.20036},
  archiveprefix= {arXiv},
  primaryclass = {astro-ph.CO},
  month        = 6,
  year         = 2026,
}

@ARTICLE{Lee:2025pzo,
  author       = {Lee, Dong Ha and Yang, Weiqiang and Di Valentino, Eleonora and Pan, Supriya and van de Bruck, Carsten},
  title        = {{Shape of dark energy: Constraining its evolution with a general parametrization}},
  eprint       = {2507.11432},
  archiveprefix= {arXiv},
  primaryclass = {astro-ph.CO},
  doi          = {10.1103/z7y2-yvhg},
  journal      = {Phys. Rev. D},
  volume       = 113,
  number       = 6,
  pages        = 063554,
  year         = 2026,
}

@ARTICLE{Lee:2022gzh,
  author       = {Lee, Nanoom and Ali-Ha{\"\i}moud, Yacine and Sch{\"o}neberg, Nils and Poulin, Vivian},
  title        = {{What It Takes to Solve the Hubble Tension through Modifications of Cosmological Recombination}},
  eprint       = {2212.04494},
  archiveprefix= {arXiv},
  primaryclass = {astro-ph.CO},
  doi          = {10.1103/PhysRevLett.130.161003},
  journal      = {Phys. Rev. Lett.},
  volume       = 130,
  number       = 16,
  pages        = 161003,
  year         = 2023,
}

@ARTICLE{Lemos:2018smw,
  author       = {Lemos, Pablo and Lee, Elizabeth and Efstathiou, George and Gratton, Steven},
  title        = {{Model independent $H(z)$ reconstruction using the cosmic inverse distance ladder}},
  eprint       = {1806.06781},
  archiveprefix= {arXiv},
  primaryclass = {astro-ph.CO},
  doi          = {10.1093/mnras/sty3082},
  journal      = {Mon. Not. Roy. Astron. Soc.},
  volume       = 483,
  number       = 4,
  pages        = {4803--4810},
  year         = 2019,
}

@ARTICLE{Li:2024yoe,
  author       = {Li, Siyang and Riess, Adam G. and Casertano, Stefano and Anand, Gagandeep S. and Scolnic, Daniel M. and Yuan, Wenlong and Breuval, Louise and Huang, Caroline D.},
  title        = {{Reconnaissance with JWST of the J-region Asymptotic Giant Branch in Distance Ladder Galaxies: From Irregular Luminosity Functions to Approximation of the Hubble Constant}},
  eprint       = {2401.04777},
  archiveprefix= {arXiv},
  primaryclass = {astro-ph.CO},
  doi          = {10.3847/1538-4357/ad2f2b},
  journal      = {Astrophys. J.},
  volume       = 966,
  number       = 1,
  pages        = 20,
  year         = 2024,
}

@ARTICLE{Li:2025ife,
  author       = {Li, Siyang and Riess, Adam G. and Scolnic, Daniel and Casertano, Stefano and Anand, Gagandeep S.},
  title        = {{JAGB 2.0: Improved Constraints on the J-region Asymptotic Giant Branch{\textendash}based Hubble Constant from an Expanded Sample of JWST Observations}},
  eprint       = {2502.05259},
  archiveprefix= {arXiv},
  primaryclass = {astro-ph.CO},
  doi          = {10.3847/1538-4357/addd0c},
  journal      = {Astrophys. J.},
  volume       = 988,
  number       = 1,
  pages        = 97,
  year         = 2025,
}

@ARTICLE{Li:2025muv,
  author       = {Li, Tian-Nuo and Du, Guo-Hong and Li, Yun-He and Li, Yichao and Ling, Jia-Le and Zhang, Jing-Fei and Zhang, Xin},
  title        = {{Updated constraints on interacting dark energy: A comprehensive analysis using multiple CMB probes, DESI DR2, and supernovae observations}},
  eprint       = {2510.11363},
  archiveprefix= {arXiv},
  primaryclass = {astro-ph.CO},
  month        = 10,
  year         = 2025,
}

@ARTICLE{Li:2025owk,
  author       = {Li, Tian-Nuo and Du, Guo-Hong and Li, Yun-He and Wu, Peng-Ju and Jin, Shang-Jie and Zhang, Jing-Fei and Zhang, Xin},
  title        = {{Probing the sign-changeable interaction between dark energy and dark matter with DESI baryon acoustic oscillations and DES supernovae data}},
  eprint       = {2501.07361},
  archiveprefix= {arXiv},
  primaryclass = {astro-ph.CO},
  doi          = {10.1007/s11433-025-2771-5},
  journal      = {Sci. China Phys. Mech. Astron.},
  volume       = 69,
  number       = 1,
  pages        = 210413,
  year         = 2026,
}

@ARTICLE{Li:2025vuh,
  author       = {Li, Tian-Nuo and Du, Guo-Hong and Zhou, Sheng-Han and Li, Yun-He and Zhang, Jing-Fei and Zhang, Xin},
  title        = {{Robust evidence for dynamical dark energy in light of DESI DR2 and joint ACT, SPT, and Planck data}},
  eprint       = {2511.22512},
  archiveprefix= {arXiv},
  primaryclass = {astro-ph.CO},
  doi          = {10.1016/j.dark.2026.102254},
  journal      = {Phys. Dark Univ.},
  volume       = 52,
  pages        = 102254,
  year         = 2026,
}

@ARTICLE{Li:2026xaz,
  author       = {Li, Tian-Nuo and Giar{\`e}, William and Du, Guo-Hong and Li, Yun-He and Di Valentino, Eleonora and Zhang, Jing-Fei and Zhang, Xin},
  title        = {{Strong Evidence for Dark Sector Interactions}},
  eprint       = {2601.07361},
  archiveprefix= {arXiv},
  primaryclass = {astro-ph.CO},
  month        = 1,
  year         = 2026,
}

@ARTICLE{Li:2024qso,
  author       = {Li, Tian-Nuo and Wu, Peng-Ju and Du, Guo-Hong and Jin, Shang-Jie and Li, Hai-Li and Zhang, Jing-Fei and Zhang, Xin},
  title        = {{Constraints on Interacting Dark Energy Models from the DESI Baryon Acoustic Oscillation and DES Supernovae Data}},
  eprint       = {2407.14934},
  archiveprefix= {arXiv},
  primaryclass = {astro-ph.CO},
  doi          = {10.3847/1538-4357/ad87f0},
  journal      = {Astrophys. J.},
  volume       = 976,
  number       = 1,
  pages        = 1,
  year         = 2024,
}

@ARTICLE{DESI:2025fii,
  author       = {Lodha, K. and others},
  collaboration= {DESI},
  title        = {{Extended dark energy analysis using DESI DR2 BAO measurements}},
  eprint       = {2503.14743},
  archiveprefix= {arXiv},
  primaryclass = {astro-ph.CO},
  reportnumber = {FERMILAB-PUB-25-0164-PPD},
  doi          = {10.1103/w4c6-1r5j},
  journal      = {Phys. Rev. D},
  volume       = 112,
  number       = 8,
  pages        = 083511,
  year         = 2025,
}

@ARTICLE{DESI:2024kob,
  author       = {Lodha, K. and others},
  collaboration= {DESI},
  title        = {{DESI 2024: Constraints on physics-focused aspects of dark energy using DESI DR1 BAO data}},
  eprint       = {2405.13588},
  archiveprefix= {arXiv},
  primaryclass = {astro-ph.CO},
  reportnumber = {FERMILAB-PUB-24-0756-PPD},
  doi          = {10.1103/PhysRevD.111.023532},
  journal      = {Phys. Rev. D},
  volume       = 111,
  number       = 2,
  pages        = 023532,
  year         = 2025,
}

@ARTICLE{ACT:2025fju,
  author       = {Louis, Thibaut and others},
  collaboration= {Atacama Cosmology Telescope},
  title        = {{The Atacama Cosmology Telescope: DR6 power spectra, likelihoods and {\ensuremath{\Lambda}}CDM parameters}},
  eprint       = {2503.14452},
  archiveprefix= {arXiv},
  primaryclass = {astro-ph.CO},
  reportnumber = {FERMILAB-PUB-25-0071-PPD},
  doi          = {10.1088/1475-7516/2025/11/062},
  journal      = {JCAP},
  volume       = 11,
  pages        = 062,
  year         = 2025,
}

@ARTICLE{Loverde:2024nfi,
  author       = {Loverde, Marilena and Weiner, Zachary J.},
  title        = {{Massive neutrinos and cosmic composition}},
  eprint       = {2410.00090},
  archiveprefix= {arXiv},
  primaryclass = {astro-ph.CO},
  doi          = {10.1088/1475-7516/2024/12/048},
  journal      = {JCAP},
  volume       = 12,
  pages        = 048,
  year         = 2024,
}

@ARTICLE{Ludwick:2017tox,
  author       = {Ludwick, Kevin J.},
  title        = {{The viability of phantom dark energy: A review}},
  eprint       = {1708.06981},
  archiveprefix= {arXiv},
  primaryclass = {astro-ph.CO},
  doi          = {10.1142/S0217732317300257},
  journal      = {Mod. Phys. Lett. A},
  volume       = 32,
  number       = 28,
  pages        = 1730025,
  year         = 2017,
}

@ARTICLE{Luongo:2024fww,
  author       = {Luongo, Orlando and Muccino, Marco},
  title        = {{Model-independent cosmographic constraints from DESI 2024}},
  eprint       = {2404.07070},
  archiveprefix= {arXiv},
  primaryclass = {astro-ph.CO},
  doi          = {10.1051/0004-6361/202450512},
  journal      = {Astron. Astrophys.},
  volume       = 690,
  pages        = A40,
  year         = 2024,
}

@ARTICLE{Lynch:2024hzh,
  author       = {Lynch, Gabriel P. and Knox, Lloyd and Chluba, Jens},
  title        = {{DESI observations and the Hubble tension in light of modified recombination}},
  eprint       = {2406.10202},
  archiveprefix= {arXiv},
  primaryclass = {astro-ph.CO},
  doi          = {10.1103/PhysRevD.110.083538},
  journal      = {Phys. Rev. D},
  volume       = 110,
  number       = 8,
  pages        = 083538,
  year         = 2024,
}

@ARTICLE{Ma:1995ey,
  author       = {Ma, Chung-Pei and Bertschinger, Edmund},
  title        = {{Cosmological perturbation theory in the synchronous and conformal Newtonian gauges}},
  eprint       = {astro-ph/9506072},
  archiveprefix= {arXiv},
  doi          = {10.1086/176550},
  journal      = {Astrophys. J.},
  volume       = 455,
  pages        = {7--25},
  year         = 1995,
}

@ARTICLE{ACT:2023kun,
  author       = {Madhavacheril, Mathew S. and others},
  collaboration= {ACT},
  title        = {{The Atacama Cosmology Telescope: DR6 Gravitational Lensing Map and Cosmological Parameters}},
  eprint       = {2304.05203},
  archiveprefix= {arXiv},
  primaryclass = {astro-ph.CO},
  reportnumber = {FERMILAB-PUB-23-206-PPD},
  doi          = {10.3847/1538-4357/acff5f},
  journal      = {Astrophys. J.},
  volume       = 962,
  number       = 2,
  pages        = 113,
  year         = 2024,
}

@ARTICLE{Manoharan:2024thb,
  author       = {Manoharan, Manosh T.},
  title        = {{Insights on Granda{\textendash}Oliveros holographic dark energy: possibility of negative dark energy at $z\gtrsim 2$}},
  doi          = {10.1140/epjc/s10052-024-12926-z},
  journal      = {Eur. Phys. J. C},
  volume       = 84,
  number       = 5,
  pages        = 552,
  year         = 2024,
}

@ARTICLE{Menci:2022wia,
  author       = {Menci, N. and Castellano, M. and Santini, P. and Merlin, E. and Fontana, A. and Shankar, F.},
  title        = {{High-redshift Galaxies from Early JWST Observations: Constraints on Dark Energy Models}},
  eprint       = {2208.11471},
  archiveprefix= {arXiv},
  primaryclass = {astro-ph.CO},
  doi          = {10.3847/2041-8213/ac96e9},
  journal      = {Astrophys. J. Lett.},
  volume       = 938,
  number       = 1,
  pages        = L5,
  year         = 2022,
}

@ARTICLE{Murakami:2023xuy,
  author       = {Murakami, Yukei S. and Riess, Adam G. and Stahl, Benjamin E. and Kenworthy, W. D'Arcy and Pluck, Dahne-More A. and Macoretta, Antonella and Brout, Dillon and Jones, David O. and Scolnic, Dan M. and Filippenko, Alexei V.},
  title        = {{Leveraging SN Ia spectroscopic similarity to improve the measurement of H $_{0}$}},
  eprint       = {2306.00070},
  archiveprefix= {arXiv},
  primaryclass = {astro-ph.CO},
  doi          = {10.1088/1475-7516/2023/11/046},
  journal      = {JCAP},
  volume       = 11,
  pages        = 046,
  year         = 2023,
}

@ARTICLE{Najafi:2026kxs,
  author       = {Najafi, Mahdi and Habibollahi, Mahdi and Reyhani, Masoume and Di Valentino, Eleonora and Pan, Supriya and Firouzjaee, Javad T. and Yang, Weiqiang},
  title        = {{When Dark Energy Turns On: Constraints on a Critical Emergence Model}},
  eprint       = {2603.13137},
  archiveprefix= {arXiv},
  primaryclass = {astro-ph.CO},
  month        = 3,
  year         = 2026,
}

@ARTICLE{Najafi:2024qzm,
  author       = {Najafi, Mahdi and Pan, Supriya and Di Valentino, Eleonora and Firouzjaee, Javad T.},
  title        = {{Dynamical dark energy confronted with multiple CMB missions}},
  eprint       = {2407.14939},
  archiveprefix= {arXiv},
  primaryclass = {astro-ph.CO},
  doi          = {10.1016/j.dark.2024.101539},
  journal      = {Phys. Dark Univ.},
  volume       = 45,
  pages        = 101539,
  year         = 2024,
}

@ARTICLE{Newman:2025gwg,
  author       = {Newman, Max J. B. and others},
  title        = {{Tip of the Red Giant Branch Distances to NGC 1316, NGC 1380, NGC 1404, {\&} NGC 4457: A Pilot Study of a Parallel Distance Ladder Using Type Ia Supernovae in Early-Type Host Galaxies}},
  eprint       = {2508.20023},
  archiveprefix= {arXiv},
  primaryclass = {astro-ph.CO},
  month        = 8,
  year         = 2025,
}

@ARTICLE{Niedermann:2019olb,
  author       = {Niedermann, Florian and Sloth, Martin S.},
  title        = {{New early dark energy}},
  eprint       = {1910.10739},
  archiveprefix= {arXiv},
  primaryclass = {astro-ph.CO},
  doi          = {10.1103/PhysRevD.103.L041303},
  journal      = {Phys. Rev. D},
  volume       = 103,
  number       = 4,
  pages        = L041303,
  year         = 2021,
}

@ARTICLE{Niedermann:2021vgd,
  author       = {Niedermann, Florian and Sloth, Martin S.},
  title        = {{Hot new early dark energy}},
  eprint       = {2112.00770},
  archiveprefix= {arXiv},
  primaryclass = {hep-ph},
  doi          = {10.1103/PhysRevD.105.063509},
  journal      = {Phys. Rev. D},
  volume       = 105,
  number       = 6,
  pages        = 063509,
  year         = 2022,
}

@ARTICLE{Ozulker:2025ehg,
  author       = {{\"O}z{\"u}lker, Emre and Di Valentino, Eleonora and Giar{\`e}, William},
  title        = {{Dark Energy Crosses the Line: Quantifying and Testing the Evidence for Phantom Crossing}},
  eprint       = {2506.19053},
  archiveprefix= {arXiv},
  primaryclass = {astro-ph.CO},
  month        = 6,
  year         = 2025,
}

@ARTICLE{Pan:2019hac,
  author       = {Pan, Supriya and Yang, Weiqiang and Di Valentino, Eleonora and Shafieloo, Arman and Chakraborty, Subenoy},
  title        = {{Reconciling $H_0$ tension in a six parameter space?}},
  eprint       = {1907.12551},
  archiveprefix= {arXiv},
  primaryclass = {astro-ph.CO},
  doi          = {10.1088/1475-7516/2020/06/062},
  journal      = {JCAP},
  volume       = 06,
  number       = 06,
  pages        = 062,
  year         = 2020,
}

@ARTICLE{Paraskevas:2024ytz,
  author       = {Paraskevas, Evangelos A. and Cam, Arman and Perivolaropoulos, Leandros and Akarsu, Ozgur},
  title        = {{Transition dynamics in the {\ensuremath{\Lambda}}sCDM model: Implications for bound cosmic structures}},
  eprint       = {2402.05908},
  archiveprefix= {arXiv},
  primaryclass = {astro-ph.CO},
  doi          = {10.1103/PhysRevD.109.103522},
  journal      = {Phys. Rev. D},
  volume       = 109,
  number       = 10,
  pages        = 103522,
  year         = 2024,
}

@ARTICLE{Chan-GyungPark:2024mlx,
  author       = {Park, Chan-Gyung and de Cruz P{\'e}rez, Javier and Ratra, Bharat},
  title        = {{Using non-DESI data to confirm and strengthen the DESI 2024 spatially flat w0waCDM cosmological parametrization result}},
  eprint       = {2405.00502},
  archiveprefix= {arXiv},
  primaryclass = {astro-ph.CO},
  doi          = {10.1103/PhysRevD.110.123533},
  journal      = {Phys. Rev. D},
  volume       = 110,
  number       = 12,
  pages        = 123533,
  year         = 2024,
}

@ARTICLE{Pedrotti:2024kpn,
  author       = {Pedrotti, Davide and Jiang, Jun-Qian and Escamilla, Luis A. and da Costa, Simony Santos and Vagnozzi, Sunny},
  title        = {{Multidimensionality of the Hubble tension: The roles of {\ensuremath{\Omega}}m and {\ensuremath{\omega}}c}},
  eprint       = {2408.04530},
  archiveprefix= {arXiv},
  primaryclass = {astro-ph.CO},
  doi          = {10.1103/PhysRevD.111.023506},
  journal      = {Phys. Rev. D},
  volume       = 111,
  number       = 2,
  pages        = 023506,
  year         = 2025,
}

@ARTICLE{Perivolaropoulos:2021jda,
  author       = {Perivolaropoulos, Leandros and Skara, Foteini},
  title        = {{Challenges for {\ensuremath{\Lambda}}CDM: An update}},
  eprint       = {2105.05208},
  archiveprefix= {arXiv},
  primaryclass = {astro-ph.CO},
  doi          = {10.1016/j.newar.2022.101659},
  journal      = {New Astron. Rev.},
  volume       = 95,
  pages        = 101659,
  year         = 2022,
}

@ARTICLE{SupernovaCosmologyProject:1998vns,
  author       = {Perlmutter, S. and others},
  collaboration= {Supernova Cosmology Project},
  title        = {{Measurements of $\Omega$ and $\Lambda$ from 42 High Redshift Supernovae}},
  eprint       = {astro-ph/9812133},
  archiveprefix= {arXiv},
  reportnumber = {LBNL-41801, LBL-41801},
  doi          = {10.1086/307221},
  journal      = {Astrophys. J.},
  volume       = 517,
  pages        = {565--586},
  year         = 1999,
}

@ARTICLE{Pesce:2020xfe,
  author       = {Pesce, D. W. and others},
  title        = {{The Megamaser Cosmology Project. XIII. Combined Hubble constant constraints}},
  eprint       = {2001.09213},
  archiveprefix= {arXiv},
  primaryclass = {astro-ph.CO},
  doi          = {10.3847/2041-8213/ab75f0},
  journal      = {Astrophys. J. Lett.},
  volume       = 891,
  number       = 1,
  pages        = L1,
  year         = 2020,
}

@ARTICLE{Pooya:2024wsq,
  author       = {Pooya, N. Nazari},
  title        = {{Growth of matter perturbations in the interacting dark energy-dark matter scenarios}},
  eprint       = {2407.03766},
  archiveprefix= {arXiv},
  primaryclass = {astro-ph.CO},
  doi          = {10.1103/PhysRevD.110.043510},
  journal      = {Phys. Rev. D},
  volume       = 110,
  number       = 4,
  pages        = 043510,
  year         = 2024,
}

@ARTICLE{DES:2025sig,
  author       = {Popovic, B. and others},
  collaboration= {DES},
  title        = {{The Dark Energy Survey Supernova Program: A Reanalysis Of Cosmology Results And Evidence For Evolving Dark Energy With An Updated Type Ia Supernova Calibration}},
  eprint       = {2511.07517},
  archiveprefix= {arXiv},
  primaryclass = {astro-ph.CO},
  reportnumber = {FERMILAB-PUB-25-0842-CSAID-PPD},
  doi          = {10.1093/mnras/stag632},
  journal      = {Mon. Not. Roy. Astron. Soc.},
  volume       = 548,
  pages        = {stag632},
  year         = 2026,
}

@ARTICLE{Poulin:2021bjr,
  author       = {Poulin, Vivian and Smith, Tristan L. and Bartlett, Alexa},
  title        = {{Dark energy at early times and ACT data: A larger Hubble constant without late-time priors}},
  eprint       = {2109.06229},
  archiveprefix= {arXiv},
  primaryclass = {astro-ph.CO},
  doi          = {10.1103/PhysRevD.104.123550},
  journal      = {Phys. Rev. D},
  volume       = 104,
  number       = 12,
  pages        = 123550,
  year         = 2021,
}

@ARTICLE{Poulin:2024ken,
  author       = {Poulin, Vivian and Smith, Tristan L. and Calder{\'o}n, Rodrigo and Simon, Th{\'e}o},
  title        = {{Implications of the cosmic calibration tension beyond H0 and the synergy between early- and late-time new physics}},
  eprint       = {2407.18292},
  archiveprefix= {arXiv},
  primaryclass = {astro-ph.CO},
  doi          = {10.1103/PhysRevD.111.083552},
  journal      = {Phys. Rev. D},
  volume       = 111,
  number       = 8,
  pages        = 083552,
  year         = 2025,
}

@ARTICLE{Poulin:2025nfb,
  author       = {Poulin, Vivian and Smith, Tristan L. and Calder{\'o}n, Rodrigo and Simon, Th{\'e}o},
  title        = {{Impact of ACT DR6 and DESI DR2 for early dark energy and the Hubble tension}},
  eprint       = {2505.08051},
  archiveprefix= {arXiv},
  primaryclass = {astro-ph.CO},
  doi          = {10.1103/bx25-1g5d},
  journal      = {Phys. Rev. D},
  volume       = 113,
  number       = 6,
  pages        = 063519,
  year         = 2026,
}

@ARTICLE{Poulin:2023lkg,
  author       = {Poulin, Vivian and Smith, Tristan L. and Karwal, Tanvi},
  title        = {{The Ups and Downs of Early Dark Energy solutions to the Hubble tension: A review of models, hints and constraints circa 2023}},
  eprint       = {2302.09032},
  archiveprefix= {arXiv},
  primaryclass = {astro-ph.CO},
  doi          = {10.1016/j.dark.2023.101348},
  journal      = {Phys. Dark Univ.},
  volume       = 42,
  pages        = 101348,
  year         = 2023,
}

@ARTICLE{Poulin:2018cxd,
  author       = {Poulin, Vivian and Smith, Tristan L. and Karwal, Tanvi and Kamionkowski, Marc},
  title        = {{Early Dark Energy Can Resolve The Hubble Tension}},
  eprint       = {1811.04083},
  archiveprefix= {arXiv},
  primaryclass = {astro-ph.CO},
  doi          = {10.1103/PhysRevLett.122.221301},
  journal      = {Phys. Rev. Lett.},
  volume       = 122,
  number       = 22,
  pages        = 221301,
  year         = 2019,
}

@ARTICLE{Pulido-Hernandez:2026hcs,
  author       = {Pulido-Hern{\'a}ndez, Hayyim and Cervantes-Cota, Jorge L.},
  title        = {{Negative Masses and Spatial Curvature: Alleviating Neutrino Mass Tensions in LambdaCDM and Extended Cosmologies}},
  eprint       = {2603.13208},
  archiveprefix= {arXiv},
  primaryclass = {astro-ph.CO},
  month        = 3,
  year         = 2026,
}

@ARTICLE{ACT:2023dou,
  author       = {Qu, Frank J. and others},
  collaboration= {ACT},
  title        = {{The Atacama Cosmology Telescope: A Measurement of the DR6 CMB Lensing Power Spectrum and Its Implications for Structure Growth}},
  eprint       = {2304.05202},
  archiveprefix= {arXiv},
  primaryclass = {astro-ph.CO},
  reportnumber = {FERMILAB-PUB-23-237-PPD, FERMILAB-PUB-23-237-PPD},
  doi          = {10.3847/1538-4357/acfe06},
  journal      = {Astrophys. J.},
  volume       = 962,
  number       = 2,
  pages        = 112,
  year         = 2024,
}

@ARTICLE{Reboucas:2024smm,
  author       = {Rebou{\c{c}}as, Jo{\~a}o and de Souza, Diogo H. F. and Zhong, Kunhao and Miranda, Vivian and Rosenfeld, Rogerio},
  title        = {{Investigating late-time dark energy and massive neutrinos in light of DESI Y1 BAO}},
  eprint       = {2408.14628},
  archiveprefix= {arXiv},
  primaryclass = {astro-ph.CO},
  doi          = {10.1088/1475-7516/2025/02/024},
  journal      = {JCAP},
  volume       = 02,
  pages        = 024,
  year         = 2025,
}

@ARTICLE{SupernovaSearchTeam:1998fmf,
  author       = {Riess, Adam G. and others},
  collaboration= {Supernova Search Team},
  title        = {{Observational evidence from supernovae for an accelerating universe and a cosmological constant}},
  eprint       = {astro-ph/9805201},
  archiveprefix= {arXiv},
  doi          = {10.1086/300499},
  journal      = {Astron. J.},
  volume       = 116,
  pages        = {1009--1038},
  year         = 1998,
}

@ARTICLE{Riess:2021jrx,
  author       = {Riess, Adam G. and others},
  title        = {{A Comprehensive Measurement of the Local Value of the Hubble Constant with 1 km s$^{−1}$ Mpc$^{−1}$ Uncertainty from the Hubble Space Telescope and the SH0ES Team}},
  eprint       = {2112.04510},
  archiveprefix= {arXiv},
  primaryclass = {astro-ph.CO},
  doi          = {10.3847/2041-8213/ac5c5b},
  journal      = {Astrophys. J. Lett.},
  volume       = 934,
  number       = 1,
  pages        = L7,
  year         = 2022,
}

@ARTICLE{Riess:2024vfa,
  author       = {Riess, Adam G. and others},
  title        = {{JWST Validates HST Distance Measurements: Selection of Supernova Subsample Explains Differences in JWST Estimates of Local H $_{0}$}},
  eprint       = {2408.11770},
  archiveprefix= {arXiv},
  primaryclass = {astro-ph.CO},
  doi          = {10.3847/1538-4357/ad8c21},
  journal      = {Astrophys. J.},
  volume       = 977,
  number       = 1,
  pages        = 120,
  year         = 2024,
}

@ARTICLE{Riess:2025chq,
  author       = {Riess, Adam G. and others},
  title        = {{The Perfect Host: JWST Cepheid Observations in a Background-free Type Ia Supernova Host Confirm No Bias in Hubble-constant Measurements}},
  eprint       = {2509.01667},
  archiveprefix= {arXiv},
  primaryclass = {astro-ph.CO},
  doi          = {10.3847/2041-8213/ae0ad6},
  journal      = {Astrophys. J. Lett.},
  volume       = 992,
  number       = 2,
  pages        = L34,
  year         = 2025,
}

@ARTICLE{RoyChoudhury:2025dhe,
  author       = {Roy Choudhury, Shouvik},
  title        = {{Cosmology in Extended Parameter Space with DESI Data Release 2 Baryon Acoustic Oscillations: A 2{\ensuremath{\sigma}}+ Detection of Nonzero Neutrino Masses with an Update on Dynamical Dark Energy and Lensing Anomaly}},
  eprint       = {2504.15340},
  archiveprefix= {arXiv},
  primaryclass = {astro-ph.CO},
  doi          = {10.3847/2041-8213/ade1cc},
  journal      = {Astrophys. J. Lett.},
  volume       = 986,
  number       = 2,
  pages        = L31,
  year         = 2025,
  note         = {[Erratum: Astrophys.J.Lett. 1001, L25 (2026), Erratum: Astrophys.J. 1001, L25 (2026)]},
}

@ARTICLE{RoyChoudhury:2024wri,
  author       = {Roy Choudhury, Shouvik and Okumura, Teppei},
  title        = {{Updated Cosmological Constraints in Extended Parameter Space with Planck PR4, DESI Baryon Acoustic Oscillations, and Supernovae: Dynamical Dark Energy, Neutrino Masses, Lensing Anomaly, and the Hubble Tension}},
  eprint       = {2409.13022},
  archiveprefix= {arXiv},
  primaryclass = {astro-ph.CO},
  doi          = {10.3847/2041-8213/ad8c26},
  journal      = {Astrophys. J. Lett.},
  volume       = 976,
  number       = 1,
  pages        = L11,
  year         = 2024,
}

@ARTICLE{Ruchika:2024ymt,
  author       = {Ruchika and Perivolaropoulos, Leandros and Melchiorri, Alessandro},
  title        = {{Effects of a local physics change on the SH0ES determination of H0}},
  eprint       = {2408.03875},
  archiveprefix= {arXiv},
  primaryclass = {astro-ph.CO},
  doi          = {10.1103/19pn-3bvs},
  journal      = {Phys. Rev. D},
  volume       = 111,
  number       = 12,
  pages        = 123526,
  year         = 2025,
}

@ARTICLE{Said:2024pwm,
  author       = {Said, Khaled and others},
  title        = {{DESI peculiar velocity survey {\textendash} Fundamental Plane}},
  eprint       = {2408.13842},
  archiveprefix= {arXiv},
  primaryclass = {astro-ph.CO},
  doi          = {10.1093/mnras/staf700},
  journal      = {Mon. Not. Roy. Astron. Soc.},
  volume       = 539,
  number       = 4,
  pages        = {3627--3644},
  year         = 2025,
}

@ARTICLE{Sailer:2025lxj,
  author       = {Sailer, Noah and Farren, Gerrit S. and Ferraro, Simone and White, Martin},
  title        = {{Addressing Tensions in {\ensuremath{\Lambda}}CDM Cosmology by an Increase in the Optical Depth to Reionization}},
  eprint       = {2504.16932},
  archiveprefix= {arXiv},
  primaryclass = {astro-ph.CO},
  doi          = {10.1103/6r54-8lv4},
  journal      = {Phys. Rev. Lett.},
  volume       = 136,
  number       = 8,
  pages        = 081002,
  year         = 2026,
}

@ARTICLE{Santos:2022atq,
  author       = {Santos, F. B. M. dos},
  title        = {{Updating constraints on phantom crossing f(T) gravity}},
  eprint       = {2211.16370},
  archiveprefix= {arXiv},
  primaryclass = {astro-ph.CO},
  doi          = {10.1088/1475-7516/2023/06/039},
  journal      = {JCAP},
  volume       = 06,
  pages        = 039,
  year         = 2023,
}

@ARTICLE{Santos:2025wiv,
  author       = {Santos, Felipe Bruno Medeiros dos and Morais, Jonathan and Pan, Supriya and Yang, Weiqiang and Di Valentino, Eleonora},
  title        = {{A New Window on Dynamical Dark Energy: Combining DESI-DR2 BAO with future Gravitational Wave Observations}},
  eprint       = {2504.04646},
  archiveprefix= {arXiv},
  primaryclass = {astro-ph.CO},
  month        = 4,
  year         = 2025,
}

@ARTICLE{Scherer:2025esj,
  author       = {Scherer, Mateus and Sabogal, Miguel A. and Nunes, Rafael C. and De Felice, Antonio},
  title        = {{Challenging the {\ensuremath{\Lambda}}CDM model: 5{\ensuremath{\sigma}} evidence for a dynamical dark energy late-time transition}},
  eprint       = {2504.20664},
  archiveprefix= {arXiv},
  primaryclass = {astro-ph.CO},
  doi          = {10.1103/n86r-sjgm},
  journal      = {Phys. Rev. D},
  volume       = 112,
  number       = 4,
  pages        = 043513,
  year         = 2025,
}

@ARTICLE{Schombert:2020pxm,
  author       = {Schombert, James and McGaugh, Stacy and Lelli, Federico},
  title        = {{Using the Baryonic Tully{\textendash}Fisher Relation to Measure H o}},
  eprint       = {2006.08615},
  archiveprefix= {arXiv},
  primaryclass = {astro-ph.CO},
  doi          = {10.3847/1538-3881/ab9d88},
  journal      = {Astron. J.},
  volume       = 160,
  number       = 2,
  pages        = 71,
  year         = 2020,
}

@ARTICLE{Schoneberg:2024ifp,
  author       = {Sch{\"o}neberg, Nils},
  title        = {{The 2024 BBN baryon abundance update}},
  eprint       = {2401.15054},
  archiveprefix= {arXiv},
  primaryclass = {astro-ph.CO},
  doi          = {10.1088/1475-7516/2024/06/006},
  journal      = {JCAP},
  volume       = 06,
  pages        = 006,
  year         = 2024,
}

@ARTICLE{Schoneberg:2021qvd,
  author       = {Sch{\"o}neberg, Nils and Franco Abell{\'a}n, Guillermo and P{\'e}rez S{\'a}nchez, Andrea and Witte, Samuel J. and Poulin, Vivian and Lesgourgues, Julien},
  title        = {{The H0 Olympics: A fair ranking of proposed models}},
  eprint       = {2107.10291},
  archiveprefix= {arXiv},
  primaryclass = {astro-ph.CO},
  doi          = {10.1016/j.physrep.2022.07.001},
  journal      = {Phys. Rept.},
  volume       = 984,
  pages        = {1--55},
  year         = 2022,
}

@ARTICLE{Schoneberg:2024ynd,
  author       = {Sch{\"o}neberg, Nils and Vacher, L{\'e}o},
  title        = {{The mass effect {\textemdash} variations of the electron mass and their impact on cosmology}},
  eprint       = {2407.16845},
  archiveprefix= {arXiv},
  primaryclass = {astro-ph.CO},
  doi          = {10.1088/1475-7516/2025/03/004},
  journal      = {JCAP},
  volume       = 03,
  pages        = 004,
  year         = 2025,
}

@ARTICLE{Scolnic:2023mrv,
  author       = {Scolnic, D. and Riess, A. G. and Wu, J. and Li, S. and Anand, G. S. and Beaton, R. and Casertano, S. and Anderson, R. I. and Dhawan, S. and Ke, X.},
  title        = {{CATS: The Hubble Constant from Standardized TRGB and Type Ia Supernova Measurements}},
  eprint       = {2304.06693},
  archiveprefix= {arXiv},
  primaryclass = {astro-ph.CO},
  doi          = {10.3847/2041-8213/ace978},
  journal      = {Astrophys. J. Lett.},
  volume       = 954,
  number       = 1,
  pages        = L31,
  year         = 2023,
}

@ARTICLE{Scolnic:2024oth,
  author       = {Scolnic, Daniel and Boubel, Paula and Byrne, Jakob and Riess, Adam G. and Anand, Gagandeep S.},
  title        = {{Calibrating the Tully-Fisher Relation to Measure the Hubble Constant}},
  eprint       = {2412.08449},
  archiveprefix= {arXiv},
  primaryclass = {astro-ph.CO},
  month        = 12,
  year         = 2024,
}

@ARTICLE{Scolnic:2024hbh,
  author       = {Scolnic, Daniel and others},
  title        = {{The Hubble Tension in Our Own Backyard: DESI and the Nearness of the Coma Cluster}},
  eprint       = {2409.14546},
  archiveprefix= {arXiv},
  primaryclass = {astro-ph.CO},
  doi          = {10.3847/2041-8213/ada0bd},
  journal      = {Astrophys. J. Lett.},
  volume       = 979,
  number       = 1,
  pages        = L9,
  year         = 2025,
}

@ARTICLE{Scolnic:2021amr,
  author       = {Scolnic, Dan and others},
  title        = {{The Pantheon+ Analysis: The Full Data Set and Light-curve Release}},
  eprint       = {2112.03863},
  archiveprefix= {arXiv},
  primaryclass = {astro-ph.CO},
  doi          = {10.3847/1538-4357/ac8b7a},
  journal      = {Astrophys. J.},
  volume       = 938,
  number       = 2,
  pages        = 113,
  year         = 2022,
}

@ARTICLE{Sekiguchi:2020teg,
  author       = {Sekiguchi, Toyokazu and Takahashi, Tomo},
  title        = {{Early recombination as a solution to the $H_0$ tension}},
  eprint       = {2007.03381},
  archiveprefix= {arXiv},
  primaryclass = {astro-ph.CO},
  reportnumber = {KEK-TH-2238},
  doi          = {10.1103/PhysRevD.103.083507},
  journal      = {Phys. Rev. D},
  volume       = 103,
  number       = 8,
  pages        = 083507,
  year         = 2021,
}

@ARTICLE{Seto:2024cgo,
  author       = {Seto, Osamu and Toda, Yo},
  title        = {{DESI constraints on the varying electron mass model and axionlike early dark energy}},
  eprint       = {2405.11869},
  archiveprefix= {arXiv},
  primaryclass = {astro-ph.CO},
  reportnumber = {EPHOU-24-006},
  doi          = {10.1103/PhysRevD.110.083501},
  journal      = {Phys. Rev. D},
  volume       = 110,
  number       = 8,
  pages        = 083501,
  year         = 2024,
}

@ARTICLE{Shlivko:2024llw,
  author       = {Shlivko, David and Steinhardt, Paul J.},
  title        = {{Assessing observational constraints on dark energy}},
  eprint       = {2405.03933},
  archiveprefix= {arXiv},
  primaryclass = {astro-ph.CO},
  doi          = {10.1016/j.physletb.2024.138826},
  journal      = {Phys. Lett. B},
  volume       = 855,
  pages        = 138826,
  year         = 2024,
}

@ARTICLE{Silva:2025hxw,
  author       = {Silva, Emanuelly and Sabogal, Miguel A. and Scherer, Mateus and Nunes, Rafael C. and Di Valentino, Eleonora and Kumar, Suresh},
  title        = {{New constraints on interacting dark energy from DESI DR2 BAO observations}},
  eprint       = {2503.23225},
  archiveprefix= {arXiv},
  primaryclass = {astro-ph.CO},
  doi          = {10.1103/qqc6-76z4},
  journal      = {Phys. Rev. D},
  volume       = 111,
  number       = 12,
  pages        = 123511,
  year         = 2025,
}

@ARTICLE{Silva:2024ift,
  author       = {Silva, Emanuelly and Z{\'u}{\~n}iga-Bola{\~n}o, Ubaldo and Nunes, Rafael C. and Di Valentino, Eleonora},
  title        = {{Non-linear matter power spectrum modeling in interacting dark energy cosmologies}},
  eprint       = {2403.19590},
  archiveprefix= {arXiv},
  primaryclass = {astro-ph.CO},
  doi          = {10.1140/epjc/s10052-024-13487-x},
  journal      = {Eur. Phys. J. C},
  volume       = 84,
  number       = 10,
  pages        = 1104,
  year         = 2024,
}

@ARTICLE{Simon:2024jmu,
  author       = {Simon, Th{\'e}o and Adi, Tal and Bernal, Jos{\'e} Luis and Kovetz, Ely D. and Poulin, Vivian and Smith, Tristan L.},
  title        = {{Toward alleviating the H0 and S8 tensions with early dark energy-dark matter drag}},
  eprint       = {2410.21459},
  archiveprefix= {arXiv},
  primaryclass = {astro-ph.CO},
  doi          = {10.1103/PhysRevD.111.023523},
  journal      = {Phys. Rev. D},
  volume       = 111,
  number       = 2,
  pages        = 023523,
  year         = 2025,
}

@ARTICLE{Smith:2025uaq,
  author       = {Smith, Adam and Mylova, Maria and van de Bruck, Carsten and Burgess, C. P. and Di Valentino, Eleonora},
  title        = {{The Serendipitous Axiodilaton: A Self-Consistent Recombination-Era Solution to the Hubble Tension}},
  eprint       = {2512.13544},
  archiveprefix= {arXiv},
  primaryclass = {astro-ph.CO},
  month        = 12,
  year         = 2025,
}

@ARTICLE{Smith:2025icl,
  author       = {Smith, Adam and {\"O}z{\"u}lker, Emre and Di Valentino, Eleonora and van de Bruck, Carsten},
  title        = {{Dynamical Dark Energy Meets Varying Electron Mass: Implications for Phantom Crossing and the Hubble Constant}},
  eprint       = {2510.21931},
  archiveprefix= {arXiv},
  primaryclass = {astro-ph.CO},
  month        = 10,
  year         = 2025,
}

@ARTICLE{Smith:2019ihp,
  author       = {Smith, Tristan L. and Poulin, Vivian and Amin, Mustafa A.},
  title        = {{Oscillating scalar fields and the Hubble tension: a resolution with novel signatures}},
  eprint       = {1908.06995},
  archiveprefix= {arXiv},
  primaryclass = {astro-ph.CO},
  doi          = {10.1103/PhysRevD.101.063523},
  journal      = {Phys. Rev. D},
  volume       = 101,
  number       = 6,
  pages        = 063523,
  year         = 2020,
}

@ARTICLE{Smith:2025zsg,
  author       = {Smith, Tristan L. and Sch{\"o}neberg, Nils},
  title        = {{Predictions for new physics in the CMB damping tail}},
  eprint       = {2503.20002},
  archiveprefix= {arXiv},
  primaryclass = {astro-ph.CO},
  doi          = {10.1103/sp34-j91r},
  journal      = {Phys. Rev. D},
  volume       = 112,
  number       = 8,
  pages        = 083559,
  year         = 2025,
}

@ARTICLE{Soriano:2025gxd,
  author       = {Soriano, Jorge F. and Wohlberg, Shimon and Anchordoqui, Luis A.},
  title        = {{New insights on a sign-switching \(\Lambda\)}},
  eprint       = {2502.19239},
  archiveprefix= {arXiv},
  primaryclass = {astro-ph.CO},
  doi          = {10.1016/j.dark.2025.101911},
  journal      = {Phys. Dark Univ.},
  volume       = 48,
  pages        = 101911,
  year         = 2025,
}

@ARTICLE{Souza:2024qwd,
  author       = {Souza, Mateus S. and Barcelos, Ana M. and Nunes, Rafael C. and Akarsu, {\"O}zg{\"u}r and Kumar, Suresh},
  title        = {{Mapping the {\ensuremath{\Lambda}}$_{s}$CDM Scenario to f(T) Modified Gravity: Effects on Structure Growth Rate}},
  eprint       = {2501.18031},
  archiveprefix= {arXiv},
  primaryclass = {astro-ph.CO},
  doi          = {10.3390/universe11010002},
  journal      = {Universe},
  volume       = 11,
  number       = 1,
  pages        = 2,
  year         = 2025,
}

@ARTICLE{Specogna:2025guo,
  author       = {Specogna, Enrico and Adil, Shahnawaz A. and Ozulker, Emre and Di Valentino, Eleonora and Nunes, Rafael C. and Akarsu, Ozgur and Sen, Anjan A.},
  title        = {{Updated constraints on omnipotent dark energy: A comprehensive analysis with CMB and BAO data}},
  eprint       = {2504.17859},
  archiveprefix= {arXiv},
  primaryclass = {gr-qc},
  doi          = {10.1103/b7ht-lx26},
  journal      = {Phys. Rev. D},
  volume       = 113,
  number       = 10,
  pages        = 103549,
  year         = 2026,
}

@ARTICLE{Stiskalek:2025ktq,
  author       = {Stiskalek, Richard and Desmond, Harry and Tsaprazi, Eleni and Heavens, Alan and Lavaux, Guilhem and McAlpine, Stuart and Jasche, Jens},
  title        = {{1.8~per{\,}cent measurement of H0 from Cepheids alone}},
  eprint       = {2509.09665},
  archiveprefix= {arXiv},
  primaryclass = {astro-ph.CO},
  doi          = {10.1093/mnras/staf2260},
  journal      = {Mon. Not. Roy. Astron. Soc.},
  volume       = 546,
  number       = 2,
  pages        = {staf2260},
  year         = 2026,
}

@ARTICLE{Tada:2024znt,
  author       = {Tada, Yuichiro and Terada, Takahiro},
  title        = {{Quintessential interpretation of the evolving dark energy in light of DESI observations}},
  eprint       = {2404.05722},
  archiveprefix= {arXiv},
  primaryclass = {astro-ph.CO},
  doi          = {10.1103/PhysRevD.109.L121305},
  journal      = {Phys. Rev. D},
  volume       = 109,
  number       = 12,
  pages        = L121305,
  year         = 2024,
}

@ARTICLE{Tang:2024gtq,
  author       = {Tang, Xin and Ma, Yin-Zhe and Dai, Wei-Ming and He, Hong-Jian},
  title        = {{Constraining holographic dark energy and analyzing cosmological tensions}},
  eprint       = {2407.08427},
  archiveprefix= {arXiv},
  primaryclass = {astro-ph.CO},
  doi          = {10.1016/j.dark.2024.101568},
  journal      = {Phys. Dark Univ.},
  volume       = 46,
  pages        = 101568,
  year         = 2024,
}

@ARTICLE{Toda:2024ncp,
  author       = {Toda, Yo and Giar{\`e}, William and {\"O}z{\"u}lker, Emre and Di Valentino, Eleonora and Vagnozzi, Sunny},
  title        = {{Combining pre- and post-recombination new physics to address cosmological tensions: Case study with varying electron mass and sign-switching cosmological constant}},
  eprint       = {2407.01173},
  archiveprefix= {arXiv},
  primaryclass = {astro-ph.CO},
  doi          = {10.1016/j.dark.2024.101676},
  journal      = {Phys. Dark Univ.},
  volume       = 46,
  pages        = 101676,
  year         = 2024,
}

@ARTICLE{Toda:2025kcq,
  author       = {Toda, Yo and Seto, Osamu},
  title        = {{Constraints on the varying electron mass and early dark energy in light of ACT DR6 and DESI DR2 and the implications for inflation}},
  eprint       = {2508.09025},
  archiveprefix= {arXiv},
  primaryclass = {astro-ph.CO},
  reportnumber = {EPHOU-25-015},
  doi          = {10.1088/1475-7516/2026/02/019},
  journal      = {JCAP},
  volume       = 02,
  pages        = 019,
  year         = 2026,
}

@ARTICLE{Uddin:2023iob,
  author       = {Uddin, Syed A. and others},
  title        = {{Carnegie Supernova Project I and II: Measurements of H $_{0}$ Using Cepheid, Tip of the Red Giant Branch, and Surface Brightness Fluctuation Distance Calibration to Type Ia Supernovae*}},
  eprint       = {2308.01875},
  archiveprefix= {arXiv},
  primaryclass = {astro-ph.CO},
  doi          = {10.3847/1538-4357/ad3e63},
  journal      = {Astrophys. J.},
  volume       = 970,
  number       = 1,
  pages        = 72,
  year         = 2024,
}
